\documentclass[aps,prd,longbibliography,onecolumn,showpacs,amsmath,amssymb,superscriptaddress,nofootinbib,floatfix,preprintnumbers,secnumarabic]{revtex4-1}

\usepackage[usenames,dvipsnames]{color}
\usepackage{multirow}
\usepackage{graphicx}
\usepackage{adjustbox}
\usepackage{booktabs}
\usepackage{appendix}
\usepackage{slashed}
\usepackage{url}
\usepackage{xspace}
\usepackage{braket}
\usepackage{tabularx}
\usepackage[normalem]{ulem} 
\usepackage{siunitx}
\usepackage[utf8]{inputenc}
\usepackage{tikzsymbols}
\usepackage{xspace}
\usepackage{bm}
\usepackage[ISO]{diffcoeff}
\definecolor{darkblue}{rgb}{0,0,0.5}
\definecolor{darkgreen}{rgb}{0.0,0.5,0.2}
\definecolor{darkred}{rgb}{0.6,0,0}
\usepackage[pdfstartview=XYZ,
bookmarks=true,
colorlinks=true,
linkcolor=darkred,
urlcolor=blue,
citecolor=darkgreen,
pdftex,
bookmarks=true,
breaklinks=true,
linktocpage=true, 
hyperindex=true
]{hyperref}
\usepackage[capitalise]{cleveref}
\usepackage{orcidlink}
\usepackage{fullpage}
\usepackage{geometry}      
\AtBeginDocument{
  \heavyrulewidth=.08em
  \lightrulewidth=.05em
  \cmidrulewidth=.03em
  \belowrulesep=.65ex
  \belowbottomsep=0pt
  \aboverulesep=.4ex
  \abovetopsep=0pt
  \cmidrulesep=\doublerulesep
  \cmidrulekern=.5em
  \defaultaddspace= .5em
  \setlength{\tabcolsep}{0.5em}
}

\newcommand{\cevns}{CE$\nu$NS\xspace}
\newcommand{\nue}{E$\nu$ES\xspace}
\newcommand{\eves}{E$\nu$ES\xspace}
\newcommand{\C}[1]{\mathcal{#1}}
\newcommand{\bb}[1]{\boldsymbol{#1}}
\newcommand{\knr}{\mathrm{keV}_\mathrm{nr}\xspace}

\newcommand{\nsi}[1]{\varepsilon^{\eta,\varphi}_{#1}\xspace}
\newcommand{\dCP}{\delta_\mathrm{CP}}
\newcommand{\code}{\texttt{SNuDD}\xspace}

\renewcommand{\vec}[1]{\bm{#1}}

\newcommand\funop[1]{\mathop{{}#1}}

\DeclareSIUnit\fb{\femto\barn}
\DeclareSIUnit\ton{ton}
\DeclareSIUnit\year{yr} 
\DeclareSIUnit\kevnr{\mathrm{keV}_\mathrm{nr}\xspace}
\DeclareSIUnit\kevee{\mathrm{keV}_\mathrm{ee}\xspace}
\DeclareSIUnit\parsec{pc}

\definecolor{lz22}{HTML}{02ABD2}
\definecolor{global_fits}{HTML}{c44d51}
\definecolor{borexino_fit}{HTML}{218a43}
\definecolor{xnt22}{HTML}{FB6A24}

\makeatletter
\def\@ssect@ltx#1#2#3#4#5#6[#7]#8{%
  \def\H@svsec{\phantomsection}%
  \@tempskipa #5\relax
  \@ifdim{\@tempskipa>\z@}{%
    \begingroup
      \interlinepenalty \@M
      #6{%
       \@ifundefined{@hangfroms@#1}{\@hang@froms}{\csname @hangfroms@#1\endcsname}%
       {\hskip#3\relax\H@svsec}{#8}%
      }%
      \@@par
    \endgroup
    \@ifundefined{#1smark}{\@gobble}{\csname #1smark\endcsname}{#7}%
  }{%
    \def\@svsechd{%
      #6{%
       \@ifundefined{@runin@tos@#1}{\@runin@tos}{\csname @runin@tos@#1\endcsname}%
       {\hskip#3\relax\H@svsec}{#8}%
      }%
      \@ifundefined{#1smark}{\@gobble}{\csname #1smark\endcsname}{#7}%
      \addcontentsline{toc}{#1}{\protect\numberline{}#8}%
    }%
  }%
  \@xsect{#5}%
}%
\makeatother

\begin{document}

\preprint{IPPP/23/08}
\preprint{IFT-UAM/CSIC-23-19}
\preprint{FT-UAM-23-1}

\vspace*{0.7cm}

\title{A direct detection view of the neutrino NSI landscape}
\date{February 28, 2023}

\author{Dorian Amaral\,\orcidlink{0000-0002-1414-932X}}
\email{dorian.amaral@rice.edu}
\affiliation{Institute for Particle Physics Phenomenology, Durham University, Durham DH1 3LE, United Kingdom}
\affiliation{Department of Physics and Astronomy, Rice University, Houston, TX 77005, USA}
\author{David Cerde\~no\,\orcidlink{0000-0002-7649-1956}}
\email{davidg.cerdeno@uam.es}
\affiliation{Instituto de F\' isica Te\'orica, Universidad Aut\'onoma de Madrid, 28049 Madrid, Spain} 
\affiliation{Departamento de F\' isica Te\'orica, Universidad Aut\'onoma de Madrid, 28049 Madrid, Spain} 
\author{Andrew Cheek\,\orcidlink{0000-0002-8773-831X}}
\email{acheek@camk.edu.pl}
\affiliation{Astrocent, Nicolaus Copernicus Astronomical Center of the Polish Academy of Sciences, ul.Rektorska 4, 00-614 Warsaw, Poland} 
\author{Patrick Foldenauer\,\orcidlink{0000-0003-4334-4228}}
\email{patrick.foldenauer@csic.es}
\affiliation{Institute for Particle Physics Phenomenology, Durham University, Durham DH1 3LE, United Kingdom} 
\affiliation{Instituto de F\' isica Te\'orica, Universidad Aut\'onoma de Madrid, 28049 Madrid, Spain}

\begin{abstract}

\vspace*{2ex}
In this article, we study the potential of direct detection experiments to explore the parameter space of general non-standard neutrino interactions (NSI) via solar neutrino scattering. Due to their sensitivity to neutrino-electron and neutrino-nucleus scattering, direct detection provides a complementary view of the NSI landscape to that of spallation sources and neutrino oscillation experiments. In particular, the large admixture of tau neutrinos in the solar flux makes direct detection experiments well-suited to probe the full flavour space of NSI. To study this, we develop a re-parametrisation of the NSI framework that explicitly includes a variable electron contribution and allows for a clear visualisation of the complementarity of the different experimental sources. Using this new parametrisation, we explore how previous bounds from spallation source and neutrino oscillation experiments are impacted. For the first time, we compute limits on NSI from the first results of the XENONnT and LUX-ZEPLIN experiments, and we obtain projections for future xenon-based experiments.
These computations have been performed with our newly developed software package, \href{https://github.com/SNuDD/SNuDD.git}{\texttt{SNuDD}}.
Our results demonstrate the importance of using a more general NSI parametrisation and indicate that next generation direct detection experiments will become powerful probes of neutrino NSI.
\end{abstract}

\maketitle
 
\tableofcontents

\section{Introduction}
\label{sec:introduction}

Neutrinos are among the most mysterious particles of the Standard Model (SM). The discovery of their flavour oscillations remains one of the strongest pieces of evidence for new physics, since it requires neutrinos to be massive~\cite{Davis:1968cp,SNO:2001kpb}. In the SM, however, neutrinos are described by a purely left-handed spinorial field forming part of an $SU(2)_L$ doublet~\cite{Weinberg:1967tq}, which, by the principles of gauge invariance, disallows a mass term for neutrinos at the renormalisable level. Hence, the neutrino sector has inspired a myriad of extensions beyond the SM (BSM) (see, for example, Ref.~\cite{deGouvea:2016qpx} for a review).

A convenient parametrisation of new physics in neutrino interactions  has been established in terms of the low-energy  effective field theory (EFT) of \textit{non-standard interactions} (NSI)~\cite{Wolfenstein:1977ue,Guzzo:1991hi,Guzzo:1991cp,Gonzalez-Garcia:1998ryc,Bergmann:2000gp,Guzzo:2000kx,Guzzo:2001mi,GonzalezGarcia:2004wg}. This formalism contemplates modifications to neutrino  interactions with SM particles while respecting the SM vector current structure. Over the last decades, a variety of experimental bounds have been derived for the NSI couplings~\cite{Fornengo:2001pm,Huber:2002bi,Davidson:2003ha,Maltoni:2004ei,Kopp:2007mi,Kopp:2007ne,Kopp:2008ds,Antusch:2008tz,Bolanos:2008km,Biggio:2009kv,Biggio:2009nt,Escrihuela:2009up,Kopp:2010qt,Escrihuela:2011cf,Agarwalla:2014bsa,Gonzalez-Garcia:2015qrr,Liao:2016orc,Liao:2016reh,Farzan:2017xzy,Coloma:2017ncl,Liao:2017awz,Coloma:2017egw,Liao:2017uzy,Esteban:2018ppq,Denton:2018xmq,AristizabalSierra:2018eqm,Coloma:2019mbs,Han:2019zkz,Esteban:2019lfo,Proceedings:2019qno,Miranda:2020tif,Denton:2020hop,Martinez-Mirave:2021cvh,Masud:2021ves,Chatterjee:2021wac,Chaves:2021pey,deSalas:2021aeh,IceCubeCollaboration:2021euf,Denton:2022nol,Denton:2022pxt,DeRomeri:2022twg,Brahma:2023wlf}, most importantly from neutrino oscillation and spallation source experiments. The latter have recently succeeded in observing the coherent elastic scattering of neutrinos with nuclei (\cevns) \cite{COHERENT:2015mry,COHERENT:2017ipa} with a rate consistent with the SM prediction \cite{Freedman:1973yd,Drukier:1984vhf}, providing stronger constraints on new physics contributions (see, for example, Refs.~\cite{Kosmas:2017zbh,Miranda:2020syh,Abdullah:2018ykz,Bauer:2018onh,Miranda:2020zji,Bauer:2020itv,Kosmas:2015sqa}).

Meanwhile, dark matter (DM) direct detection (DD) experiments have experienced remarkable progress. Current detectors have significantly increased their target size and sensitivity, to the point where they will be able to observe the scattering of solar neutrinos. This constitutes a new background for DM searches, which leads to the so-called neutrino floor (or fog) \cite{Billard:2013qya, Billard:2021uyg,OHare:2021utq}, but it also offers the unique opportunity to probe new physics with these instruments \cite{Cerdeno:2016sfi,Dutta:2017nht,Gelmini:2018gqa,Essig:2018tss, Amaral:2020tga, Dutta:2020che, Amaral:2021rzw,Munoz:2021sad,deGouvea:2021ymm}. Neutrinos can be observed in DD experiment through elastic neutrino-electron scatting (E$\nu$ES) or their coherent elastic scattering with nuclei. Due to their larger target size, liquid noble gas detectors like LUX-ZEPLIN (LZ)~\cite{Aalbers:2022fxq}, PandaX~\cite{PandaX-4T:2021bab}, and XENONnT~\cite{XENONCollaboration:2022kmb}, are better positioned than other DD techniques to carry out this kind of search. Indeed, the larger xenon-based DD experiments have thus far succeeded in placing upper bounds on the $\mathrm{^8B}$ neutrino flux \cite{XENON:2020rca, PandaX:2022aac}.
The sensitivity of DD experiments to these processes has already led to  studies in which the expected solar neutrino scattering rate has been used as a laboratory for gaining a deeper understanding of the nature of solar physics, neutrino oscillations, and BSM neutrino physics \cite{Harnik:2012ni, Baudis:2013qla, Billard:2014yka, Cerdeno:2016sfi,Cui:2017ytb,Dutta:2017nht,Cerdeno:2017xxl, Newstead:2018muu,Essig:2018tss,AristizabalSierra:2019ykk,Dutta:2019oaj,Dutta:2020che,deGouvea:2021ymm,A:2022acy}.

In this work, we set out to exploit the sensitivity of DD experiments to solar neutrino scattering with the aim of exploring their impact on the NSI landscape. In the context of dedicated neutrino experiments, NSI studies are numerous; however, the potential of DD experiments has not been fully investigated. Previous works have pointed out that non-zero NSI parameters could produce appreciable signals for both \cevns and E$\nu$ES \cite{Dutta:2017nht, Dutta:2020che} in DD, as well as potentially modify the neutrino fog~\cite{AristizabalSierra:2017joc,Gonzalez-Garcia:2018dep}.

To this end, we will introduce a convenient parametrisation of NSI, extending the framework of Ref.~\cite{Esteban:2018ppq} to include an explicit separation between NSI in the electron and proton directions. This is needed to interpret the results of DD experiments. Ignoring the electron contribution is a valid choice as long as one is mostly interested in matter effects for oscillation experiments\footnote{Non-standard matter effects enter the matter Hamiltonian via a contribution from the neutron and an overall charged contribution from both the proton and the electron.}, but this is a non-general treatment. Striving for greater generality, we allow for the possibility that the `charged' neutrino NSI is shared between both the proton and the electron. While the total charged contribution can be designed to leave neutrino oscillations unchanged, allowing for electron NSI can instead lead to changes in the \nue cross section. This, in turn, can affect the bounds set by oscillation experiments, which could instead be dominated by NSI effects at the detection point~\cite{Esteban:2018azc}.

Furthermore, as was recently pointed out in Ref.~\cite{Coloma:2022umy}, when new physics introduces potential flavour-changing neutral currents (FCNC), the full flavour structure of the cross section must be retained when dealing with a neutrino flux composed of an admixture of flavour eigenstates. This is in contrast with the SM, where interactions are diagonal in the flavour basis. Thus, in the general NSI case, it is no longer appropriate to project the neutrino state that reaches Earth  onto any one particular flavour state and convolve the result with flavour-pure cross sections, as neutrinos arrive in a superposition of flavour eigenstates.\footnote{We stress that the simplified treatment of calculating the number of neutrino scattering events in the presence of new physics, given by $N_\nu \propto \sum_\alpha P_{e\alpha} \,{\mathrm{d}\sigma_{\nu_\alpha T}}/{\mathrm{d}E_R}$, where $P_{e\alpha}$ is the transition probability to a neutrino of flavour $\alpha$, is only appropriate in two cases. Firstly, if the flux of neutrinos incident on a target is only composed of one flavour. Secondly, if the new physics contribution is flavour-conserving.} Instead, we must consider the full flavour-structure of both the cross section and the density matrix describing the evolution of the initial neutrino state.

The rate of neutrino  events in a generic neutrino scattering experiment is then described by the expression~\cite{Coloma:2022umy},
\begin{equation}\label{eq:dr_gen}
    \diff{R}{E_R}=  N_T \int_{E_\nu^\mathrm{min}} \diff{\phi_\nu}{E_\nu}\,  \mathrm{Tr}\left[\bb{\rho}\, \diff{\bb{\zeta}}{ E_R}\right] \, \dl E_\nu\, ,
\end{equation}
where $N_T$ is the number of targets, $\phi_\nu$ is the neutrino flux at the source\footnote{Of particular relevance to experiments sensitive to solar neutrinos is the fact that electron neutrino production in the Sun proceeds through a series of charged-current interactions. Since we are only considering neutral-current NSI, the electron neutrino flux produced in the Sun is unchanged.}, $\bb\rho$ is the neutrino density matrix at the experiment and $\bb\zeta$ is a generalised scattering cross section in the neutrino-flavour space, encoding correlations between scattering amplitudes of neutrinos with different flavours. Here, $E_\nu$ is the energy of the incident neutrinos and $E_\nu^\mathrm{min}$ is the minimum $E_\nu$ required  to produce a target recoil energy of $E_R$.

Using this generalised framework, in this paper we study how DD experiments will constitute a valuable complementary probe of the NSI landscape. To do this, we first explore how previous limits derived from oscillation and spallation source  experiments map onto the full NSI parameter space. Then, we derive new limits from the recent data from LZ and XENONnT onto the NSI parameters and make projections for their full exposure runs, as well as for the future DARWIN experiment~\cite{DARWIN:2016hyl}. We demonstrate that xenon-based DD experiments like XENON~\cite{XENON:2018voc,XENON:2020kmp,XENON:2020rca}, DARWIN~\cite{DARWIN:2016hyl}, PandaX~\cite{PandaX-II:2017hlx,PandaX:2018wtu,PandaX-4T:2021bab} and LZ~\cite{Mount:2017qzi,LUX-ZEPLIN:2018poe,LZ:2021xov} will be sensitive to generic NSI couplings in a competitive and complementary way to oscillation, beam, and spallation source experiments. We do this by comparing our results and projections to those derived in Refs.~\cite{Farzan:2017xzy,Esteban:2018ppq,Coloma:2019mbs,Dutta:2020che,Coloma:2022umy}. Given that DD will be sensitive to both \cevns and \eves, we explore their limits and projections in our extended parametrisation, emphasising the complementarity of both signals. Indeed, due to the high flux of solar neutrinos and the excellent background reduction of DD experiments, their sensitivity to electron NSI is remarkable and is competitive with that of conventional neutrino oscillation experiments.

\bigskip

This article is organised as follows. In \cref{sec:framework}, we introduce the framework of non-standard neutrino interactions, explicitly incorporating interactions with electrons as well as the impact of such NSI on solar neutrino physics. We then derive the relevant formalism for computing the density matrix and the generalised cross section, both required to compute the expected solar neutrino scattering rates. In~\cref{sec:spall_and_osc}, we shed light on the current landscape of NSI constraints derived from oscillation and spallation source experiments, as well as their sensitivity to interactions with electrons. We present and discuss the results of our sensitivity study of DD experiments to NSI as the main results of this work in~\cref{sec:dd}. Finally, we  draw  our conclusions in \cref{sec:conclusions}.

\section{Solar Neutrino Physics and Non-Standard Interactions}
\label{sec:framework}

In this section, we introduce the framework of neutrino NSI and study their impact on solar neutrino physics, both in propagation effects and in scattering  with nuclei and electrons. In doing so, we derive the relevant expressions for the solar neutrino density matrix, $\boldsymbol{\rho}$, and the generalised cross section, $\boldsymbol{\zeta}$, for \cevns and \eves entering the rate in~\cref{eq:dr_gen}. For an explanation of the origin of this rate equation, we refer to~\cref{sec:amplitude}.

\subsection{NSI parametrisation}
\label{sec:parametrisation}

In order to understand how potential new neutrino interactions enter the scattering rate in~\cref{eq:dr_gen}, we need to specify a BSM model. Since we want to remain as general as possible about the origin of such new physics, we will work in terms of a low-energy effective theory. Making  use of the framework of neutrino NSI~\cite{Wolfenstein:1977ue, Miranda:2015dra, Davidson:2003ha,GonzalezGarcia:2004wg}, we can parametrise  new physics effects in the neutrino sector by contact terms of the form\footnote{Note that this parametrisation is not $SU(2)_L$ invariant and is mainly motivated by the structure of the SM weak current. In order to systematically capture all gauge invariant dimension-six operators modifying neutrino interactions, it is more suitable to consider a complete basis of EFT operators and map them onto the enlarged basis of general neutrino interactions~\cite{Gavela:2008ra,Lindner:2016wff,Bischer:2019ttk,Terol-Calvo:2019vck}.}
\begin{equation}\label{eq:eff_op}
    \C{L}_\text{NSI} = -2\sqrt{2}\, G_F \sum_{\substack{f=e,u,d\\\alpha,\beta=e,\mu,\tau}} \varepsilon^{fP}_{\alpha\beta} \ \left[\bar\nu_\alpha \gamma_\rho P_L \nu_\beta\right] \, \left[\bar f \gamma^\rho P f \right]\,,
\end{equation}
where  $G_F$ denotes the Fermi constant and $P\in\{P_L,P_R\}$. The NSI parameters $\varepsilon^{fP}_{\alpha\beta}$, which are in general flavour-violating, quantify the strength of the interaction between the neutrinos $\nu_\alpha$ and $\nu_\beta$ and the pair of fermions $f$ relative to the SM weak interaction, characterised by $G_F$. In this work, we will not consider any new source of CP-violation and hence assume the parameters  $\varepsilon^{fP}_{\alpha\beta}$ to be real. Furthermore, we have assumed that the charged fermions $f$ are identical, resembling the SM neutral current (NC) interaction. However, charged current (CC) NSI could also exist, where the neutrinos couple to two different charged fermions, $f$ and $f'$. Since these are, in general, subject to much harsher constraints and DD experiments do not probe CC interactions in NRs, we do not consider them here but rather direct the reader to, for example, Refs.~\cite{Kopp:2010qt,Akhmedov:2010vy,Ohlsson:2012kf,Miranda:2015dra,Farzan:2017xzy,Khan:2021wzy,Falkowski:2021bkq}.

\begin{figure}[t]
    \centering
    \tikzset{every picture/.style={line width=0.75pt}} 

\begin{tikzpicture}[x=0.75pt,y=0.75pt,yscale=-1,xscale=1]

\draw  [draw opacity=0][fill={rgb, 255:red, 128; green, 128; blue, 128 }  ,fill opacity=0.17 ] (229.89,189.77) .. controls (229.89,129.02) and (279.13,79.77) .. (339.89,79.77) .. controls (400.64,79.77) and (449.89,129.02) .. (449.89,189.77) .. controls (449.89,250.52) and (400.64,299.77) .. (339.89,299.77) .. controls (279.13,299.77) and (229.89,250.52) .. (229.89,189.77) -- cycle ;
\draw  [draw opacity=0][fill={rgb, 255:red, 74; green, 144; blue, 226 }  ,fill opacity=0.66 ][dash pattern={on 4.5pt off 4.5pt}] (431.51,174.72) .. controls (432.15,179.86) and (432.47,185.11) .. (432.47,190.46) .. controls (432.47,250.64) and (391.07,299.46) .. (339.89,299.77) -- (339.39,190.46) -- cycle ; \draw  [color={rgb, 255:red, 74; green, 144; blue, 226 }  ,draw opacity=1 ][dash pattern={on 4.5pt off 4.5pt}] (431.51,174.72) .. controls (432.15,179.86) and (432.47,185.11) .. (432.47,190.46) .. controls (432.47,250.64) and (391.07,299.46) .. (339.89,299.77) ;
\draw    (340,320) -- (340,74.5) -- (340,62) ;
\draw [shift={(340,60)}, rotate = 90] [color={rgb, 255:red, 0; green, 0; blue, 0 }  ][line width=0.75]    (7.65,-2.3) .. controls (4.86,-0.97) and (2.31,-0.21) .. (0,0) .. controls (2.31,0.21) and (4.86,0.98) .. (7.65,2.3)   ;
\draw    (210,190) -- (468,190) ;
\draw [shift={(470,190)}, rotate = 180] [color={rgb, 255:red, 0; green, 0; blue, 0 }  ][line width=0.75]    (7.65,-2.3) .. controls (4.86,-0.97) and (2.31,-0.21) .. (0,0) .. controls (2.31,0.21) and (4.86,0.98) .. (7.65,2.3)   ;
\draw    (229,233.5) -- (453.64,146.22) ;
\draw [shift={(455.5,145.5)}, rotate = 158.77] [color={rgb, 255:red, 0; green, 0; blue, 0 }  ][line width=0.75]    (7.65,-2.3) .. controls (4.86,-0.97) and (2.31,-0.21) .. (0,0) .. controls (2.31,0.21) and (4.86,0.98) .. (7.65,2.3)   ;
\draw  [draw opacity=0] (449.55,189.77) .. controls (449.58,190.04) and (449.59,190.31) .. (449.59,190.58) .. controls (449.59,207.15) and (400.32,220.58) .. (339.55,220.58) .. controls (316.38,220.58) and (294.89,218.63) .. (277.15,215.29) -- (339.55,190.58) -- cycle ; \draw  [color={rgb, 255:red, 208; green, 2; blue, 27 }  ,draw opacity=1 ] (449.55,189.77) .. controls (449.58,190.04) and (449.59,190.31) .. (449.59,190.58) .. controls (449.59,207.15) and (400.32,220.58) .. (339.55,220.58) .. controls (316.38,220.58) and (294.89,218.63) .. (277.15,215.29) ;
\draw  [draw opacity=0][dash pattern={on 4.5pt off 4.5pt}] (402.99,165.96) .. controls (431.19,171.39) and (449.59,180.34) .. (449.59,190.46) .. controls (449.59,190.48) and (449.59,190.5) .. (449.59,190.52) -- (339.39,190.46) -- cycle ; \draw  [color={rgb, 255:red, 208; green, 2; blue, 27 }  ,draw opacity=1 ][dash pattern={on 4.5pt off 4.5pt}] (402.99,165.96) .. controls (431.19,171.39) and (449.59,180.34) .. (449.59,190.46) .. controls (449.59,190.48) and (449.59,190.5) .. (449.59,190.52) ;
\draw  [draw opacity=0][fill={rgb, 255:red, 209; green, 59; blue, 75 }  ,fill opacity=0.8 ] (402.99,165.96) .. controls (431.66,171.37) and (450.43,180.41) .. (450.43,190.64) .. controls (450.43,207.21) and (401.18,220.64) .. (340.43,220.64) .. controls (317.18,220.64) and (295.62,218.67) .. (277.85,215.31) -- (340.43,190.64) -- cycle ; \draw  [color={rgb, 255:red, 0; green, 0; blue, 0 }  ,draw opacity=0 ] (402.99,165.96) .. controls (431.66,171.37) and (450.43,180.41) .. (450.43,190.64) .. controls (450.43,207.21) and (401.18,220.64) .. (340.43,220.64) .. controls (317.18,220.64) and (295.62,218.67) .. (277.85,215.31) ;
\draw  [draw opacity=0][fill={rgb, 255:red, 74; green, 144; blue, 226 }  ,fill opacity=0.66 ][dash pattern={on 4.5pt off 4.5pt}] (340.44,81.16) .. controls (386.74,81.76) and (424.93,122.06) .. (431.48,174.44) -- (339.39,190.46) -- cycle ; \draw  [color={rgb, 255:red, 74; green, 144; blue, 226 }  ,draw opacity=1 ][dash pattern={on 4.5pt off 4.5pt}] (340.44,81.16) .. controls (386.74,81.76) and (424.93,122.06) .. (431.48,174.44) ;
\draw [color={rgb, 255:red, 0; green, 0; blue, 0 }  ,draw opacity=1 ]   (357.5,166.75) .. controls (365,173.25) and (364.5,181.25) .. (364.5,186) ;
\draw [color={rgb, 255:red, 0; green, 0; blue, 0 }  ,draw opacity=1 ][fill={rgb, 255:red, 0; green, 0; blue, 0 }  ,fill opacity=1 ] [dash pattern={on 4.5pt off 4.5pt}]  (340,190) -- (398.8,111.6) ;
\draw [shift={(400,110)}, rotate = 126.87] [fill={rgb, 255:red, 0; green, 0; blue, 0 }  ,fill opacity=1 ][line width=0.08]  [draw opacity=0] (12,-3) -- (0,0) -- (12,3) -- cycle    ;
\draw [color={rgb, 255:red, 0; green, 0; blue, 0 }  ,draw opacity=1 ]   (390,181.5) .. controls (392.25,183.5) and (392.5,185.75) .. (392,190) ;
\draw [color={rgb, 255:red, 0; green, 0; blue, 0 }  ,draw opacity=1 ][fill={rgb, 255:red, 208; green, 2; blue, 27 }  ,fill opacity=1 ] [dash pattern={on 0.84pt off 2.51pt}]  (400,110) -- (400,180) ;
\draw [color={rgb, 255:red, 0; green, 0; blue, 0 }  ,draw opacity=1 ] [dash pattern={on 0.84pt off 2.51pt}]  (340,190) -- (400,180) ;
\draw  [draw opacity=0][fill={rgb, 255:red, 0; green, 0; blue, 0 }  ,fill opacity=0.07 ] (277.02,214.76) .. controls (248.02,209.36) and (228.98,200.27) .. (228.98,189.97) .. controls (228.98,189.6) and (229.01,189.23) .. (229.06,188.86) -- (339.03,189.97) -- cycle ; \draw  [color={rgb, 255:red, 0; green, 0; blue, 0 }  ,draw opacity=1 ] (277.02,214.76) .. controls (248.02,209.36) and (228.98,200.27) .. (228.98,189.97) .. controls (228.98,189.6) and (229.01,189.23) .. (229.06,188.86) ;
\draw  [draw opacity=0][fill={rgb, 255:red, 0; green, 0; blue, 0 }  ,fill opacity=0.07 ][dash pattern={on 4.5pt off 4.5pt}] (228.99,189.7) .. controls (229.51,173.26) and (278.58,159.97) .. (339.03,159.97) .. controls (362.8,159.97) and (384.81,162.02) .. (402.8,165.52) -- (339.03,189.97) -- cycle ; \draw  [color={rgb, 255:red, 0; green, 0; blue, 0 }  ,draw opacity=1 ][dash pattern={on 4.5pt off 4.5pt}] (228.99,189.7) .. controls (229.51,173.26) and (278.58,159.97) .. (339.03,159.97) .. controls (362.8,159.97) and (384.81,162.02) .. (402.8,165.52) ;

\draw (462,128) node [anchor=north west][inner sep=0.75pt]   [align=left] {$\displaystyle \hat{\varepsilon} ^{e}_{\alpha\beta}$};
\draw (476,178) node [anchor=north west][inner sep=0.75pt]   [align=left] {$\displaystyle \hat{\varepsilon} ^{p}_{\alpha\beta}$};
\draw (331,38) node [anchor=north west][inner sep=0.75pt]   [align=left] {$\displaystyle \hat{\varepsilon} ^{n}_{\alpha\beta}$};
\draw (350,128) node [anchor=north west][inner sep=0.75pt]  [color={rgb, 255:red, 0; green, 0; blue, 0 }  ,opacity=1 ] [align=left] {$\displaystyle \vec{\varepsilon_{\alpha\beta}}$};
\draw (367.75,166) node [anchor=north west][inner sep=0.75pt]  [font=\footnotesize,color={rgb, 255:red, 0; green, 0; blue, 0 }  ,opacity=1 ] [align=left] {$\displaystyle \eta $};
\draw (400,178.5) node [anchor=north west][inner sep=0.75pt]  [font=\footnotesize,color={rgb, 255:red, 0; green, 0; blue, 0 }  ,opacity=1 ] [align=left] {$\displaystyle \varphi $};

\end{tikzpicture}
    \caption{Extended NSI parametrisation. A given NSI is defined by the radial component, $\sqrt{5}\, \varepsilon_{\alpha\beta}^{\eta,\varphi}$ (which can be either positive or negative), the angle between the charged $(\hat\varepsilon^{p}_{\alpha\beta}, \hat\varepsilon^{e}_{\alpha\beta})$-plane and the neutron direction, $\eta$, and the new angle $\varphi$, which defines the NSI direction along either the proton or the electron component. The domains of these angles are $\eta,\,\varphi\in [-\pi/2, \pi/2]$, as visualised by the blue and red semicircles, respectively.}
    \label{fig:parametrisation}
\end{figure}
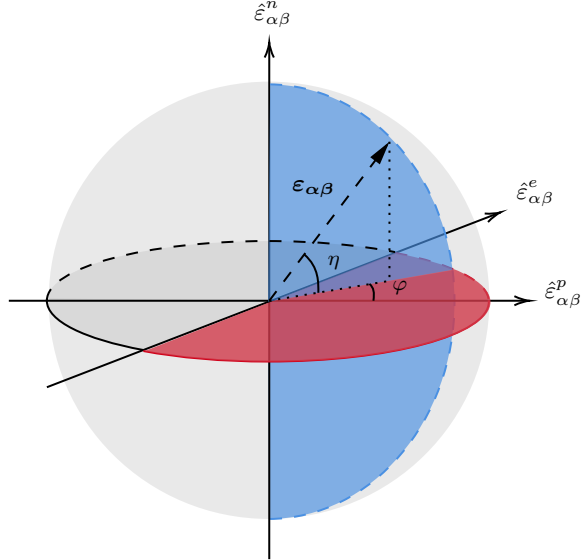

To describe neutrinos interacting with ordinary matter (made up of electrons, protons and neutrons), only interactions with the first generation of SM fermions need to be considered. If we assume that the neutrino flavour structure of the NSI is independent of the charged fermion $f$ that the neutrinos  couple to, we can factorise the NSI coupling as~\cite{Esteban:2018ppq}
\begin{equation}
    \varepsilon^{fP}_{\alpha\beta} = \varepsilon^{\eta,\varphi}_{\alpha\beta} \, \xi^{fP}\,,
\end{equation}
where $\xi^{fP}$ describes the relative strength of the interaction with the fermions $f\in \{e, u, d\}$ and $\varepsilon^{\eta,\varphi}_{\alpha\beta}$ denotes the overall strength of the NSI. We further define the vector and axial-vector NSI couplings
\begin{equation}
\begin{aligned}
     \varepsilon^{f}_{\alpha\beta} & =  \varepsilon^{fL}_{\alpha\beta} +  \varepsilon^{fR}_{\alpha\beta} =  \varepsilon^{\eta,\varphi}_{\alpha\beta} \, \xi^{f}\,, 
     \\
     \tilde\varepsilon^{f}_{\alpha\beta} & =  \varepsilon^{fL}_{\alpha\beta} -  \varepsilon^{fR}_{\alpha\beta} =  \tilde\varepsilon^{\eta,\varphi}_{\alpha\beta} \, \tilde\xi^{f}\,, 
\end{aligned}
\end{equation}
with $\xi^{f} = \xi^{fL} + \xi^{fR}$ and $\tilde\xi^{f} = \xi^{fL} - \xi^{fR}$.
As matter effects are only sensitive to the vector part of the interaction, we focus only on vector NSI in this work, setting $\tilde\varepsilon_{\alpha\beta}$ to zero. Since  we are ultimately testing neutrino interactions with matter, it is convenient to parametrise the NSI with quarks in terms of proton and neutron NSI,
\begin{equation}
\begin{aligned}
    \varepsilon^p_{\alpha\beta} =  2\, \varepsilon_{\alpha\beta}^u + \varepsilon_{\alpha\beta}^d \,,  
    \\
    \varepsilon^n_{\alpha\beta} = \varepsilon_{\alpha\beta}^u + 2\, \varepsilon_{\alpha\beta}^d \,.
    \label{eq:nucl_nsi}
\end{aligned}
\end{equation}
Extending the parametrisation of Ref.~\cite{Esteban:2018ppq} by re-introducing the electron direction via a second angle $\varphi$, the relative strengths of the electron, proton, and neutron NSI are written as\footnote{The normalisation factor of $\sqrt{5}$ was originally introduced in Ref.~\cite{Esteban:2018ppq} to have unit vectors $\xi^u$ and $\xi^d$ if the NSI are entirely in the up- and down-quark direction, respectively. We cohere to this normalisation for comparability of our results with the literature on NSI global fits.},
\begin{align}
    \xi^e &=\sqrt{5}\, \cos\eta\, \sin\varphi \,, \nonumber
    \\
    \xi^p &=\sqrt{5}\, \cos\eta\, \cos\varphi \,,  
    \\
    \xi^n &=\sqrt{5}\, \sin\eta \,. \nonumber
\end{align}
In~\cref{fig:parametrisation}, we illustrate our parametrisation of the three base NSI directions $\hat\varepsilon^p_{\alpha\beta}, \hat\varepsilon^n_{\alpha\beta}$ and $\hat\varepsilon^e_{\alpha\beta}$. We define the angle $\eta$ as the angle of general NSI coupling $\vec{\varepsilon_{\alpha\beta}} = (\varepsilon^p_{\alpha\beta},\varepsilon^n_{\alpha\beta},\varepsilon^e_{\alpha\beta})$ with the plane of charged NSI $(\hat\varepsilon^p_{\alpha\beta},\hat\varepsilon^e_{\alpha\beta})$. The second angle $\varphi$ is defined as the angle between the general NSI element $\vec{\varepsilon_{\alpha\beta}}$ and the plane of hadronic NSI $(\hat\varepsilon^p_{\alpha\beta},\hat\varepsilon^n_{\alpha\beta})$. In order to match our notation with the literature on global NSI fits~\cite{Esteban:2018ppq,Coloma:2019mbs}, we allow for both {\it positive} and {\it negative} values for $\varepsilon^{\eta,\varphi}_{\alpha\beta}$. Thus, the azimuthal angle $\eta$ only runs in the interval $[-\pi/2,\pi/2]$ to span the full two-dimensional plane of hadronic NSI $(\varepsilon^p_{\alpha\beta},\varepsilon^n_{\alpha\beta})$. The second, polar angle $\varphi$ (taken from the hadronic NSI plane) also runs in the interval $[-\pi/2,\pi/2]$ to cover the full sphere. For example, $\eta = 0$ and $\varphi=0$ corresponds to NSI only in the proton direction $\hat\varepsilon^p_{\alpha\beta}$, $\eta = 0$ and $\varphi=\pi/2$ to NSI only in the electron direction $\hat\varepsilon^e_{\alpha\beta}$, and $\eta = \pi/2$ to NSI only in the neutron direction $\hat\varepsilon^n_{\alpha\beta}$.

\subsection{Three-flavour neutrino oscillations in the presence of NSI}
\label{subsec:three_nu_osc}

With this extended framework, we can describe the evolution of neutrino and antineutrino states during propagation in the Hamiltonian formalism by 
\begin{equation}\label{eq:hamiltonian3}
\begin{aligned}
    H^{\nu} &= H_\mathrm{vac} + H_\mathrm{mat} \,, \\ 
    H^{\bar\nu} &= (H_\mathrm{vac} - H_\mathrm{mat})^* \,,
\end{aligned}
\end{equation}
where the standard vacuum Hamiltonian is given by 
\begin{equation}
H_\mathrm{vac} = U_\mathrm{PMNS}\ \frac{1}{2 E_\nu}\left(\begin{matrix}
0 & 0 & 0 \\
0 & \Delta m_{21}^{2} & 0 \\
0 & 0 & \Delta m_{31}^{2}
\end{matrix}\right) U_\mathrm{PMNS}^{\dagger} \,,
\end{equation}
with $\Delta m_{ij}^2 \equiv m_i^2 - m_j^2$ and  $U_\mathrm{PMNS}$ being the PMNS matrix, defined as
\begin{equation}
    U_\mathrm{PMNS} = 
    \underbrace{
    \begin{pmatrix}
    1 && 0 && 0 \\
    0 && c_{23} && s_{23}\\
    0 && -s_{23} && c_{23}
    \end{pmatrix} }_{\equiv~R_{23}}
    \underbrace{
    \begin{pmatrix}
    c_{13} && 0 && s_{13} \\
    0 && 1 && 0\\
    -s_{13} && 0 && c_{13}
    \end{pmatrix} }_{\equiv~R_{13}}
    \underbrace{
    \begin{pmatrix}
    c_{12} && s_{12}\, e^{i\, \delta_\mathrm{CP}} && 0\\
    -s_{12}\, e^{-i\, \delta_\mathrm{CP}} && c_{12} && 0\\
    0 && 0 && 1
    \end{pmatrix} }_{\equiv~U_{12}} \,.
\end{equation}
Here, $\delta_\mathrm{CP}$ is the CP-phase, and $c_{ij}$ and $s_{ij}$ refer to $\cos \theta_{ij}$ and $\sin\theta_{ij}$, respectively.

The matter Hamiltonian, consisting of both the SM charged current and the NSI neutral current contributions, is given by 
\begin{equation}
 H_\mathrm{mat} = \sqrt{2 } G_F \, N_e(x) \, 
 \left(
 \begin{matrix}
1 + \mathcal{E}_{ee}(x) & \mathcal{E}_{e\mu}(x) & \mathcal{E}_{e\tau}(x) \\
\mathcal{E}_{e\mu}^*(x) & \mathcal{E}_{\mu\mu}(x) & \mathcal{E}_{\mu\tau}(x) \\
\mathcal{E}_{e\tau}^*(x) & \mathcal{E}_{\mu\tau}^*(x) & \mathcal{E}_{\tau\tau}(x) \\
 \end{matrix}
 \right), 
\end{equation}
with 
\begin{equation}
    \mathcal{E}_{\alpha\beta} = \sum_{f} \frac{N_f(x)}{N_e(x)} \, \varepsilon^{f}_{\alpha\beta}\,,
\end{equation}
where $N_f(x)$ is the spatial fermion density in matter. 
With the definition of the nuclear NSI couplings in~\cref{eq:nucl_nsi} and the fact that in neutral matter $N_p(x)=N_e(x)$, we can express the dimensionless NSI matter Hamiltonian elements as
\begin{equation}\label{eq:full_nsi}
    \mathcal{E}_{\alpha\beta} = \varepsilon^e_{\alpha\beta} + \varepsilon^p_{\alpha\beta} + Y_n(x)\, \varepsilon^n_{\alpha\beta}  = 
    \left[\xi^e + \xi^p  + Y_n(x)\, \xi^n\right] \, \varepsilon^{\eta,\varphi}_{\alpha\beta} \,,
\end{equation}
where $Y_n(x)={N_n(x)}/{N_e(x)}$ denotes  the fractional neutron density. In studying solar neutrino propagation effects, we take $Y_n(x)$ from Ref.~\cite{Bahcall:2005va}.

In the context of solar neutrino physics, it is convenient to switch from the conventional neutrino flavour basis to a new basis $ \boldsymbol{\hat\nu}  = O^\dagger \boldsymbol{\nu}$, which we will refer to as the \textit{solar neutrino flavour basis}, via the rotation $O=R_{23}\,R_{13}$. In this basis, the full Hamiltonian reads,
\begin{equation}\label{eq:rot_ham}
     H^\nu = \frac{1}{2E_\nu}
    \begin{pmatrix}
    c^2_{13}\,  A_{\text{cc}}+ s^2_{12}\,\Delta m_{21}^2 &  s_{12} \, c_{12}\, e^{i \dCP } \,\Delta m_{21}^2  &  s_{13}\,c_{13}\,A_{\text{cc}} \\
    s_{12}\, c_{12}\, e^{-i \dCP }\, \Delta m_{21}^2 &  c^2_{12} \,\Delta m_{21}^2 & 0 \\
    s_{13} c_{13}  \,A_{\text{cc}}  & 0 & s^2_{13}\,A_{\text{cc}}+\Delta m_{31}^2 \\
    \end{pmatrix}
    +\sqrt{2 } G_F \, N_e(x) \  O^\dagger 
    \boldsymbol{ \mathcal{E}}\,
    O\,,
\end{equation}
where we have defined the matter potential $A_\text{cc}= 2\,E_\nu\, V_\text{cc}= 2\,E_\nu\sqrt{2} G_F N_e(x)$. From the structure of the Hamiltonian above, we see that if $\Delta m_{31}^2 \gg \Delta m_{21}^2,  \,A_{\mathrm{cc}}, \,2E_\nu\,G_F \sum_f N_f(x)\varepsilon_{\alpha\beta}^f$, the Hamiltonian is dominated by the third eigenvalue, $\Delta m_{31}^2$. In this case, it is effectively block-diagonal, turning our $3\nu$ problem into a $2\nu$  one. In this rotated basis, the third mass eigenstate decouples from the rest of the system and evolves adiabatically. Throughout its journey from the Sun to the Earth, this third eigenstate can be well-approximated by its vacuum mass eigenstate. Within this approximation, the Hamiltonian in \cref{eq:rot_ham} is transformed to an effective $2\times2$ picture, where we only have to track the evolution of the two lighter matter mass eigenstates.

The first condition, $\Delta m_{31}^2 \gg \Delta m_{21}^2$, is satisfied by current best-fits to oscillation parameters \cite{Esteban:2020cvm}. The second is satisfied for solar neutrinos across the full range of solar neutrino energies,  $E_\nu \lesssim \SI{20}{\mega\electronvolt}$. The third condition can be interpreted as one on the value of $\nsi{\alpha\beta}$,
\begin{equation}
    \nsi{\alpha\beta} \ll  \frac{\sqrt{2}\, \Delta m_{31}^2 }{\funop{A_\mathrm{cc}}(x)\left[\xi^e + \xi^p  + Y_n(x)\, \xi^n\right]}\,.
\end{equation}
Taking the maximum of all these quantities, which occurs at the solar core, and using $E_\nu \sim \SI{20}{\mega\electronvolt}$, we find that $\nsi{\alpha\beta} \lesssim 3$. We treat this as an upper bound on the value of $\nsi{\alpha\beta}$, and we do not interpret our results above this value throughout our analyses. Ultimately, for $\nsi{\alpha\beta} \sim 3$ at these higher neutrino energies, which are relevant for NRs in DD experiments, a full numerical simulation should be performed to more accurately model neutrino oscillations. For the purposes of our sensitivity study, however, our approach is sufficient.

Following the conventions of Ref.~\cite{Esteban:2018ppq} and setting $\dCP = 0$ in this work, we can write the effective Hamiltonian as $H^{\mathrm{eff}} \equiv H_\mathrm{vac}^\mathrm{eff} + H_\mathrm{mat}^{\mathrm{eff}}$, where
\begin{equation}\label{eq:ham_eff_vac}
    H_\mathrm{vac}^\mathrm{eff} \equiv \frac{\Delta m_{21}^{2}}{4 E_{\nu}}\left(\begin{matrix}
    -\cos 2 \theta_{12} & \sin 2 \theta_{12} \\
    \sin 2 \theta_{12} & \cos 2 \theta_{12}
    \end{matrix}\right) \,,
\end{equation}
and
\begin{equation}\label{eq:ham_eff_mat}
    H_\mathrm{mat}^\mathrm{eff} \equiv \sqrt{2} G_{F} N_{e}(x)\left[\left(\begin{matrix}
    c_{13}^{2} & 0 \\
    0 & 0
    \end{matrix}\right)+\left[\xi^e + \xi^p  + Y_n(x)\, \xi^n\right]\left(\begin{matrix}
    -\varepsilon_{D}^{\eta, \varphi} & \varepsilon_{N}^{\eta, \varphi} \\
    \varepsilon_{N}^{\eta, \varphi} & \varepsilon_{D}^{\eta, \varphi}
    \end{matrix}\right)\right]\,.
\end{equation}
The coefficients $\varepsilon_{N}^{\eta, \varphi}$ and $\varepsilon_{D}^{\eta, \varphi}$ are related to our parametrisation by
\begin{equation}
    \begin{aligned}
    \varepsilon_{D}^{\eta, \varphi} \equiv\,& c_{13}\, s_{13} \left(s_{23}\, \varepsilon_{e \mu}^{\eta, \varphi}+c_{23} \,\varepsilon_{e \tau}^{\eta, \varphi}\right)-\left(1+s_{13}^{2}\right) c_{23}\, s_{23}\, \varepsilon_{\mu \tau}^{\eta, \varphi} \\
    &-\frac{c_{13}^{2}}{2}\left(\varepsilon_{e e}^{\eta, \varphi}-\varepsilon_{\mu \mu}^{\eta, \varphi}\right)+\frac{s_{23}^{2}-s_{13}^{2}\, c_{23}^{2}}{2}\left(\varepsilon_{\tau \tau}^{\eta, \varphi}-\varepsilon_{\mu \mu}^{\eta, \varphi}\right)\,,
    \end{aligned}
\end{equation}
and
\begin{equation}
    \varepsilon_{N}^{\eta, \varphi} \equiv c_{13}\left(c_{23} \,\varepsilon_{e \mu}^{\eta, \varphi}-s_{23}\, \varepsilon_{e \tau}^{\eta, \varphi}\right)+s_{13}\left[s_{23}^{2}\, \varepsilon_{\mu \tau}^{\eta, \varphi}-c_{23}^{2}\, \varepsilon_{\mu \tau}^{\eta, \varphi}+c_{23}\, s_{23}\left(\varepsilon_{\tau \tau}^{\eta, \varphi}-\varepsilon_{\mu \mu}^{\eta, \varphi}\right)\right]\,.
\end{equation}
Diagonalising $H^{\mathrm{eff}}$ then allows us to find the matrix $U^m_{12}$ such that $U^{m\dagger}_{12} H^{\mathrm{eff}} U^m_{12} = \operatorname{diag}(E^m_{1}, E^m_{2})$. Typically, $U^m_{12}$ is parametrised as
\begin{equation}
    U^m_{12} = 
    \begin{pmatrix}
    \cos \theta_{12}^m &  \sin \theta_{12}^m \\
    - \sin \theta_{12}^m & \cos \theta_{12}^m
    \end{pmatrix}\,,
\end{equation}
for some matter mixing angle $\theta_{12}^m$. We find that the eigenvalues of the effective matter Hamiltonian $H^\mathrm{eff}$ are given by
\begin{equation}
\begin{aligned}
    E^m_{1,\, 2}  &=  c_{13}^2 A_{\mathrm{cc}} \mp \frac{\Delta m_{21}^2}{4 E_\nu}\,\sqrt{p^2 + q^2} \,,
\end{aligned}
\end{equation}
where  we have defined the two quantities
\begin{equation}
\begin{aligned}
    p \equiv&~\sin 2\theta_{12}   + 2\,\varepsilon^{\eta, \varphi}_N \left[\xi^e + \xi^p + Y_n(x)\,\xi^n\right] \, \frac{A_{\mathrm{cc}}}{\Delta m_{21}^2}\,, \\
    q \equiv&~ \cos 2\theta_{12} + \left( 2\,\varepsilon^{\eta, \varphi}_D \left[\xi^e + \xi^p + Y_n(x)\,\xi^n\right] - c_{13}^2\right)\, \frac{A_{\mathrm{cc}}}{\Delta m_{21}^2}\,.
\end{aligned}
\end{equation}
Thus, the energy difference between the two energy eigenvalues in matter, responsible for the coherent mixing of the two matter mass eigenstates, is given by
\begin{equation}\label{eq:en_diff}
    \Delta E_{21}^m \equiv E^m_{2} - E^m_{1} =  \frac{\Delta m_{21}^2}{2 E_\nu}\sqrt{p^2 + q^2}\,.\end{equation}
Moreover, we find that the matter mixing angle, $\theta^{m}_{12}$,  obeys the relations
\begin{equation}
\begin{aligned}
    \sin 2\theta_{12}^m &= \frac{p}{\sqrt{p^2+q^2}}\,, \\ 
    \cos 2\theta_{12}^m &= \frac{q}{\sqrt{p^2+q^2}}\,, \\ 
    \tan 2\theta_{12}^m &= \frac{p}{q}\,.
    \label{eq:mat_angles}
\end{aligned}
\end{equation}

With these expressions, we are in the position to describe the neutrino evolution in the full $3\times 3 $ picture. Using the notation of Ref.~\cite{Coloma:2022umy} and the fact that solar neutrinos are relativistic (such that $t \simeq x$), we can write the evolution equation in the solar neutrino flavour basis as 
\begin{equation}
    i\, \diff{}{x}
    \begin{pmatrix}
    \hat \nu_e \\
    \hat \nu_\mu \\
    \hat \nu_\tau
    \end{pmatrix}
    = \underbrace{
    \begin{pmatrix}
    \mathrm{Evol}[H^\mathrm{eff}] && 0 \\
    0 && \exp[-i \, \frac{\Delta m^2_{31}}{2\, E_\nu} L]
    \end{pmatrix} 
    }_{\equiv~\tilde S}
    \begin{pmatrix}
    \hat \nu_e \\
    \hat \nu_\mu \\
    \hat \nu_\tau
    \end{pmatrix} \,.
\end{equation}
To obtain the evolved matter Hamiltonian, we can split up the distance of propagation within the Sun into $N$ equidistant slabs of thickness $\Delta x$ with approximate homogeneous matter density. We can then obtain the evolved Hamiltonian by formally  taking the limit
\begin{equation}
    \mathrm{Evol}[H^\mathrm{eff}] = \lim_{\Delta x \to 0} \ \prod_{n=0}^{N} U^m_\mathrm{PMNS}(x_n)\ \exp\left[-i \left(U^{m\dagger}_{12}H^\mathrm{eff}U^m_{12}- i\, U^{m\dagger}_{12}\, \dot U^m_{12}\right)\Delta x \right] \ U^m_\mathrm{PMNS}(x_n)^\dagger\,,
\end{equation}
where $\dot U^m_{12} = (\mathrm{d}/{\mathrm{d}x})\,U^m_{12}$ and $x_n=x_{n-1}+\Delta x$.
From this, we can  write the full evolution equation in the conventional vacuum-flavour basis as 
\begin{equation}
    i\, \diff{}{ x}
    \begin{pmatrix}
    \nu_e \\
    \nu_\mu \\
    \nu_\tau
    \end{pmatrix}
    = \underbrace{O\, \tilde S\, O^\dagger}_{S}
    \begin{pmatrix}
    \nu_e \\
    \nu_\mu \\
    \nu_\tau
    \end{pmatrix} \,,
\end{equation}
with the rotation matrix $O= R_{23}R_{13}$. In this notation, the full $S$-matrix is thus given by
\begin{equation}\label{eq:sfull}
    S = \underbrace{O\, U_{12}}_{U_\mathrm{PMNS}} \,
    \begin{pmatrix}
    \exp\left[-i \,\int_0^L D(x)\,\dl x\right] && 0 \\
    0 && \exp[-i \,\Phi_{33}]
    \end{pmatrix} \,
    \underbrace{U^m_{12} (x_0)^\dagger O^\dagger}_{U^m_\mathrm{PMNS}(x_0)^\dagger}\,,
\end{equation}
with  $\Phi_{33} =  \Delta m^2_{31} L /(2 E_\nu)$, where we evolve the neutrinos from their production point within the Sun, $x_0$, to their detection point at an experiment over the distance $L$.  Finally, the $2\times2$ time-evolution matrix is given by
\begin{equation}
    D(x) = 
    \begin{pmatrix}
        E^m_1 & - i \, \dot \theta^m_{12} \\
        i \, \dot \theta^m_{12}  & E^m_2 \\
    \end{pmatrix}   \,.
\end{equation}

To simplify our analysis, we make the assumption that the two light matter mass eigenstates of $H^{\mathrm{eff}}$, $\ket{\nu_{1m}}$ and $\ket{\nu_{2m}}$,  propagate adiabatically within the Sun. As such, the two eigenstates do not mix with one another as they travel to the surface of the Sun, remaining eigenstates of $H^{\mathrm{eff}}$ throughout their evolution.  This assumption is appropriate because the matter density within the Sun, described by $N_f(x)$, varies slowly enough to allow the matter eigenstates to adapt to the medium as they propagate through it. The adiabatic approximation is valid if the adiabaticity parameter, $\gamma$, satisfies 
\begin{equation}\label{eq:def_adiab}
    \gamma \equiv \frac{|\Delta E_{21}^m|}{2|\dot{\theta}_{12}^m|} \gg 1\,,
\end{equation}
where $\Delta E_{21}^m$ is given by \cref{eq:en_diff}. In the adiabatic approximation, the matrix $D(x)$ is thus approximately diagonal, and after a common rephasing of the neutrino matter mass eigenstates, the upper $2\times2$ block in~\cref{eq:sfull} can be expressed as 
\begin{equation}
     \exp\left[-i \,\int_0^L D(x) \,\dl x\right] \approx 
     \begin{pmatrix}
         e^{i\,\phi} & 0 \\
         0 & e^{- i\,\phi}
     \end{pmatrix} \,,
\end{equation}
with $\phi = \int_{0}^L \Delta E_{21}^m(x)\, \dl x $.
Since the neutrinos exiting the Sun will free-stream to the Earth, there is no further evolution effect to be taken into account for the Sun-Earth propagation.

However, in principle, there is a further propagation effect when neutrinos pass thorough the Earth at night, which should be taken into account for a complete treatment. For high-energy $\mathrm{^8B}$ neutrinos, for which $E_\nu \sim \SI{10}{\mega\electronvolt}$, this effect typically changes oscillation probabilities only at the percent level~\cite{Bahcall:1997jc,Akhmedov:2000cs,Chiang:2000kh}.
In particular, Super-Kamiokande has determined the day-night asymmetry to about $-3.3\%$ in $\mathrm{^8B}$ neutrinos~\cite{Super-Kamiokande:2013mie}, while Borexino has found no asymmetry in  $\mathrm{^7Be}$ neutrinos~\cite{Borexino:2011bhn}.
Therefore, in this work  we neglect Earth matter effects for simplicity.

\subsection{Solar neutrino density matrix}
\label{subsec:density_mat}

From the expression of the $S$-matrix in~\cref{eq:sfull}, we can derive the expression for the full three-flavour density matrix for solar neutrinos reaching the Earth. With the projector onto the electron-neutrino flavour state, $\pi^{(e)}=\mathrm{diag}(1,0,0)$, the density matrix reads,
\begin{align}
    \rho^{(e)} = S\, \pi^{(e)}\, S^\dagger = 
    \begin{pmatrix}
        |S_{11}|^2 &  S_{11}\, S_{21}^* & S_{11}\, S_{31}^* \\
        S_{11}^*\, S_{21} &  |S_{21}|^2 & S_{21}\, S_{31}^* \\
        S_{11}^*\, S_{31} &  S_{21}^*\, S_{31} & |S_{31}|^2 
    \end{pmatrix} \,.
\end{align}
Since the density matrix is Hermitian, $\rho_{\alpha\beta} =\rho_{\beta\alpha}^{*}$, the solar neutrino density matrix $\rho^{(e)}$  is completely characterised by the three independent $S$-matrix components,
\begin{align}
    S_{11} & =   e^{-i\,\Phi_{33}}\, s_{13}^2  + c_{13}^2 \, \left( e^{i\,\phi}  \, c_{12} \,  c_{m}  +  e^{-i\,\phi}   \, s_{12} \,  s_{m} \right) \,,\\ 
    S_{21} & = c_{13} \left[   s_{13}\, s_{23}\,(e^{i\,\Phi_{33}} - e^{i\,\phi}\, c_{12} \, c_{m} -  e^{-i \,\phi} \,s_{12} \, s_{m}) + e^{-i\,\phi} \, c_{23}\,( c_{12} \, s_{m} -  e^{ 2 i\,\phi} \,s_{12} \, c_{m} ) \right] \,, \\
    S_{31} & = c_{13} \left[ s_{13}\, c_{23}\,(e^{i\,\Phi_{33}} - e^{i\,\phi}\, c_{12} \, c_{m} -  e^{-i\,\phi} \,s_{12} \, s_{m}) - e^{-i\,\phi} \, s_{23}\,( c_{12} \, s_{m} - e^{i\,2\phi} \,s_{12} \, c_{m} )\right] \,,
\end{align}
where $c_{m}$ and $s_{m}$ refer to $\cos \theta_{12}^m$ and $\sin\theta_{12}^m$, respectively. Given that we do not know precisely where neutrinos are produced in the solar core, we must average over the neutrino production positions. This effectively removes terms dependent on $\phi$ and $\Phi_{33}$ from the density matrix. The six independent density matrix elements then read, 
\begin{align}
    \rho_{ee} =&~  s_{13}^4 +  c_{13}^4\, P_{\text{ee}}^{2 \nu }  \,, \\
    \rho_{\mu\mu} =&~ 
    c_{13}^2\, \Big[c_{23}^2\, \left(1-P_{\text{ee}}^{2 \nu }\right)+s_{13}^2 \,s_{23}^2 \,\left(1+P_{\text{ee}}^{2 \nu }\right) +\Delta\Big]  \,, 
    \\
    \rho_{\tau\tau} =&~ 
    c_{13}^2\, \Big[s_{23}^2\, \left(1-P_{\text{ee}}^{2 \nu }\right)+s_{13}^2 \,c_{23}^2 \,\left(1+P_{\text{ee}}^{2 \nu }\right) -\Delta\Big] \,, 
    \\
    \rho_{e\mu} =&~ c_{13}\, s_{13}^3\, s_{23}-\frac{1}{2}\, c_{13}^3\, \bigg[2\, s_{13}\, s_{23} \, P_{\text{ee}}^{2 \nu } +  c_{23}\,   \sin \left(2 \theta _{12}\right)\, \cos \left(2 \theta^m_{12} \right) \bigg] \,, 
    \\
    \rho_{e\tau} =&~ c_{13}\, s_{13}^3 \, c_{23} - \frac{1}{2}\, c_{13}^3\, \bigg[2\, s_{13}\, c_{23} \, P_{\text{ee}}^{2 \nu }-   s_{23}\,  \sin \left(2 \theta _{12}\right)\, \cos \left(2 \theta^m_{12}\right) \bigg] \,, 
    \\
    \rho_{\mu\tau} =&~ \frac{1}{2}\, c_{13}^2 \left[ \sin \left(2 \theta _{23}\right)  \Big(\left(1+s_{13}^2\right)\,P_{\text{ee}}^{2 \nu } -c_{13}^2 \Big) + 2\,  \cot \left(2 \theta _{23}\right) \, \Delta \right] \,,
\end{align}
where we have defined 
\begin{align}
    P^{2\nu}_{ee} &= \frac{1}{2}\, \Big(1+\cos \left(2 \theta _{12}\right) \,\cos \left(2 \theta^m_{12}\right)\Big)\,, \\
    \Delta &= \frac{1}{2}\  \sin \left(\theta _{13}\right) \, \sin \left(2 \theta _{12}\right)\, \sin \left(2 \theta _{23}\right)\, \cos \left(2 \theta^m_{12}\right)\,. 
\end{align}
Since neutrinos are produced within a finite volume of the Sun and the matter mixing angle, $\funop{\theta^m_{12}}(x)$, depends on position, there is some ambiguity in what to take for the value of $\cos (2\theta^m_{12})$. We have taken its spatial average over the radius of the Sun as a representative value, given by
\begin{equation}
  \label{eq:cos_average}
  \langle \cos 2\theta_{12}^{m}\rangle_p \equiv \int_{0}^{1}\cos 2\theta_{12}^{m}(x) \,f_p(x)\,\dl x\,,
\end{equation}
where $x$ is the fractional solar radius, $p$ denotes a particular solar neutrino population, and $f_p(x)$ is the spatial distribution function describing where in the Sun that population is produced. These populations are labelled according to the reaction that generated them, with $p \in \{pp,\, \mathrm{^8B},\,\ldots\}$. The distributions $f_p(x)$ are SSM-dependent; we have used the BP16-GS98 predictions calculated by Ref.~\cite{Vinyoles:2016djt}. We have taken the values for each oscillation parameter from the latest \texttt{NuFIT} results~\cite{Esteban:2020cvm}.

\subsection{Generalised neutrino cross sections}
\label{subsec:gen_cs}

Following our  discussion of neutrino propagation in the presence of NSI and the relevant formalism needed to derive the neutrino density matrix, $\bb{\rho}$, we move on to find expressions for the generalised scattering cross sections, ${\mathrm{d}\bb{\zeta}}/{\mathrm{d} E_R}$, for both neutrino-nucleus and neutrino-electron scattering.

Considering the process of elastic  scattering of a neutrino $\nu$ off a target $T$ with mass $m_T$ via the matrix element $\mathcal{M}$, the general expression for the cross section reads
\begin{equation}
    \diff{\sigma_{\nu T}}{t} = \frac{1}{16\,\pi}\, \frac{\mathcal{M}^* \mathcal{M}}{(s-m^2_T)^2}\,.
\end{equation}
From this, we can define the generalised cross section correlating the matrix elements of the flavour specific scattering processes $\nu_\alpha\,T\to f\, T$ and $\nu_\beta\,T\to f\,T$ as
\begin{equation}\label{eq:gen_xsec}
    \left(\diff{\zeta}{E_R}\right)_{\alpha\beta}= \left(\diff{\zeta}{t}\right)_{\alpha\beta} \diff{t}{E_R}= \frac{\mathcal{M}^*(\nu_\alpha\to f)\, \mathcal{M}(\nu_\beta\to f) }{32\pi\, m_T \, E_\nu^2}\,,
\end{equation}
where we have made use of the relations $t=- 2\,m_T\, E_R$ and $s=m_T^2+2\,m_T\, E_\nu$ for relativistic neutrino scattering.
Note that the diagonal elements of the generalised cross section are the conventional scattering cross sections of a neutrino of flavour $\alpha$ off the target material $T$,
\begin{equation}
    \left(\diff{\zeta}{E_R}\right)_{\alpha\alpha} = \diff{\sigma_{\nu_\alpha T}}{E_R} \,.
\end{equation}
With the general expression of~\cref{eq:gen_xsec},  we can now  derive the corresponding expressions for the generalised \cevns and \eves cross sections.

\subsubsection{\cevns cross section}

Following Ref.~\cite{Papoulias:2015vxa}, we can derive the expression for the generalised coherent elastic neutrino-nucleus scattering cross section using the NSI formalism introduced in~\cref{sec:parametrisation}. The cross section  reads
\begin{align}\label{eq:sig_CEVNS_gen}
    \left(\diff{\zeta_{\nu N}}{E_R}\right)_{\alpha\beta} &  = \frac{G_F^2\, M_N}{\pi}\left(1-\frac{M_N\, E_R}{2 E_\nu^2}\right)\
    \sum_\gamma\ \langle gs|| \hat G^\mathrm{SM}\,\delta_{\alpha\gamma} + \hat G^\mathrm{NSI}_{\alpha\gamma} ||gs\rangle\langle gs|| \hat G^{\mathrm{SM}}\,\delta_{\gamma\beta} + \hat G^{\mathrm{NSI}\dagger}_{\gamma\beta} ||gs\rangle\,, \notag\\
    & = \frac{G_F^2\, M_N}{\pi}\left(1-\frac{M_N\, E_R}{2 E_\nu^2}\right)\ \left[ \frac{1}{4} \, Q_{\nu N}^2\, \delta_{\alpha\beta} - Q_{\nu N} \,G^\mathrm{NSI}_{\alpha\beta} + \sum_\gamma G^\mathrm{NSI}_{\alpha\gamma}G^{\mathrm{NSI}}_{\gamma\beta}\right]\, F^2(E_R) \,,
\end{align}
where $F(E_R)$ is the Helm form factor \cite{Helm:1956zz, Lewin:1995rx}, and  ${Q_{\nu N} = N - (1 -4\,\sin^2\theta_W)\,Z}$ is the SM coherence factor. Furthermore, we have used the Hermiticity of the NSI nucleus coupling, defined by
\begin{align}\label{eq:g_cevns}
G^\mathrm{NSI}_{\alpha\beta} &\equiv \left(2\,\varepsilon^{u}_{\alpha\beta} + \varepsilon^{d}_{\alpha\beta} \right) Z +  \left(\varepsilon^{u}_{\alpha\beta} + 2\,\varepsilon^{d}_{\alpha\beta} \right) N\,, \notag \\
 &= (\xi^p\, Z + \xi^n \, N)\ \varepsilon_{\alpha\beta}^{\eta,\varphi} \,.
\end{align}

As observed in Ref.~\cite{Amaral:2021rzw}, the BSM contribution can destructively interfere with the SM one. Thus, there are regions in the NSI parameter space where, despite having a non-zero NSI $\varepsilon_{\alpha\beta}$, the cross-section is  the same as for the SM. In these {\em blind spots},  the presence of new physics cannot be distinguished from the SM. Since the \cevns rate is determined by the trace over density matrix times cross section, the cancellation conditions are non-trivial. We discuss them in greater detail in~\cref{sec:nuc_plane}.

\subsubsection{\eves cross section}

Similarly, following Ref.~\cite{Coloma:2022umy} and by use of~\cref{eq:gen_xsec}, we can derive the expression for the generalised neutrino-electron scattering cross section in the presence of NSI,
\begin{multline}\label{eq:xsec_el}
        \left(\diff{\zeta_{\nu e}}{E_R}\right)_{\alpha\beta} = \frac{2 \, G_F^2\, m_e}{\pi}\ \sum_\gamma\,\left\{G^L_{\alpha\gamma}G^{L}_{\gamma\beta} + G^R_{\alpha\gamma}G^{R}_{\gamma\beta}\left(1-\frac{E_R}{E_\nu}\right)^2
    - \left(G^L_{\alpha\gamma}G^{R}_{\gamma\beta} + G^R_{\alpha\gamma}G^{L}_{\gamma\beta} \right) \frac{m_e\, E_R}{2 E_\nu^2}\right\} \,,
\end{multline}
where we have defined the generalised neutrino-electron couplings as
\begin{equation}
    G^P_{\alpha\beta} =
    g^e_{P\alpha}\delta_{\alpha\beta} + \varepsilon^{eP}_{\alpha\beta}\,.
\end{equation}
The SM electroweak neutrino-electron couplings are given by
\begin{equation}
    g^e_{P\alpha} = \begin{cases}  1 + g^e_L \,, & \text{if}\ \alpha=e \ \text{and} \ P=L \,, \\
    g^e_P \,, & \text{otherwise} \,,
    \end{cases} 
\end{equation}
with $g^f_P = T^3_f - \sin^2\theta_w\, Q^\mathrm{EM}_f$.
In order to express the generalised neutrino-electrons coupling in terms of their vector and axial-vector components with the parameterisation of~\cref{sec:parametrisation}, we introduce
\begin{align}
    G^V_{\alpha\beta} &= G^L_{\alpha\beta} + G^R_{\alpha\beta} \,, &&
    G^A_{\alpha\beta} = G^L_{\alpha\beta} - G^R_{\alpha\beta} \,, \\[.3cm]
    G^L_{\alpha\beta} &= \frac{1}{2} (G^V_{\alpha\beta} + G^A_{\alpha\beta}) \,, &&
    G^R_{\alpha\beta} = \frac{1}{2} (G^V_{\alpha\beta} - G^A_{\alpha\beta}) \,.
\end{align}
Effectively, this means that in~\cref{eq:xsec_el} we can  make the replacements
\begin{align}
    G^L_{\alpha\beta} &
    = ( \delta_{e\alpha}+g^e_L)\, \delta_{\alpha\beta} + \frac{1}{2}\left(\varepsilon^{\eta,\varphi}_{\alpha\beta}\,\xi^e + \tilde\varepsilon^{\eta,\varphi}_{\alpha\beta}\,\tilde\xi^e\right)\,, \\
    G^R_{\alpha\beta} &
    = g^e_R\, \delta_{\alpha\beta} + 
    \frac{1}{2} \left(\varepsilon^{\eta,\varphi}_{\alpha\beta}\,\xi^e - \tilde\varepsilon^{\eta,\varphi}_{\alpha\beta}\,\tilde\xi^e\right) \,,
\end{align}
where $\varepsilon^{\eta,\varphi}_{\alpha\beta}$ denotes the vector component of the general NSI as before and $\tilde\varepsilon^{\eta,\varphi}_{\alpha\beta}$ denotes the axial-vector component (which does not contribute to matter effects and CE$\nu$NS). Note that if the NSI is only due to a vector interaction, we have $\varepsilon^L=\varepsilon^R$, such that the axial-vector component vanishes, $\tilde\varepsilon^{\eta,\varphi}_{\alpha\beta}=0$. As stated before,  we only focus on the vector interaction for electron scattering. We do this  because the results from oscillation and coherent experiments will have no impact on $\tilde\varepsilon^{\eta,\varphi}_{\alpha\beta}$. Furthermore, to accurately predict the signal from the axial-vector interaction, one would have to use a different ionisation form factor to that of the vector interaction.

\section{Extending current constraints to the full NSI parameter space}
\label{sec:spall_and_osc}

With our extended formalism in place, we are ready to explore how previous NSI results map onto the extended parameter space. Earlier constraints on NSI parameters derived from spallation source \cite{Coloma:2017ncl, Miranda:2020tif, Coloma:2019mbs,COHERENT:2020iec,COHERENT:2017ipa,COHERENT:2021xmm} and neutrino oscillation \cite{Coloma:2022umy, Coloma:2019mbs, Khan:2019jvr, Esteban:2018ppq} experiments have assumed that the NSI contribution in the charged plane is entirely in either the proton ($\varphi = 0$) or the electron ($\varphi = \pm \pi / 2$) directions. In the $\varphi = 0$ case, the \cevns cross section is maximally modified with no change to the \nue cross section, leading to the strongest constraints on $\nsi{\alpha\beta}$ from spallation source experiments and limits from oscillation experiments that only arise from non-standard propagation effects. In the $\varphi = \pm \pi/2$ case, constraints from oscillation source experiments arise from propagation effects and a maximal change to the \nue cross section. However, the evolution of these bounds with variable charged NSI contribution has not yet been studied, and our parametrisation provides a convenient way to visualise this. Since we have no reason to believe that charged NSI would lie preferentially in any direction, a general treatment must be sought.

In this section, we recompute the bounds from spallation source and oscillation experiments, allowing for $\varphi$ to vary along its entire allowed range. In particular, we consider the CENNS-10 LAr \cite{COHERENT:2020iec} and Borexino \cite{Borexino:2008gab} experiments as our spallation source and oscillation experiment candidates, respectively. To demonstrate the non-trivial evolution of previously computed constraints with variable $\varphi$, we take inspiration from earlier analyses, showing how the same approaches can lead to very different results.

\subsection{The CENNS-10 LAr Experiment}
\label{subsec:spallation}


The CENNS-10 LAr experiment \cite{COHERENT:2020iec} has measured the coherent elastic scattering of neutrinos with nuclei using a liquid argon scintillator target. The neutrino flux has three components: a prompt flux of muon neutrinos generated by the decay of pions, and two delayed fluxes of anti-muon and electron neutrinos, produced in the three-body decay of anti-muons. Their normalised spectra are given by
\begin{equation}
    \begin{aligned}
    f_{\nu_{\mu}}(E_\nu) &=\delta\left(E_{\nu}-\frac{m_{\pi}^{2}-m_{\mu}^{2}}{2 m_{\pi}}\right)\,, \\
    f_{\bar{\nu}_{\mu}}(E_\nu) &=\frac{64}{m_{\mu}}\left[\left(\frac{E_{\nu}}{m_{\mu}}\right)^{2}\left(\frac{3}{4}-\frac{E_{\nu}}{m_{\mu}}\right)\right]\,, \\
    f_{\nu_{e}}(E_\nu) &=\frac{192}{m_{\mu}}\left[\left(\frac{E_{\nu}}{m_{\mu}}\right)^{2}\left(\frac{1}{2}-\frac{E_{\nu}}{m_{\mu}}\right)\right]\,,
    \end{aligned}
    \label{eq:coherent_flux}
\end{equation}
where, from kinematics, $E_\nu \in [0, m_\mu/2]$. The expected neutrino flux is then given by scaling these spectra to 
account for the total beam luminosity and distance of the liquid argon target from the source. This scaling is given by $\eta \equiv r\, N_\mathrm{POT} / (4 \pi L^2)$, where $r$ is the number of neutrinos produced per proton collision, $N_\mathrm{POT}$ is the number of protons on target, and $L$ is the length of the experimental baseline. This gives us the total expected neutrino flux, $\phi_\alpha (E_\nu) \equiv \eta \,f_\alpha (E_\nu)$, where $\alpha \in \{\nu_\mu, \bar{\nu}_\mu, \nu_e\}$. For the CENNS-10 LAr experiment, $r = 0.08$, $N_\mathrm{POT} = \num{1.37e23}\,\mathrm{yr^{-1}}$, and $L = \SI{27.5}{\meter}$ \cite{COHERENT:2020iec}.

From these fluxes, we can retrieve the expected \cevns rate spectrum. Since the neutrino beam does not undergo significant decoherence over the experimental baseline, it can be treated as being composed of independent $\nu_\mu$, $\overline{\nu}_\mu$, and $\nu_e$ parts. This means that the rate is given by the integral of the neutrino flux and the appropriately flavoured cross section, as it is usually written. In our notation, this reads
\begin{equation}
    \diff{N_\alpha}{E_R} = \frac{M_\mathrm{det}}{m_N} \,\epsilon(E_R) \int_{E_\nu^\mathrm{min}}^{m_\mu/2} \phi_\alpha (E_\nu) \left(\diff{\zeta}{E_R}\right)_{\alpha\alpha}\,\dl E_\nu\,,
\end{equation}
where $M_\mathrm{det} = \SI{24}{\kg}$ is the mass of the detector, $m_N$ is the mass of an $\mathrm{^{40}Ar}$ nucleus (for which we assume $100\%$ isotopic abundance), and $\epsilon(E_R)$ is the energy-dependent efficiency function, which we have taken from Analysis A of Ref.~\cite{COHERENT:2020iec}. Since this function is given in units of electron-equivalent energy (\si{\kevee}), we convert our spectrum into $E_{ee}$ energies before folding in the efficiency function using the energy-dependent quenching factor \cite{COHERENT:2020iec}
\begin{equation}
  Q_{F}(E_{R}) = 0.246 + (7.8 \times \SI{e-4}{\per\kevnr})\,E_{R}\,.
  \label{eq:coherent_qf}
\end{equation}
Finally, the integral over neutrino energy runs from the minimum neutrino energy required to cause a recoil of energy $E_R$, $E_\nu^\mathrm{min} \approx \sqrt{m_N E_R/2}$.

\begin{figure}[p!]
    \includegraphics[]{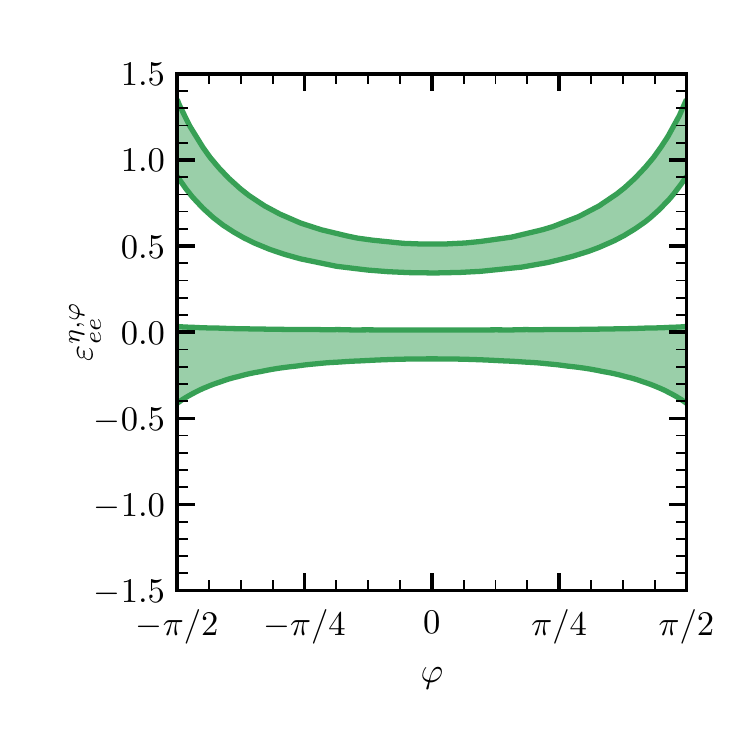}
    \includegraphics[]{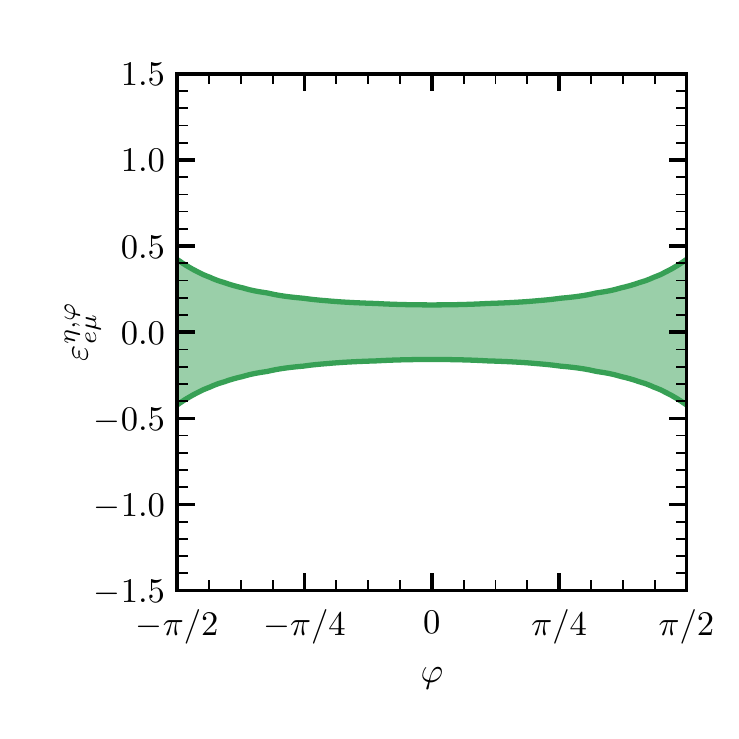}

    \vspace*{-6ex}
    \includegraphics[]{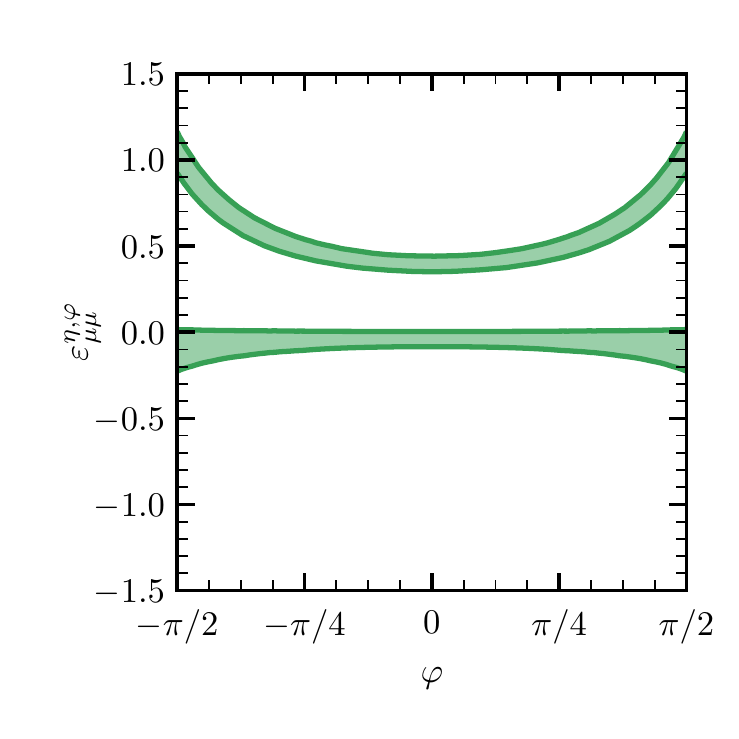}
    \includegraphics[]{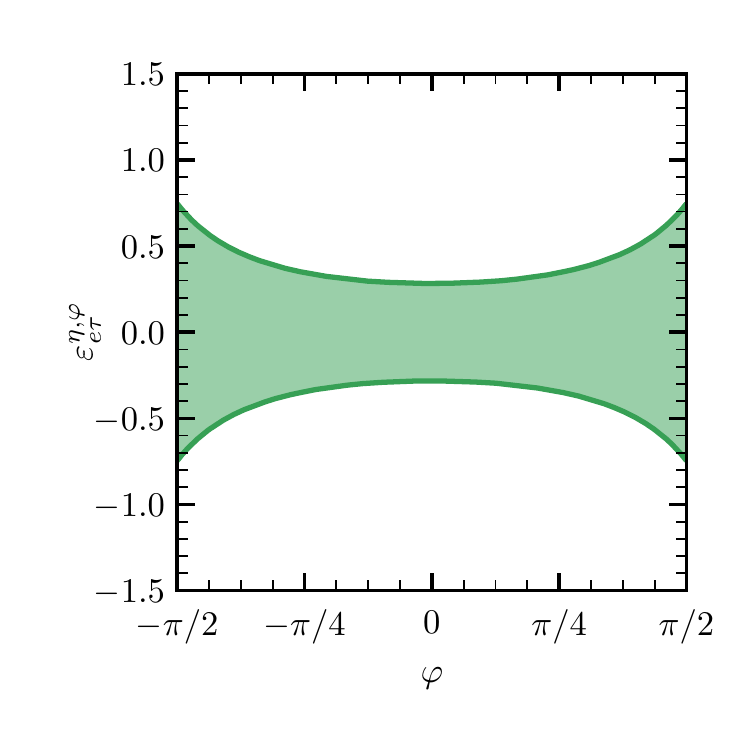}

    \vspace*{-6ex}
    \includegraphics[]{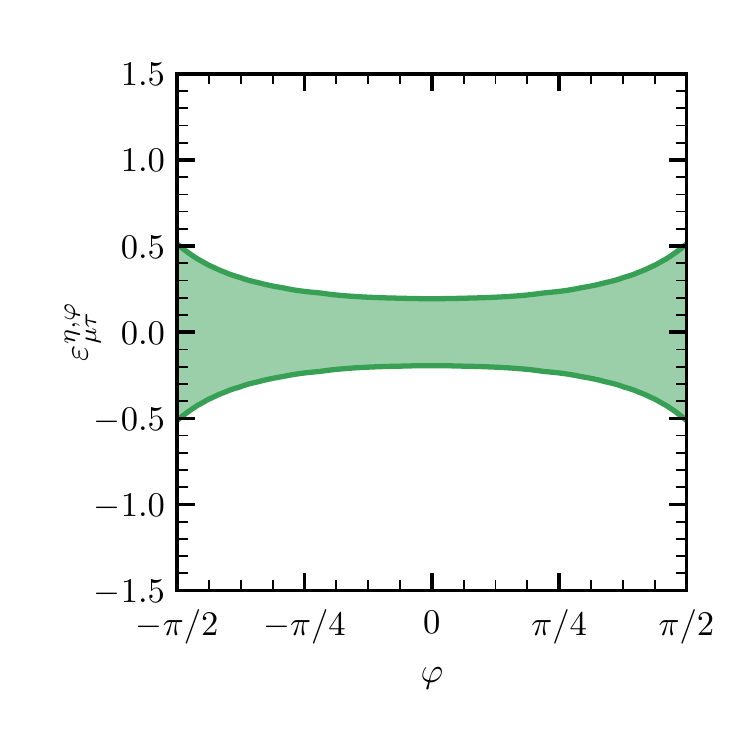}

    \vspace{-2ex}
    \caption{The 90\% CL allowed regions for each NSI parameter over $\varphi$ from the CENNS-10 LAr results \cite{COHERENT:2020iec}. The bounds usually quoted correspond to the NSI parameter values at $\varphi = 0$. We have fixed $\eta = \arctan (1/2)$, corresponding to a pure up-quark NSI when $\varphi = 0$.}
    \label{fig:nsi-cenns10}
\end{figure}

To compute the allowed regions for $\nsi{\alpha\beta}$, we perform a similar analysis to that of Ref.~\cite{Miranda:2020tif}, with the key difference that we allow for the charged NSI contribution to lie anywhere within the charged plane. Using a $\chi^2$ statistic, we compare the number of events measured by CENNS-10 LAr to the theoretical expectation given a particular choice for $\varphi$ and $\nsi{\alpha\beta}$. We fix $\eta = \tan^{-1}({1 / 2})$ in order to match the analysis of Ref.~\cite{Miranda:2020tif}, equivalent to having neutrino NSI with the up-quark only when $\varphi =0$. Our $\chi^2$ statistic is given by
\begin{equation}
  \label{eq:coherent-chi2}
  \chi^{2}(\nsi{\alpha\beta}, \varphi )=\min _{a}\left[\left(\frac{N_{\text {exp}} - (1+a)\, N_{\mathrm{CE} \nu \mathrm{NS}}(\nsi{\alpha\beta}, \varphi )}{\sqrt{N_{\text {exp }}+N_{\text {bkg}}}}\right)^{2}+\left(\frac{a}{\sigma_{a}}\right)^{2}\right]\,,
\end{equation}
where $N_\mathrm{exp} = 159$ is the number of measured events and $N_\mathrm{bkg} = 563$ is the number of background events (primarily from the beam-related neutron rate) \cite{COHERENT:2020iec}. The nuisance parameter $a$ acts as a pull parameter on the theoretical rate, allowing it to vary around its central value. This accounts for the systematic uncertainties in its calculation, and we take it to be $\sigma_a =  8.5\%$ \cite{COHERENT:2020iec}. The quadratic penalty term in \cref{eq:coherent-chi2} penalises deviations of size much greater than this.

To compute the $90\%$ CL allowed regions, we vary one NSI parameter at a time for a given angle $\varphi$ and find those values of $\nsi{\alpha\beta}$ for which $\Delta \chi^2(\nsi{\alpha\beta}) \equiv \chi^2(\nsi{\alpha\beta},\,\varphi) - \chi^2_\mathrm{min}(\varphi) \leq 2.71$, where $\chi^2_\mathrm{min}(\varphi)$ is the minimum $\chi^2$ optimised over $\nsi{\alpha\beta}$. We repeat this analysis over the full range of $\varphi$, drawing the $90\%$ CL allowed regions in \cref{fig:nsi-cenns10}. We also show our $\Delta \chi^2$ plot for the extremal cases of $\varphi = 0$ and $\varphi = \pi/2$ in \cref{fig:chi2_cenns} of \cref{sec:app_delta}. We see that, for $\nsi{ee}$ and $\nsi{\mu\mu}$, these regions allow for two solutions: one that is consistent with the SM (i.e.~$\varepsilon_{\alpha\beta}=0$) and one that is not. This first region is slightly displaced from $\nsi{\alpha\beta} = 0$ as CENNS-10 LAr observed a slight excess of events over the SM expectation. The second region is due to a cancellation between the interference and NSI-only terms in the cross section, which can be seen by inspecting \cref{eq:sig_CEVNS_gen} and is discussed in greater detail in the context of DD experiments in \cref{sec:nuc_plane}. While this second minimum occurs for all $\nsi{\alpha\beta}$, the effect is most pronounced for $\nsi{ee}$ and $\nsi{\mu\mu}$, as can be seen from \cref{fig:chi2_cenns}. Importantly,  no bounds can be placed on $\nsi{\tau\tau}$ since the CENNS-10 LAr neutrino beam has a negligible $\nu_\tau$ component.

Typically, the intervals that would be quoted correspond to the allowed values at $\varphi = 0$. This reflects the assumption that the charged NSI lies purely in the proton direction. However, we see in \cref{fig:nsi-cenns10} that these bounds generally worsen for increasing values of $|\varphi|$. While this trend is partially led by our parametrisation (whereby the strength of $\nsi{\alpha\beta}$ required for a constant contribution should scale as $1 / \cos\varphi$ in any one of the proton, neutron, or electron directions), the bounds do not vary via this same scaling. This is particularly evident from the limits drawn for the second minima in the cases of $\nsi{ee}$ and $\nsi{\mu\mu}$, both of which deteriorate more rapidly than the first minima bounds. Moreover, the constraints on the NSI contribution from the neutron, which is inherently independent of $\varphi$ in our formalism, would worsen for increasing $|\varphi|$ (at fixed $\eta$), reflecting the requirement for a stronger NSI with the neutron to account for the diminishing contribution from the proton.

\subsection{The Borexino Experiment}
\label{subsec:osc}

The Borexino experiment, located at the Laboratori Nazionali del Gran Sasso, observes solar neutrinos through their elastic scattering with electrons in its multi-ton scintillator target \cite{Borexino:2008gab}. The differential scattering rate per target electron is given most generally by \cref{eq:dr_gen}. In the case of Borexino, we consider the flux of solar neutrinos, which has contributions from different populations of electron neutrinos depending on where in the $pp$ chain or CNO cycle they are produced. We take the spectrum for each population,   $\mathrm{d}\phi_{\nu_e}^p/{\mathrm{d}E_\nu}$, from the predictions of the B16-GS98 SSM \cite{Vinyoles:2016djt}, where $p \in \{pp,\, \mathrm{^8B},\,\ldots\}$.  In the case of electron recoils, the minimum neutrino energy necessary to cause a recoil of energy $E_R$ is given by
\begin{equation}
    E_{\nu}^{\min} = \frac{1}{2}\left(E_{R}+\sqrt{E_{R}^{2}+2 m_{e} E_{R}}\right)\,.
    \label{eq:nu_min_sm}
\end{equation}
Ignoring experimental effects, such as energy resolution and efficiency functions, the scattering rate due to a particular neutrino population, $p$, is given by
\begin{equation}
    R^p_{\mathrm{Borexino}} = \int_0^{E^{p,\mathrm{max}}_{R}} \diff{R^p}{E_R} \ \dl E_R\,,
\end{equation}
where $E^{p,\mathrm{max}}_{R}$ is the maximum possible recoil energy for the population $p$ and ${\mathrm{d} R^p}/{\mathrm{d} E_R}$ is the differential rate calculated from \cref{eq:dr_gen}. We take the number of target electrons in the scintillator to be $\num{3.307e31} / (100\,\si{\ton})$ \cite{Borexino:2017rsf}.

We wish to explore the evolution of previous oscillation bounds in the full plane of charged NSI, i.e.~with variable angle $\varphi$ at a fixed angle $\eta$. To this end, we perform a similar analysis to that of Ref.~\cite{Khan:2019jvr}. Namely, we consider how Borexino's Phase-II measurements of the $pp$, $\mathrm{^7Be}$, and $pep$ solar neutrino rates \cite{Borexino:2017rsf} can be used to constrain neutrino NSI with our more general formalism. In conducting our analysis, we determine the bounds on the off-diagonal matrix elements $\nsi{\alpha\beta}$ ($\alpha \neq \beta$), which were not computed in Ref.~\cite{Khan:2019jvr}. This is only possible through the correct treatment of the differential rate in \cref{eq:dr_gen} using the density matrix formalism. We note that a direct comparison between our results and those of Ref.~\cite{Khan:2019jvr} is particularly difficult due to our different treatments of the NSI Lagrangian.

The results from the Borexino's Phase-II run, along with the results of our calculations for the theoretical rate for each respective neutrino population, are shown in \cref{table:borexino_rates}. As was done in Ref.~\cite{Khan:2019jvr}, we assume that the fractional uncertainties in the theoretical rates for each solar neutrino population are the same in our calculation as those reported by Borexino. Our results are in good agreement with the measured rate and the rate predicted by the collaboration \cite{Borexino:2017rsf}.

\begin{table}[t!]
    \begin{tabular*}{\textwidth}{@{\extracolsep{\fill}}cccc}
    \toprule
    \quad\textbf{Population} & 
    \begin{minipage}[c]{3cm}\textbf{Phase-II Rate}

    [$(100\,\mathrm{ton}\,\mathrm{day})^{-1}$]\end{minipage}
    &     
    \begin{minipage}[c]{3cm}\textbf{Theoretical Rate}

    [$(100\,\mathrm{ton}\,\mathrm{day})^{-1}$]\end{minipage}
    & 
    \begin{minipage}[c]{3cm}\textbf{Fractional}

    \textbf{Uncertainty}\end{minipage} 
    \\[0.75ex] \midrule
    $pp$ & $134 \pm 10$ & 133 &  $1.1\%$  \\
    $\mathrm{^7Be}$ & $48.3 \pm 1.1$ & 48.5 & $5.8\%$   \\
    $pep$ & $2.43 \pm 0.36$ & 2.78 & $1.5\%$ \\ \bottomrule
    \end{tabular*}
    \caption{Solar neutrino rates relevant for our Borexino analysis. Shown are the measured rates from the Phase-II run of Borexino \cite{Borexino:2017rsf}, our calculated theoretical rates, and the assumed fractional uncertainties in our calculation.}
    \label{table:borexino_rates}
\end{table}

To perform our statistical analysis, we construct a $\chi^2$ function similar to that of \cref{subsec:spallation}:
\begin{equation}
    \label{eq:chi2-borex}
    \chi^2(\nsi{\alpha\beta}, \varphi) \equiv \min_{\vec{a}}\left[
    \sum_p \left(\frac{R^p_{\mathrm{Borexino}} - (1 + a^p)\,R^p_{\mathrm{Theo}}(\nsi{\alpha\beta}, \varphi)}{\sigma_\mathrm{stat}^{p}}\right)^2 + \left(\frac{a^p}{\sigma_a^p}\right)^2\right]\,,
\end{equation}
where the sum is taken over each considered solar neutrino population, $p \in \{pp, \mathrm{^7Be},\, pep\}$. The rates $R^p_\mathrm{Borexino}$ are the measured rates from the Phase-II run, with statistical uncertainties $\sigma_\mathrm{stat}^p$, while $R^p_\mathrm{Theo}$ are our calculated rates given a choice of $\nsi{\alpha\beta}$ and $\varphi$. We have also introduced the pull parameters $a^p$ for each rate, the values of which we show in \cref{table:borexino_rates}. To compute our $\chi^2$, we profile over the nuisance parameters $\vec{a} \equiv (a^{pp},\,a^\mathrm{^7Be},\,a^{pep})^\mathrm{T}$, whose standard deviations are given in the last column of \cref{table:borexino_rates}. The $90\%$ CL regions are computed via the same prescription as in \cref{subsec:spallation}.

\begin{figure}[p!]
    \includegraphics[]{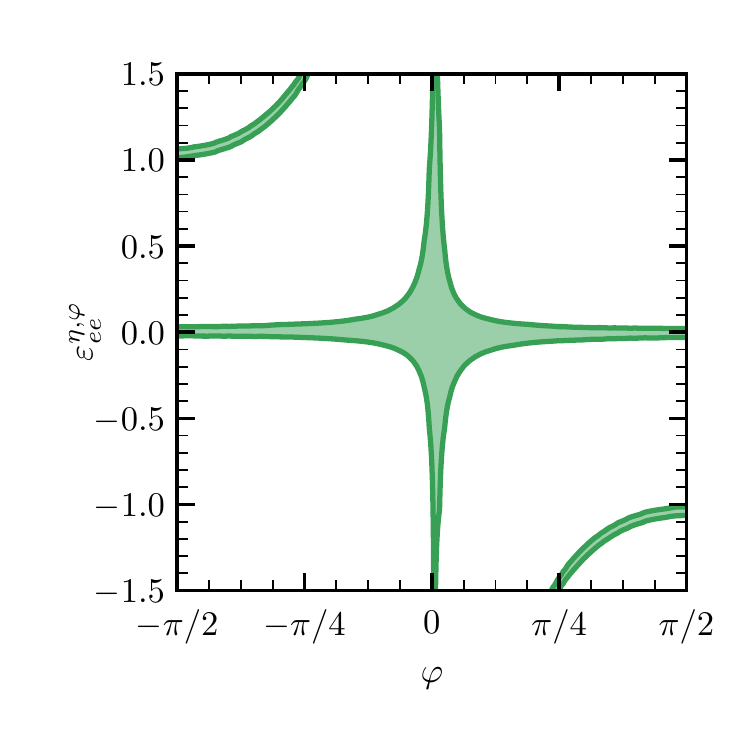}
    \includegraphics[]{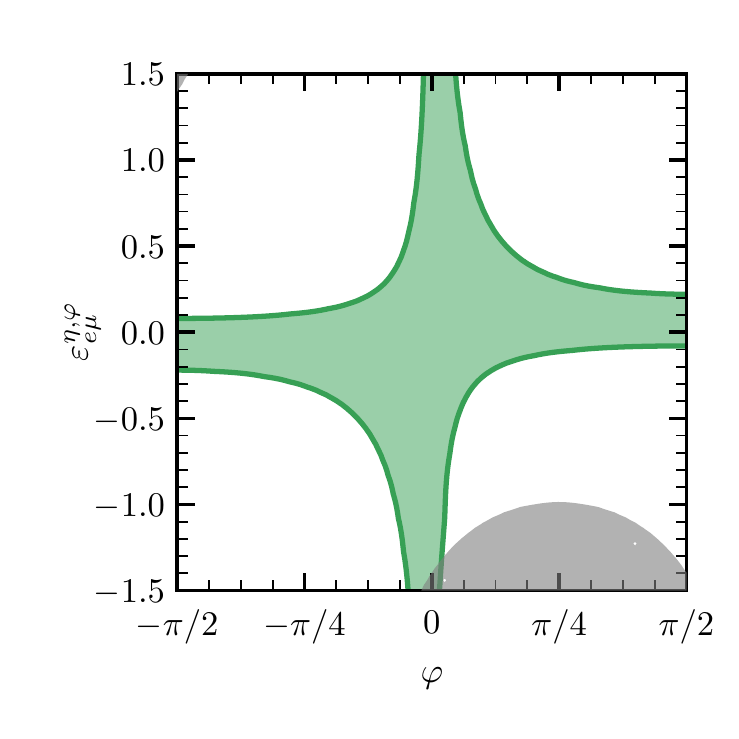}

    \vspace*{-6ex}
    \includegraphics[]{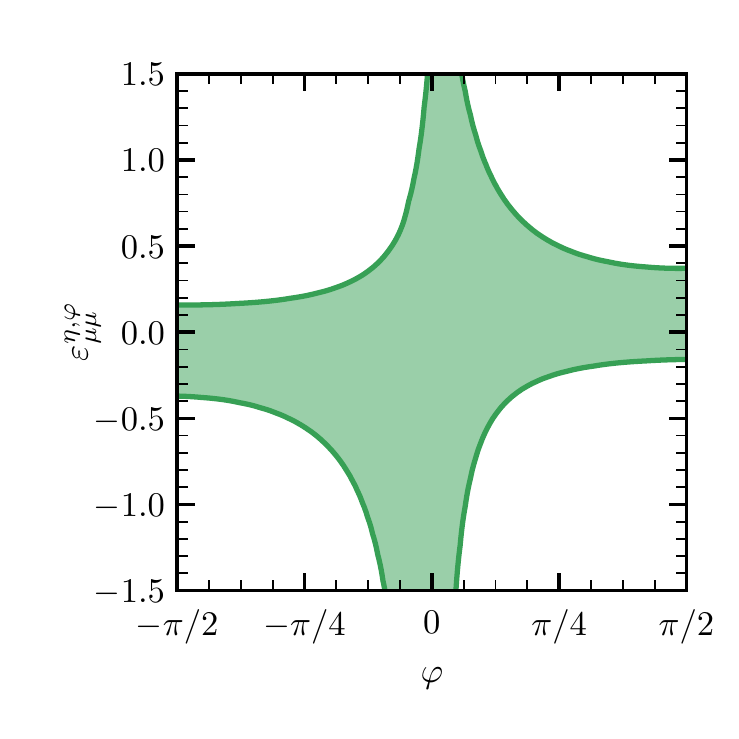}
    \includegraphics[]{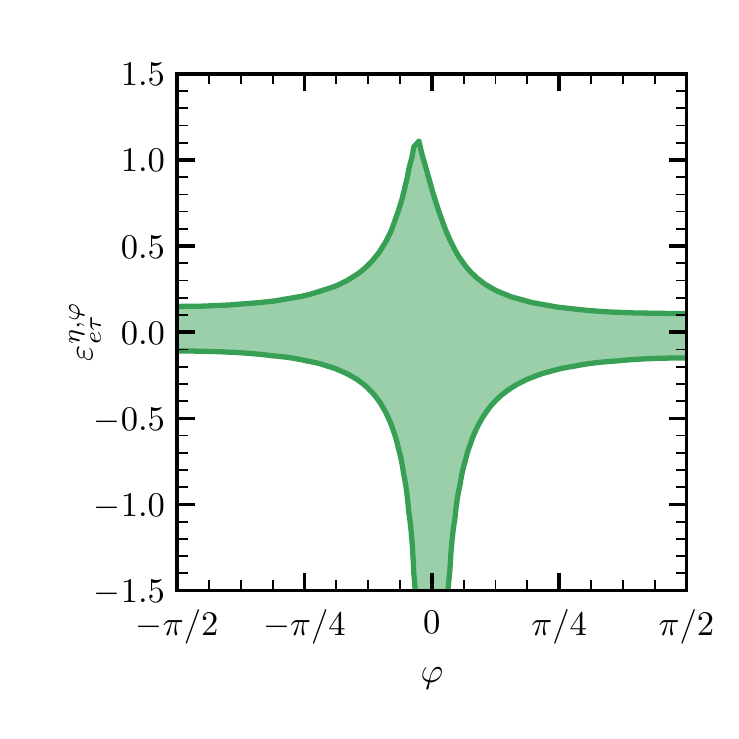}

    \vspace*{-6ex}
    \includegraphics[]{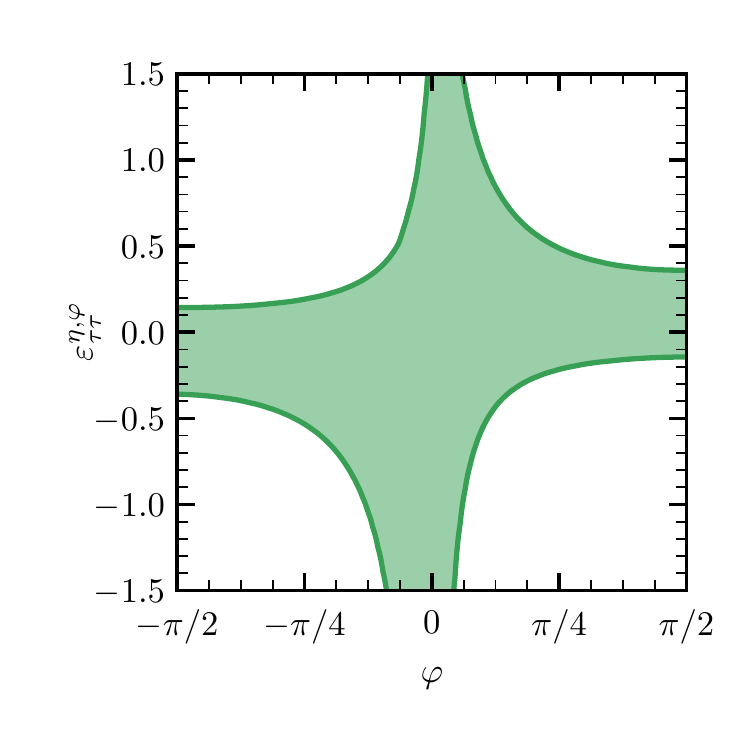}
    \includegraphics[]{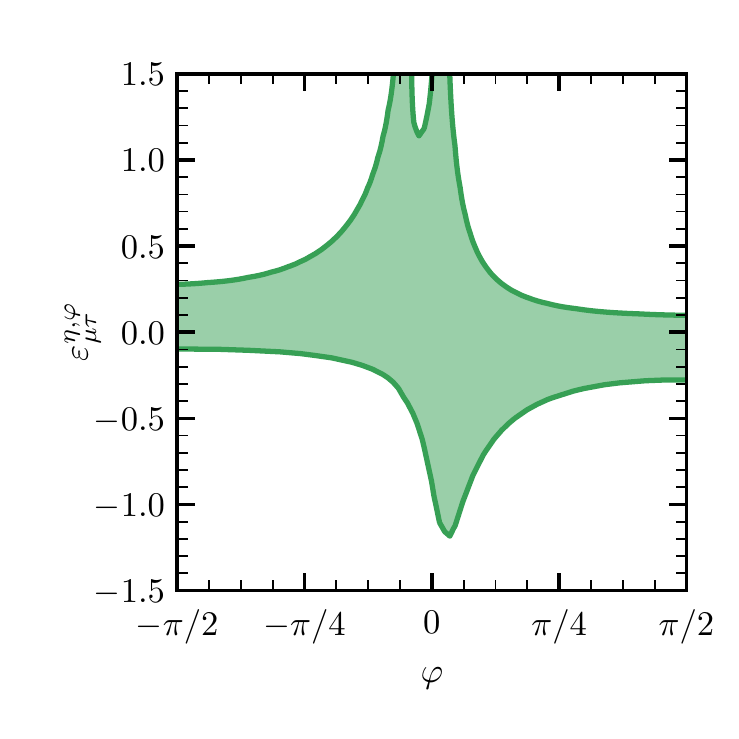}
    
    \vspace{-2ex}
    \caption{The 90\% CL allowed regions for each NSI parameter along the angle $\varphi$ from the Phase-II run of the Borexino experiment. The bounds usually quoted correspond to the NSI parameter values at $\varphi = 0$. We have fixed $\eta = 0$, corresponding to a pure electron NSI when $\varphi = \pm \pi/2$. The dark grey region shows where we conservatively assumed the adiabatic limit to break down ($\gamma < 100$).}
    \label{fig:nsi-borexino}
\end{figure}

We show the $90\%$ CL allowed regions in \cref{fig:nsi-borexino} and the corresponding $\Delta \chi^2$ values for the extremal cases of $\varphi = 0$ and $\varphi = \pi/2$ in \cref{fig:chi2_borexino} in \cref{sec:app_delta}. The shapes of these regions can be understood by expanding the trace of \cref{eq:dr_gen} using the \nue cross section of \cref{eq:xsec_el} when only one $\nsi{\alpha\beta}$ is turned on at a time. The resulting formula for the rate then contains three types of terms: a propagation-only term, which contains NSI effects only at the level of neutrino propagation; a term linear in $\xi^e \nsi{\alpha\beta}$, which can be understood as an inteference term between the SM and NSI; and a positive-definite term quadratic in $\xi^e \nsi{\alpha\beta}$, which encodes the pure NSI effect in the cross section. These terms can be explicitly seen in \cref{eq:cancel_diag_er,eq:cancel_odiag_er} in the context of our DD analysis.

At $\varphi = 0$, the NSI effect is purely due to a change in the matter potential experienced by neutrinos on their way out of the Sun, altering neutrino propagation as per the description of \cref{subsec:three_nu_osc}. This leads to constraints that are only due to propagation effects, with the neutrino-electron cross section unchanged. Around $\varphi = 0$ and for $\nsi{\alpha\beta} \lesssim 1$, NSI effects remain dominated by propagation-only effects, but non-standard cross section terms linear in $\nsi{\alpha\beta}\xi^e$ begin to contribute. While the impact on the expected rate due to propagation-only effects in this regime is approximately symmetric under the exchange $\nsi{\alpha\beta} \rightarrow -\nsi{\alpha\beta}$, the effect due to the term linear in $\nsi{\alpha\beta}\xi^e$ is approximately symmetric under the combined exchange $\{\varphi,\,\nsi{\alpha\beta}\} \rightarrow \{-\varphi,\,-\nsi{\alpha\beta}\}$. This means that, depending on the sign of $\varphi$, cross section and propagation-only effects will either positively or negatively interfere with one another. For larger values of $|\varphi|$, NSI effects are predominantly due to changes in the scattering cross section, which are dominated by the term quadratic in $\xi^e \nsi{\alpha\beta}$ in the \nue cross section for large values of $\nsi{\alpha\beta}$.

These effects are perhaps best evidenced by the lower-right panel of \cref{fig:nsi-borexino}. For $\varphi = 0$, propagation effects alone lead to rates that are irreconcilable with the data, allowing us to constrain $\nsi{\mu\tau}$ without alterations to the \nue  cross section. For small negative values of $\varphi$, both cross section and propagation effects suppress the expected neutrino rate, leading to a large overall predicted deficit and a more constrained allowed region. Beyond this, terms quadratic in $\xi^e \nsi{\mu\tau}$ begin to dominate, reducing this deficit to momentarily retrieve the SM expectation, but ultimately leading to a large predicted excess. This results in the valley at negative $\varphi$ values. On the other hand, for small positive values of $\varphi$, these two effects destructively interfere with one another, resulting in a larger allowed region. For negative values of $\nsi{\mu\tau}$, the oscillation-only and quadratic terms reinforce one another, such that this enlarged region quickly shrinks for large $\nsi{\mu\tau}$ and $\varphi$. For positive $\nsi{\mu\tau}$, these two terms instead cancel one another out.

Additionally, \cref{fig:nsi-borexino} contains a grey region for $\nsi{e\mu}$ where the adiabatic approximation used to model matter effects in the Sun may be inappropriate \cite{Friedland:2000rn,Giunti:2007ry}. Within this region, the adiabaticity parameter, defined in \cref{eq:def_adiab}, takes values $\gamma < 100$, where we have calculated $\gamma$ at $E_\nu = \SI{1}{\mega\electronvolt}$, approximately corresponding to the highest energy reached by $\mathrm{^7Be}$ neutrinos. We conservatively interpret these values to be in violation of the adiabaticity condition, $\gamma \gg 1$, such that a full numerical calculation of the density matrix elements would be required for an accurate analysis. This would be beyond the scope of our work, and since the allowed  NSI regions in \cref{fig:nsi-borexino} are almost entirely within the adiabatic regime, we do not believe a numerical treatment is necessary. We have checked that the we fulfil the adiabatic criterion for all other $\nsi{\alpha\beta}$.

Our analysis also shows that constraints are strongest for off-diagonal NSI. This is because the trace in \cref{eq:dr_gen} leads to two terms (which are equal in our case as $\delta_\mathrm{CP}$ is set to $0$)  contributing to the total NSI rate in the off-diagonal case, as opposed to a  single contribution arising from diagonal NSI. Thus, the allowed regions for off-diagonal NSI are generally tighter than those for diagonal NSI. The exception to this is in the bounds for $\nsi{ee}$, which are highly constrained due to the enhanced \nue cross section arising from the additional CC contribution via the $W$-boson exchange.

This enhanced cross section not only leads to much tighter bounds for $\nsi{ee}$ but also allows for a finely tuned second minimum at non-zero $\nsi{ee}$. With a non-zero $\nsi{ee}$, the differential rate spectrum is modified such that the total rate, given by integrating over all recoil energies, coincidentally retrieves the SM expectation. The spectrum itself, however, is significantly modified, and incorporating spectral information into our analysis would ultimately prohibit this second solution.

Our results should not be taken as a dedicated Borexino analysis. Though we have attempted to capture the variation in the calculated solar neutrino rates by introducing pull parameters, as was done in Ref.~\cite{Khan:2019jvr}, this only accounts for the theoretical uncertainty in these rates. Ultimately, a more sophisticated analysis would require a spectral fit of Borexino's data and allow for multiple NSI parameters to vary at a time. Such a fit should then allow for the various background components inherent in the data to float and permit correlations between all fit parameters. Such an analysis was recently done in the context of neutrino NSI by Ref.~\cite{Coloma:2022umy} assuming charged contributions only from the electron ($\varphi = \pi/2$). Our results should instead be taken as a demonstration of how bounds on NSI can vary dramatically depending on what one takes as the underlying charged NSI contribution.

\section{Direct Detection Experiments}
\label{sec:dd}

Finally, we turn to the main motivation of this paper: determining the potential of DD experiments within the NSI landscape using the extended framework introduced in \cref{sec:framework}.

DD experiments will provide a unique probe of neutrino interactions because they have access to both \cevns and \eves. 
This is due to the properties of the solar neutrino flux. To produce a recoil in the energy range detectable in DD experiments ($\sim 10$~eV$- 100$~keV) \eves and \cevns probe different solar neutrino populations. In particular, the main contribution to NRs comes from $^8$B neutrinos, whereas for ERs it is $pp$ and $\mathrm{^7Be}$ that contribute the most. Thus, although the scattering cross-section for \cevns is significantly larger, the much larger $pp$ neutrino flux compensates for the smaller \eves cross section.

As we saw in \cref{subsec:osc}, experiments such as Borexino have optimal sensitivity when NSI occur purely with the electron, but they rapidly lose their constraining power as $\varphi \rightarrow 0$. On the other hand, \cevns experiments such as CENNS-10 LAr, which we explored in \cref{subsec:spallation}, have excellent sensitivity when the charged NSI contribution is wholly in the proton direction, but they lose this sensitivity as this contribution turns to the electron. Moreover, having no $\nu_\tau$ component, they are completely insensitive to $\nsi{\tau\tau}$. Not only can DD experiments probe $\nsi{\tau\tau}$, but their ability to measure and discriminate between NRs and ERs means that they retain their constraining power across the full range of $\varphi$. When applied to specific BSM models, this can be crucial to identify the underlying nature of the new physics scenarios (see e.g., Ref.~\cite{Amaral:2021rzw}).

In this work, we focus on the xenon-based DD experiments   LZ~\cite{Mount:2017qzi,LUX-ZEPLIN:2018poe,LZ:2021xov}, XENON~\cite{XENON:2018voc,XENON:2020kmp,XENON:2020rca}, and DARWIN~\cite{DARWIN:2016hyl}. More specifically, we derive exclusions from the data reported by the recent LZ WIMP search \cite{LZ:2022ufs}  and the XENONnT electron-recoil excess search \cite{XENONCollaboration:2022kmb}. We also determine the expected sensitivities of LZ, XENONnT and DARWIN, based on their projections for their final experimental configurations. Similar results can be obtained with PandaX~\cite{PandaX-II:2017hlx,PandaX:2018wtu,PandaX-4T:2021bab}.

\subsection{Expected Number of Events and Statistical Procedure}

To calculate our sensitivities, we 
consider the differential rate of \cref{eq:dr_gen}
and incorporate detector effects, such as efficiency and energy resolution. The 
expected recoil rate from neutrino scattering is then given by 
\begin{equation}
    \label{eq:dr_dd_obs}
    \diff{R}{E_R} = \int_0^\infty \diff{R}{E_R'}\,\epsilon(E_R')\, \frac{1}{\sigma(E_R')\sqrt{2\pi}}\,e^{-\frac{\left(E_R - E_R'\right)^2}{2\sigma^2(E_R')}}\,\dl E_R'\,.
\end{equation}
Here, ${\mathrm{d}R}/{\mathrm{d}E_R'}$ is given by \cref{eq:dr_gen}. When computing this rate, we use the solar neutrino fluxes predicted by the B16-GS98 model \cite{Vinyoles:2016djt}, as we did for our Borexino analysis in \cref{subsec:osc}. The integral over the expected energy, $E_R'$, is the convolution that describes the effect that the detector resolution, $\sigma$, has on the observed signal; we assume this to be equivalent to a Gaussian smearing. This resolution is typically reported in terms of the measured, electron-equivalent energy, so we first convert the \cevns differential rate and NR efficiency functions into electron-equivalent energies when considering NRs. We do this by applying an energy-dependent quenching factor, which relates the two energy scales via $E_\mathrm{ee} = Q(E_\mathrm{nr})\, E_\mathrm{nr}$. We take this to be the Lindhard factor \cite{Lindhard1963INTEGRALEG} with $k=0.1735$, reflecting the $k$-value found in the fit performed by the LUX collaboration \cite{LUX:2016ezw}. Finally,  $\epsilon$ is the energy-dependent efficiency function.

The differential rate depends on the number of targets per unit mass of the detector, $n_T = N_T / m_\mathrm{det}$. For NRs, we take this to be the number density of atoms in the detector, $n_T = 1 / m_N$, where $m_N$ is the nuclear mass of the relevant xenon isotope. For ERs, this corresponds to the number of ionisable electrons given a recoil of energy $E_R$, scaled to agree with the \textit{ab initio} calculations from the relativistic random-phase approximation for xenon \cite{Chen:2016eab}. This takes into account the many-body dynamics involved in such collisions, and they have shown a consistent suppression of the rate at low recoil energies \cite{Chen:2016eab, Hsieh:2019hug}.

To compute the number of expected events within the $i^\mathrm{th}$ bin, we integrate the differential rate in the energy window defined by the edges of the bin, $[E^i_1,\, E^i_2]$, and sum the contributions from each nuclear isotope $A$ multiplied by its corresponding relative isotopic abundances, $X_A$:
\begin{equation}
    \label{eq:n_dd}
    N^i_\nu = \varepsilon \sum_A X_A \int_{E^i_1}^{E^i_2} 
    \diff{R_A}{E_R}\,\dl E_R\,.
\end{equation}
Here, $\varepsilon$ is the experimental exposure and 
${\mathrm{d}R_A}/{\mathrm{d}E_R}$ is the differential rate in \cref{eq:dr_dd_obs} due to isotope $A$.

We determine our sensitivities using a series of log-likelihood-ratio tests in which we vary only one NSI parameter at a time, fixing all others to zero\footnote{A global analysis in which all parameters are allowed to vary is beyond the scope of this article, where our aim is to motivate next-generation and far-future DD experiments to be included in future such analyses.}. We construct our likelihoods from a Poisson part and a Gaussian part, which we use to capture the effect of uncertainties on nuisance parameters. For this latter part, we consider Gaussian distributed pull parameters serving to scale the number of expected neutrino events, as we did for CENNS-10 LAr and Borexino, and the number of expected background events. We respectively label these parameters as $a$ and $b$, having standard deviations $\sigma_a$ and $\sigma_b$. Given some number of observed events in bin $i$, $N^i_\mathrm{obs}$, we define the likelihood function
\begin{align}
    \label{eq:likelihood}
    \funop{\mathcal{L}}(\nsi{\alpha\beta},\,\eta,\,\varphi,\,a,\, b) \equiv &\prod_i^{N_\mathrm{bins}} \funop{\mathrm{Po}}\left[N^i_\mathrm{obs}\,|\,(1 + a) N^i_\nu(\nsi{\alpha\beta},\,\eta,\,\varphi) + (1 + b)N^i_\mathrm{bkg} \right] \\ \nonumber
    &\quad\times \mathrm{Gauss}\left(a |\, 0,\, \sigma_{a}\,\right)\,
    \mathrm{Gauss}\left(b |\, 0,\, \sigma_{b}\,\right)\,,
\end{align}
where $N^i_\mathrm{bkg}$ is the number of expected background events in the $i^\mathrm{th}$ bin. The product is over $N_\mathrm{bins}$ bins. The number of observed events for each of our analyses depends on whether we compute exclusions based on data or derive projected limits. If the former, we take the number of observed neutrino events reported in each bin. If the latter, we assume an Asimov data set \cite{Cowan:2010js}, such that $N^i_\mathrm{obs}$ is set to the number of expected SM neutrino events ($\nsi{\alpha\beta} = 0$) in the $i^\mathrm{th}$ bin.

Finally, to derive our limits, we use \cref{eq:likelihood} to define the test statistic 
\begin{equation}
    \label{eq:test_stat}
    q_\varepsilon \equiv -2 \ln\left[\frac{\funop{\mathcal{L}}(\nsi{\alpha\beta};\, \eta=\eta_0,\, \varphi=\varphi_0,\, \hat{a},\, \hat{b})}{\funop{\mathcal{L}}(\hat{\hat{\varepsilon}}^{\eta, \varphi}_{\alpha\beta};\, \eta=\eta_0,\, \varphi=\varphi_0,\, \hat{\hat{a}},\,\hat{\hat{b}})}\right]\,,
\end{equation}
where hatted variables indicate quantities that maximise the likelihood given the parameter $\nsi{\alpha\beta}$ and double-hatted variables indicate those quantities that maximise the overall, unconstrained likelihood. By fixing the angles to take some values $\eta = \eta_0$ and $\varphi = \varphi_0$, we constrain only the parameter of interest, $\nsi{\alpha\beta}$, in each of our analyses. In practice, we perform our analysis by only considering the most dominant nuisance parameter (either $a$ or $b$), which  depends on the experiment and is discussed in detail below. The $90\%$ CL limits are then calculated by finding the value for $q_\varepsilon$, $q^\mathrm{lim}_\varepsilon$, for which
\begin{equation}
    \int_{q^\mathrm{lim}_\varepsilon}^\infty \funop{f}(q_\varepsilon)\,\dl q_\varepsilon = 0.90\,,
\end{equation}
where $\funop{f}(q_\varepsilon)$ is the distribution of the test statistic. In the limit of high statistics and that the true parameter value does not lie on the boundary of our parameter space, Wilks' theorem tells us that this distribution asymptotically follows a $\chi^2$-distribution with number of degrees of freedom $k=1$\footnote{We have checked that this holds true in all of our analyses.}. This leads to $q_\varepsilon^\mathrm{lim} = 2.71$, and our limits then follow from finding that $\nsi{\alpha\beta}$ which yields this value for the test statistic.

To implement the analysis in the sections that follow, we make use of \code~\cite{SNuDD-code}\footnote{\href{https://github.com/SNuDD/SNuDD.git}{https://github.com/SNuDD/SNuDD.git}} (Solar NeUtrinos for Direct Detection), a novel code-base that we have developed.  \code is a Python package that calculates the generalised cross section of \cref{subsec:gen_cs} and the density matrix elements of \cref{subsec:density_mat}, combining them to compute the trace of \cref{eq:dr_gen} and arrive at a prediction for the expected solar neutrino rate at a DD experiment while folding in detector efficiency and resolution effects. We will release it in a separate publication. We hope that \code facilitates future DD analyses in the NSI landscape.

Lastly, we have checked that our limits are not greatly impacted by uncertainties in the neutrino oscillation parameters when these uncertainties are computed using standard neutrino oscillations and interactions. While including NSI can significantly impact the best-fit values of and uncertainties in these parameters, especially those derived from solar neutrino experiments \cite{Esteban:2018ppq}, we note that the future medium-baseline reactor experiment JUNO \cite{JUNO:2015sjr} will be largely insensitive to the effects of NSI. In particular, the fits to the solar neutrino parameters, $\theta_{12}$ and $\Delta m_{12}^2$, and the reactor neutrino angle, $\theta_{13}$, will be robust in the presence of NSI \cite{Martinez-Mirave:2021cvh}. Moreover, the atmospheric angle ($\theta_{23}$), which impacts the muon and tau neutrino fractions, is not greatly changed by the inclusion of NSI \cite{Esteban:2018ppq}. Thus, we take the liberty of ignoring these systematics in this first study, taking the best-fit values for these parameters from Ref.~\cite{Esteban:2020cvm}; a more comprehensive analysis that includes these systematics is left for future work.

\subsection{Sensitivities in the nucleon NSI plane}
\label{sec:nuc_plane}

From the many existing DD NR constraints, we consider only the recent, leading LZ result~\cite{LZ:2022ufs} to derive a bound in the NSI landscape. We take the efficiency function given in Ref.~\cite{LZ:2022ufs} and we model the energy resolution according to Ref.~\cite{LUX:2016rfb}. For our signal region, we use the 90$\%$ quantile of the nuclear recoil band in the S1$c$ (scintillation) and S2$c$ (ionisation) event reconstruction space as shown in Fig.~4 of Ref.~\cite{LZ:2022ufs}. For our analysis, we focus only on the low-energy region ($[5,\, 15]\,{\rm keV}$), which is sensitive to solar neutrinos, and we integrate over it to constitute one bin. We take the number of expected background and observed events to be 1 and 0, respectively (see Ref.~\cite{Chang:2022jgo}). We have validated our procedure by reproducing the WIMP-nucleon cross-section limit reported by the collaboration (Fig.~5 of Ref.~\cite{LZ:2022ufs}) to a good agreement in the low and high DM mass values.

To assess the future prospects of detecting an NR signal, we plot the projected sensitivities for LZ, XENONnT, and DARWIN with the full exposures of $15.34$, $20$, and $200$~\si{\ton\year}, respectively. We have taken a background-free scenario in the NR search, which is consistent with experimental aims (for our analysis, the neutrino signal is not considered a background).  With no backgrounds, we then conservatively perform a one-bin analysis, using the total number of expected solar neutrino NR events as our observation.Additionally, since the expected background is so low, the nuisance parameter that will have the largest impact on our sensitivities will be associated with the solar neutrino flux, and hence we set $b = 0$ in this case. For our remaining pull parameter in \cref{eq:likelihood}, we assume an uncertainty of $\sigma_a = 12\%$, reflecting the $12\%$ uncertainty in the theoretical value of the total $\mathrm{^8B}$ flux in the B16-GS98 SSM \cite{Vinyoles:2016djt}. We take the resolution function for LZ at full exposure to be the same as that of their first result \cite{LZ:2022ufs, LUX:2016rfb}, whereas for XENONnT and DARWIN we use the resolution function given in Ref.~\cite{XENON:2020rca}. The NR efficiency functions, as presented by the collaborations, reach $50\%$ at $3.8\,\knr$ for LZ \cite{LZ:2018qzl} and $5.7\,\knr$ for XENONnT and DARWIN~\cite{XENON:2018voc}. However, to explore how DD experiments could feasibly probe NSI in the future, we take the liberty of further lowering these thresholds. This is to take advantage of the higher $\mathrm{^8B}$ rate at these energies.

In particular, we augment the efficiency functions such that, for each future experiment, the efficiency instead reaches $50\%$ at \SI{3}{\kevnr}, which we consider to be a feasible future goal. For instance, the xenon-based LUX experiment was able to reach thresholds as low as \SI{1.1}{\kevnr} while retaining NR/ER discrimination \cite{LUX:2015abn}. The LUX collaboration has also developed techniques allowing for single-photon sensitivities, resulting in sensitivities to much lower recoil energies at the cost of a lower overall detection efficiency. Finally, the XENON1T collaboration has recently performed a dedicated $\mathrm{^8B}$ search by lowering their threshold to \SI{1.6}{\kevnr}, achieved by relaxing the necessity for a three-fold S1 coincidence in the PMTs to a two-fold one \cite{XENON:2020gfr}. Furthermore, taking the systematic $^8$B uncertainty to be 12\%, we find that lowering the threshold further provides little-to-no benefit in terms of NSI sensitivity. For each future experiment, we take $E^{\max}_{R}=30\,{\rm keV}$.

\begin{figure}[p!]
    \vspace*{-10ex}
    \includegraphics[]{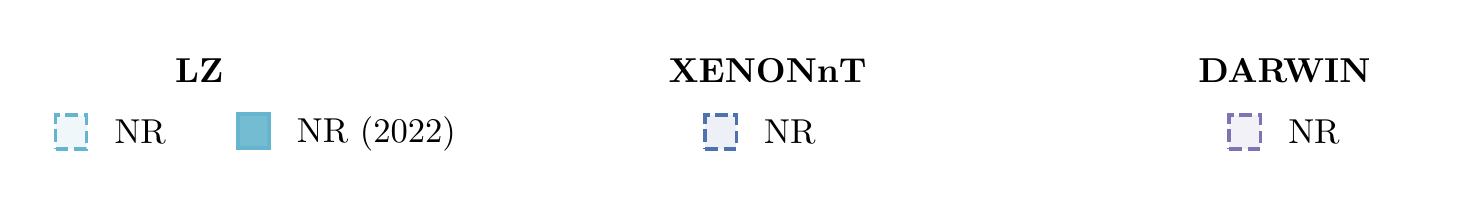}

    \vspace*{-6ex}
    \includegraphics[]{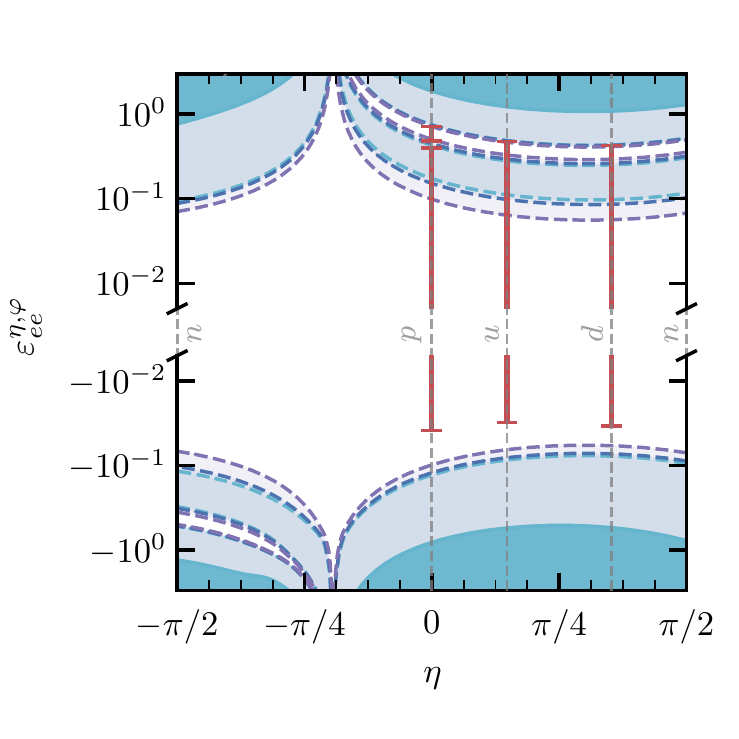}
    \includegraphics[]{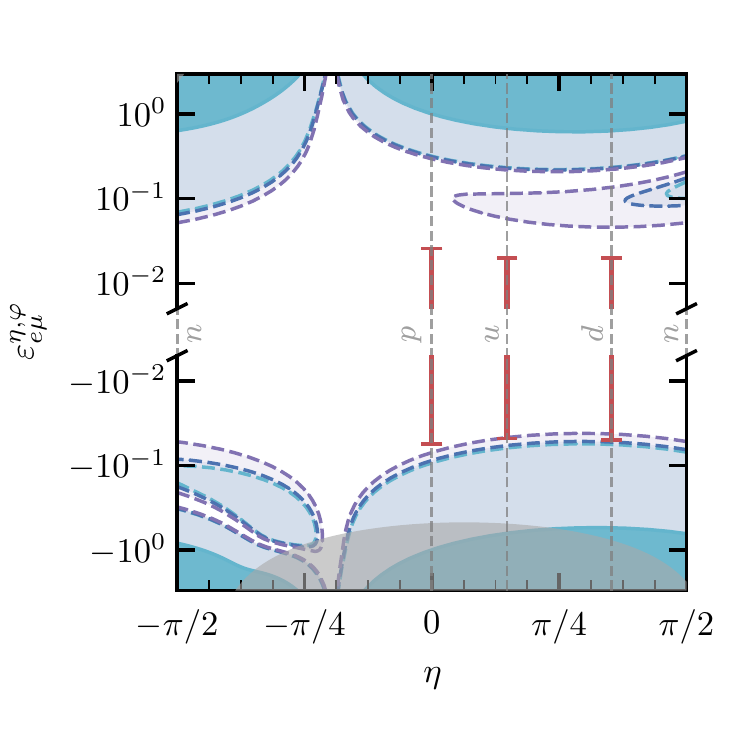}

    \vspace*{-6ex}
    \includegraphics[]{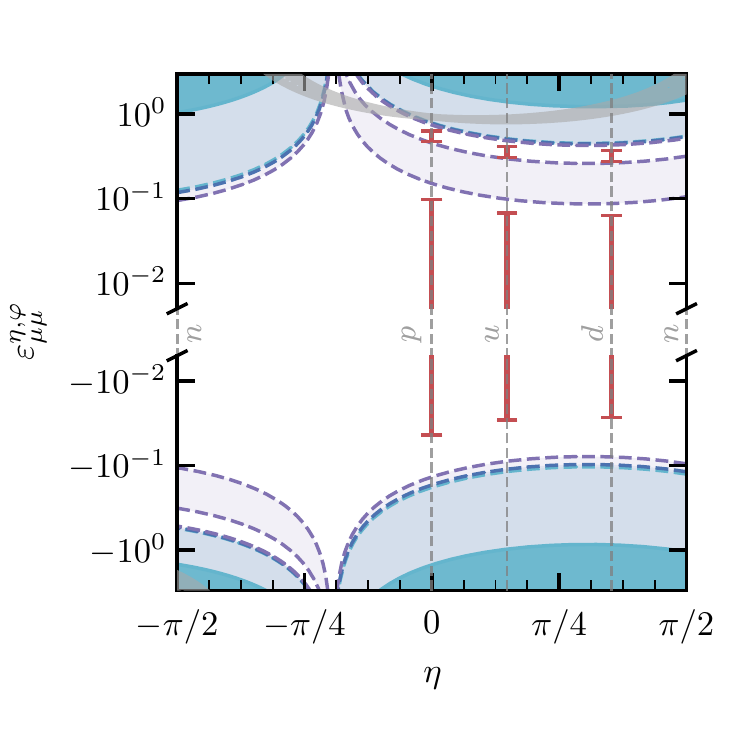}
    \includegraphics[]{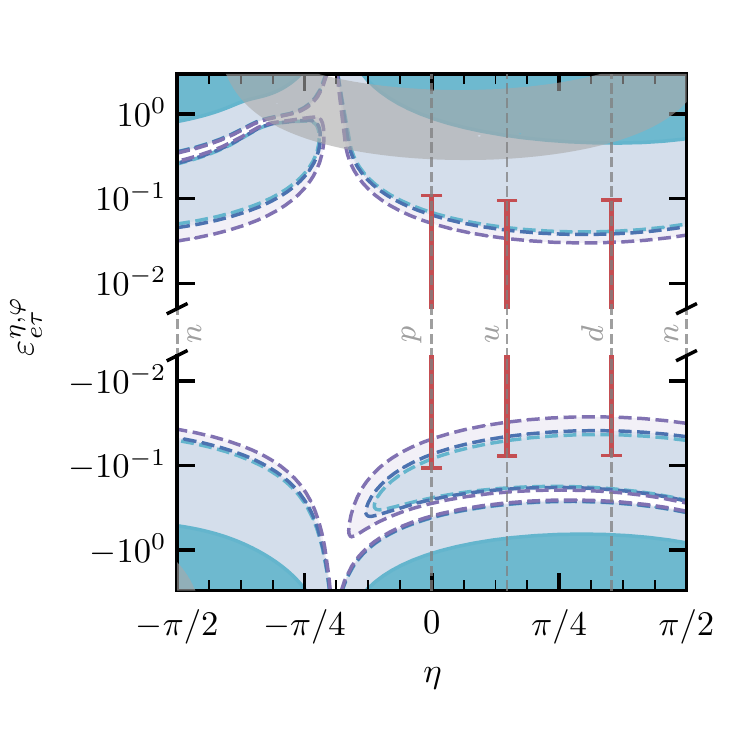}

    \vspace*{-6ex}
    \includegraphics[]{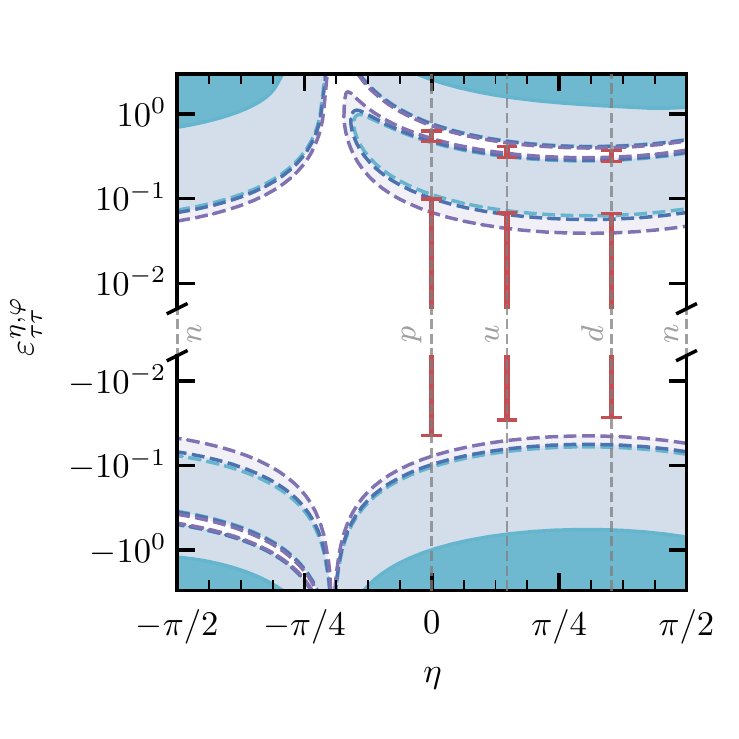}
    \includegraphics[]{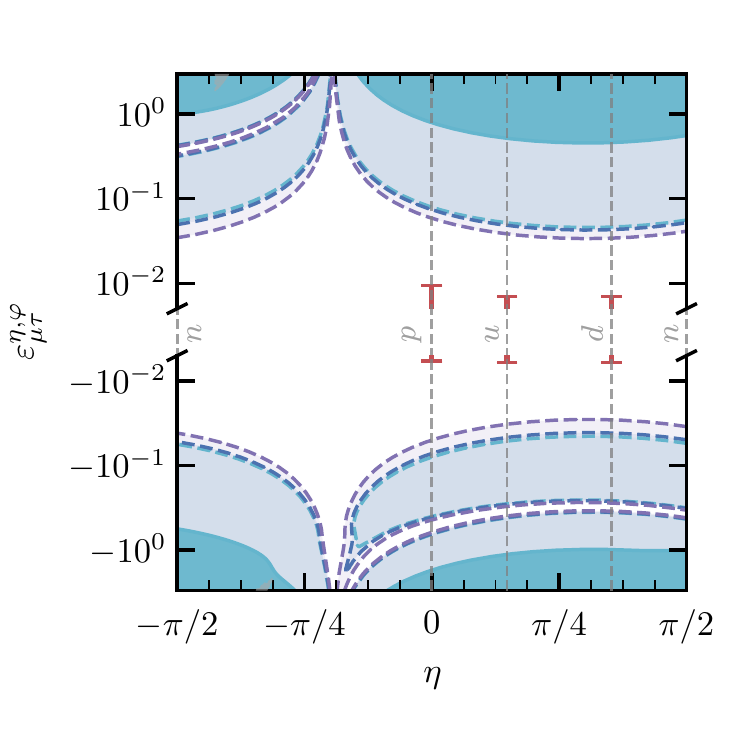}

    \vspace{-2ex}
    \caption{The 90\% CL limits set by multi-ton LXe DD experiments in the NSI parameter space using NRs. Shown are the limits from the first results of LZ~\cite{LZ:2022ufs} (turquoise), the full LZ exposure (baby blue), XENONnT (dark blue), and DARWIN (purple) in the typically assumed case that $\varphi = 0$. The bounds from the global analysis of Ref.~\cite{Coloma:2019mbs} are shown for comparison (red bars). The grey regions indicate where the adiabaticity parameter is such that $\gamma < 100$, where we consider the adiabatic approximation to begin to falter \cite{Giunti:2007ry}.}
    \label{fig:dd-nsi-cevns_NR}
\end{figure}

We show our results in \cref{fig:dd-nsi-cevns_NR}. 
The shaded areas represent the 90\% CL limits set by the different experimental configurations. From less  constraining (smaller areas) to more constraining (larger areas), we show the limits derived from the first LZ results (turquoise with solid boundary) and the expected sensitivities of the full exposure of LZ (baby blue, dashed), XENONnT (dark blue, dashed) and the proposed DARWIN (purple, dashed). For comparison, we also show with red bars the NSI limits derived from the global study of Ref.~\cite{Coloma:2019mbs}, which included the results from COHERENT and a variety of oscillation experiments, when NSI take place purely with the proton ($\eta=0$), up-quark ($\eta=\tan^{-1}(1/2)$) and down-quark ($\eta=\tan^{-1}(2)$). We extract the limits we have computed using the LZ WIMP search data and tabulate them in \cref{tab:lz_results}, contrasting them with the results from the global fits of \cite{Coloma:2019mbs}.  We see that, currently, DD experiments are not sensitive to globally allowed NSI values, but our projections indicate that they will be in the near future.

\begin{table}[t]
\begin{tabular*}{\textwidth}{@{\extracolsep{\fill}}c c c}
\toprule
\multicolumn{1}{c}{} & \textcolor{lz22}{LZ 2022 (\textbf{this work})} & \textcolor{global_fits}{Global Fits} \cite{Coloma:2019mbs}\\ \midrule
\quad $\varepsilon_{ee}^{u}$ & $[-0.545,\, 1.222]$ & $[-0.031,\, 0.476]$ \\
\quad $\varepsilon_{\mu\mu}^{u}$ & $[-0.971,\, 1.397]$ & $[-0.029,\, 0.068] \oplus [0.309,\, 0.415]$ \\
\quad $\varepsilon_{\tau\tau}^{u}$ & $[-0.645,\, 1.598]$ & $[-0.029,\, 0.068] \oplus [0.309,\, 0.414]$ \\
\quad $\varepsilon_{e\mu}^{u}$ & $[-0.630,\, 0.679]$ & $[-0.048,\, 0.020]$ \\
\quad $\varepsilon_{e\tau}^{u}$ & $[-0.721,\, 0.558]$ & $[-0.077,\, 0.095]$ \\
\quad $\varepsilon_{\mu\tau}^{u}$ & $[-1.120,\, 0.518]$ & $[-0.006,\, 0.007]$ \\ \midrule[0.1pt]
\quad $\varepsilon_{ee}^{d}$ & $[-0.540,\, 1.084]$ & $[-0.034,\, 0.426]$ \\
\quad $\varepsilon_{\mu\mu}^{d}$ & $[-0.863,\, 1.233]$ & $[-0.027,\, 0.063] \oplus [0.275,\, 0.371]$ \\
\quad $\varepsilon_{\tau\tau}^{d}$ & $[-0.576,\, 1.241]$ & $[-0.027,\, 0.067] \oplus [0.274,\, 0.372]$ \\
\quad $\varepsilon_{e\mu}^{d}$ & $[-0.542,\, 0.635]$ & $[-0.050,\, 0.020]$ \\
\quad $\varepsilon_{e\tau}^{d}$ & $[-0.655,\, 0.455]$ & $[-0.076,\, 0.097]$ \\
\quad $\varepsilon_{\mu\tau}^{d}$ & $[-0.982,\, 0.461]$ & $[-0.006,\, 0.007]$ \\ \midrule[0.1pt]
\quad $\varepsilon_{ee}^{p}$ & $[-1.805,  4.195]$ & $[-0.086,\, 0.884] \oplus [1.083,\, 1.605]$ \\
\quad $\varepsilon_{\mu\mu}^{p}$ & $[-3.330,  4.791]$ & $[-0.097,\, 0.220] \oplus [1.063,\, 1.410]$ \\
\quad $\varepsilon_{\tau\tau}^{p}$ & $[-2.209,  5.710]$ & $[-0.098,\, 0.221] \oplus [1.063,\, 1.408]$ \\
\quad $\varepsilon_{e\mu}^{p}$ & $[-2.209,  2.249]$ & $[-0.124,\, 0.058]$ \\
\quad $\varepsilon_{e\tau}^{p}$ & $[-2.434,  2.006]$ & $[-0.239,\, 0.244]$ \\
\quad $\varepsilon_{\mu\tau}^{p}$ & $[-3.849,  1.772]$ & $[-0.013,\, 0.021]$ \\ \midrule[0.1pt]
\quad $\varepsilon_{ee}^{n}$ & $[-1.714,  2.915]$ & --- \\
\quad $\varepsilon_{\mu\mu}^{n}$ & $[-2.331,  3.282]$ & --- \\
\quad $\varepsilon_{\tau\tau}^{n}$ & $[-1.564,  2.705]$ & --- \\
\quad $\varepsilon_{e\mu}^{n}$ & $[-1.426,  1.846]$ & --- \\
\quad $\varepsilon_{e\tau}^{n}$ & $[-1.829,  1.147]$ & --- \\
\quad $\varepsilon_{\mu\tau}^{n}$ & $[-2.275,  1.250]$ & --- \\\bottomrule
\end{tabular*}
\caption{$90\%$ CL allowed intervals for NSI in the up quark, down quark, proton, and neutron directions. Shown are the results from our analysis of the LZ 2022 data \cite{LZ:2022ufs} and those of the global fit study of Ref.~\cite{Coloma:2019mbs}. Note that the latter do not quote NSI in the neutron direction.
\label{tab:lz_results}}
\end{table}

Like in \cref{fig:nsi-borexino} of our Borexino analysis in \cref{subsec:osc}, \cref{fig:dd-nsi-cevns_NR} contains grey regions, indicating those points in the parameter space where the adiabatic approximation may be invalid. Within these regions, $\gamma < 100$, where in this case we have calculated $\gamma$ at $E_\nu = \SI{16}{\mega\electronvolt}$, approximately corresponding to the highest energy reached for $\mathrm{^8 B}$ neutrinos.  As  current global fits show that the allowed values of the NSI parameters are firmly within the adiabatic regime, we believe that our analytical approach is valid for the regions of interest.  However, it is important to keep in mind that our sensitivities may be inaccurate within the grey bands.

Our limits exhibit many interesting non-trivial features. Specifically, we see that there are regions in each NSI parameter space where every DD experiment loses sensitivity. The two most remarkable of these regions are, firstly, the strong cancellation in the angle $\eta$ occurring at $\eta \approx -35^\circ$ and, secondly, the band of insensitivity in $\nsi{\alpha\beta}$ across the entire range of $\eta$ values (made manifest by the gaps in the projected sensitivity areas). These blind spots present a challenge for DD experiments, and they should be understood if DD experiments are to maximise their constraining power in the NSI landscape.

\bigskip 

We first consider the cancellation in $\eta$, which occurs at the same point regardless of the nature of the NSI. From~\cref{eq:sig_CEVNS_gen} and~\cref{eq:g_cevns}, we see that the non-standard contribution to the \cevns cross section vanishes when $\xi^p Z + \xi^n N=0$ , recovering the SM cross section regardless of the value of $\nsi{\alpha\beta}$. For a given nuclear isotope, this occurs when\footnote{This cancellation was used in direct dark matter detection to argue that dark matter particles might escape detection in some specific targets (see e.g., Ref.\,\cite{Feng:2013vod}).}
\begin{equation}
    \label{eq:blind-asymp}
    \eta = \tan^{-1}\left(-\frac{Z}{N} \cos\varphi\right)\,.
\end{equation}
This condition depends on the choice of target material. For composite materials or targets, or non-mononuclidic targets, the cancellation is not exact, as the contributions from the different isotopes must be added up in \cref{eq:n_dd}. Yet, for xenon the ratio $Z/N$ is very similar in all its natural isotopes, and the observed rate is greatly reduced for $\eta \approx -35^\circ$ when $\varphi = 0$. Since stable nuclei tend to have similar $Z/N$ fractions, the position of the blind spot does not  vary greatly for different target choices. Interestingly, argon (a target employed in current detectors \cite{DarkSide:2014llq} and planned tonne-scale ones \cite{DarkSide-20k:2017zyg}) leads to a  considerable shift,  with $\eta \approx -39^\circ$. This could lead to a noticeable effect if a full spectral analysis is performed of the observed signal, thus strengthening the notion of complementarity among  DD targets. This would be even more noticeable in light nuclei, such as He, F, Na, or Si, for which $Z/N\sim 1$ and the cancellation takes place for $\eta\rightarrow 45^\circ$ (although a large detector would still be needed).

\bigskip

The second blind spot occurs at intermediate values of $\nsi{\alpha\beta}$, stretching over the full range of $\eta$ values. These insensitivity bands arise due to interference effects, where the NSI contribution is cancelled and thus the SM \cevns differential rate is restored. The exact location of these bands differs for flavour-conserving and flavour-violating NSI.

In the case of flavour-conserving NSI, we can derive a simple analytical formula for the values of the NSI parameters leading to a non-trivial realisation of the SM differential rate. This relation, which defines the centres of each of these insensitivity bands where the NSI contribution exactly cancels, has previously been pointed out in Ref.~\cite{Dutta:2020che} and in our framework is given by
\begin{equation}
    \label{eq:blind-band-on}
    \varepsilon_{\alpha\alpha}^{\eta,\varphi} = \frac{Q_{\nu N}}{\xi^{p}Z + \xi^{n}N}\,.
\end{equation}
The dependence on $\eta$, encoded in $\xi^p$ and $\xi^n$, gives us the band over different values of $\nsi{\alpha\alpha}$ as a function of $\eta$. Note that, as with the first blind spot, the locations of these bands depend on the choice of the target material due to the dependence on $Z$ and $N$. For $\eta = 0$, for instance, \cref{eq:blind-band-on} gives $\nsi{\alpha\alpha} \approx 0.6$ for xenon, whereas it yields the lower $\nsi{\alpha\alpha}\approx 0.5$ for argon. Since the non-trivial cancellation occurs at  different values for different targets, this could be important in determining whether global minima are driven by data or just artefacts of the blind spot that nuclei have, see for instance Ref.~\cite{Chaves:2021pey}. Considering different materials thus gives us one possible avenue to mitigate this particular loss of sensitivity, though we note that the blind spots of xenon and argon move closer together as $\eta\rightarrow\pi/2$.

In the case of flavour-changing NSI, the cancellation condition becomes more complicated and is only retrieved in the (correct) basis-independent formulation of the scattering rate in terms of the trace $\mathrm{Tr}\left[\bb{\rho}\ {\mathrm{d}\bb{\zeta}}/{\mathrm{d} E_R}\right]$ in~\cref{eq:dr_gen}. Due to the flavour-coherence effects, we still expect regions where the SM-NSI interference term cancels the  NSI-only term; however, these regions now also depend on the density matrix elements. To investigate this behaviour, we consider, as a simplification,  the values of $\varepsilon_{\alpha\beta}^{\eta,\varphi}$ for which the differential rate spectrum returns to its expected SM value for a given recoil energy $E_R$. This prescription removes the need to integrate over $E_R$ to find the number of events. From~\cref{eq:dr_gen,eq:sig_CEVNS_gen}, we find that in general, for $\alpha \neq \beta$, the condition for restoring the SM rate reads
\begin{equation}
  \label{eq:blind-band-off}
  \int_{E_{\nu}^{\mathrm{min}}}\diff{\phi_{\nu_{e}}}{E_{\nu}}\left(1 - \frac{m_{N}E_{R}}{2E_{\nu}^{2}}\right) \left[(\xi^{p}Z + \xi^{n}N)(\rho_{\alpha\alpha} + \rho_{\beta\beta})\,\varepsilon_{\alpha\beta}^{\eta,\varphi} - 2 \,Q_{\nu N}\,\rho_{\alpha\beta}\right]\,\dl E_{\nu}= 0\,.
\end{equation}
The difference in the forms of the relations in \cref{eq:blind-band-on,eq:blind-band-off} is why the positions of these bands are different for flavour-conserving and flavour-violating NSI. In particular, we note that, unlike in the case of the former, for the latter we have a flavour dependence through the appearance of the density matrix elements. This is why, for instance, we see a sign flip of the bands in the case of $\nsi{e\tau}$ and $\nsi{\mu\tau}$ with respect to  $\nsi{e\mu}$, as the relevant off-diagonal density matrix elements $\rho_{e\tau}$ and $\rho_{\mu\tau}$ are negative in contrast to $\rho_{e\mu}$.

As we can see from~\cref{fig:dd-nsi-cevns_NR}, the insensitivity bands for the off-diagonal NSI elements $\nsi{e\mu}$ and $\nsi{e\tau}$ exhibit some interesting features at $\eta\approx -5/16\, \pi$, where they seem to expose a \textit{kink}. The origin of these kinks in the off-diagonal NSI insensitivity bands can be traced back to the appearance of the off-diagonal density matrix elements, $\rho_{\alpha\beta}$, in the \cevns cancellation condition. These kinks arise because the last term in~\cref{eq:blind-band-off} proportional to $\rho_{\alpha\beta}$ undergoes a qualitative change of behaviour at $\eta\approx -5/16\, \pi$.

We will discuss the behaviour of the kinks using the insensitivity band for $\varepsilon^{\eta,\varphi}_{e\mu}$ in the top right plot of~\cref{fig:dd-nsi-cevns_NR} as an example. At very negative angles $\eta \approx - \pi/2$, the off-diagonal interference term in~\cref{eq:blind-band-off} proportional to $\rho_{\alpha\beta}$ has an extremum for positive $\varepsilon^{\eta,\varphi}_{\alpha\beta}$.  In this regime, the behaviour of the cancellation line (which occurs at negative $\varepsilon^{\eta,\varphi}_{\alpha\beta}$) is entirely dominated by the NSI-only term proportional to the diagonal density matrix elements, $\rho_{\alpha\alpha}$. However, at larger angles, $\eta \approx -5/16\, \pi$, the extremum in the off-diagonal term  shifts from positive values of $\varepsilon^{\eta,\varphi}_{\alpha\beta}$ to negative values, and it hence begins to dominate the behaviour of the cancellation bands in~\cref{eq:blind-band-off}. This change of behaviour in the cancellation integral leads to the appearance of the kinks in the insensitivity band.

For $\varepsilon^{\eta,\varphi}_{e\tau}$, the same effect leads to the appearance of a kink, but with opposite signs of $\varepsilon^{\eta,\varphi}_{\alpha\beta}$. In principle, the same reasoning holds for $\varepsilon^{\eta,\varphi}_{\mu\tau}$; however, in this case, the off-diagonal term exhibits an almost negligible extremum, such that there is no visible kink. Finally, the fact that the behaviour of the off-diagonal density matrix element $\rho_{\alpha\beta}$ is responsible for the appearance of these kinks also explains why they are absent for the diagonal NSI elements, $\varepsilon^{\eta,\varphi}_{\alpha\alpha}$, since there is no contribution from $\rho_{\alpha\beta}$.

\medskip

\subsection{Sensitivities in the charged NSI plane}

\begin{figure}[p!]
    \vspace*{-10ex}
    \includegraphics[]{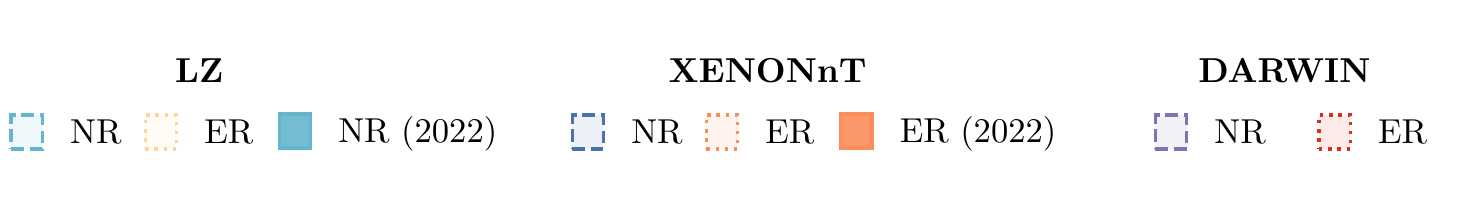}

    \vspace*{-6ex}
    \includegraphics[]{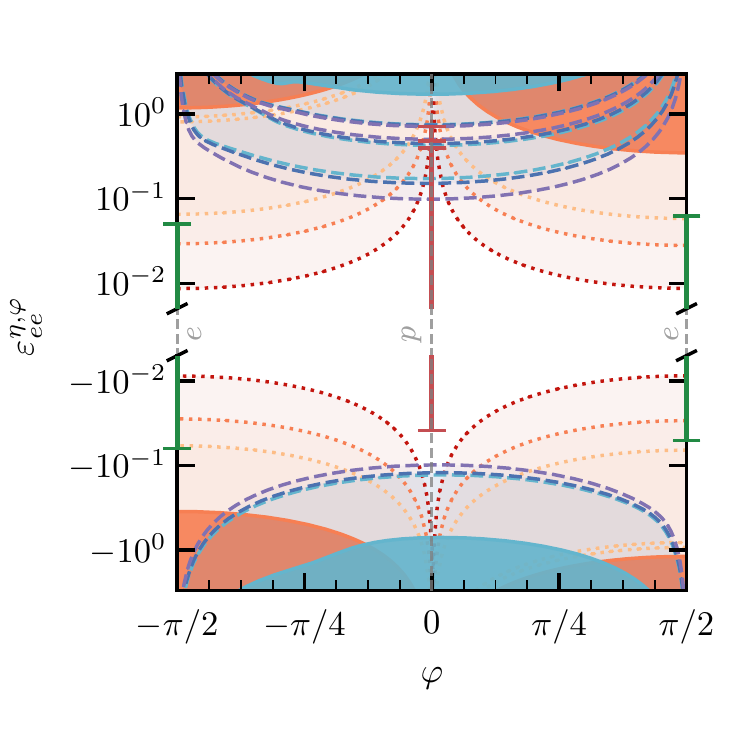}
    \includegraphics[]{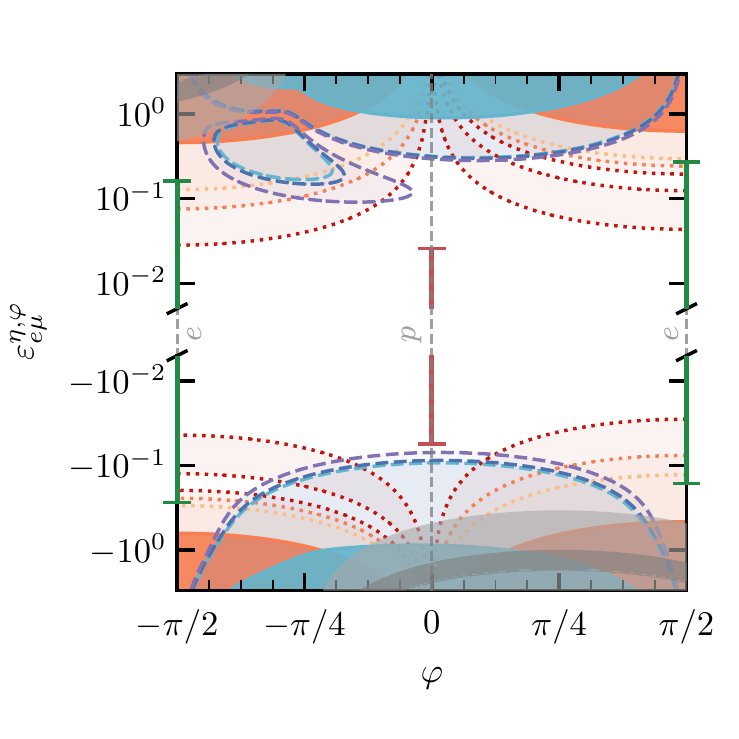}

    \vspace*{-6ex}
    \includegraphics[]{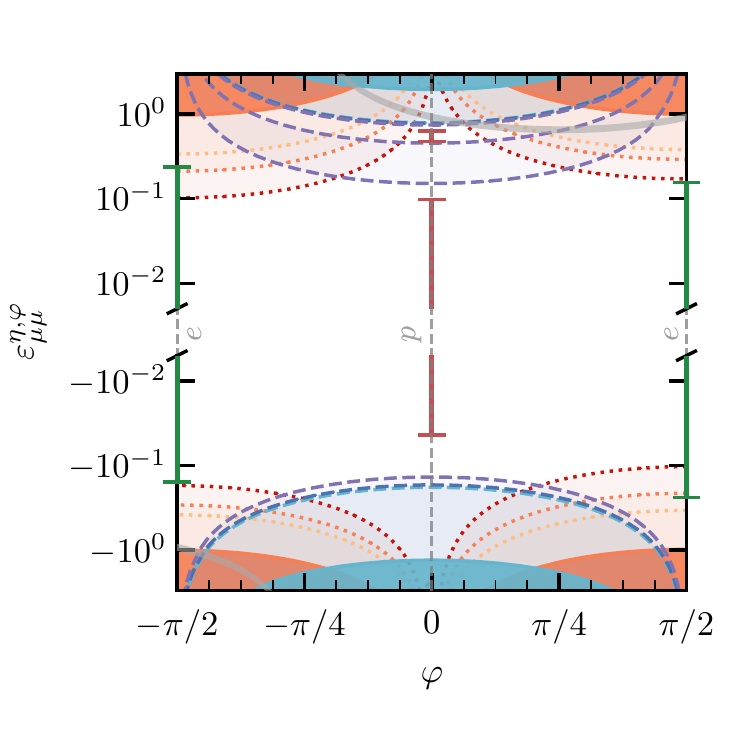}
    \includegraphics[]{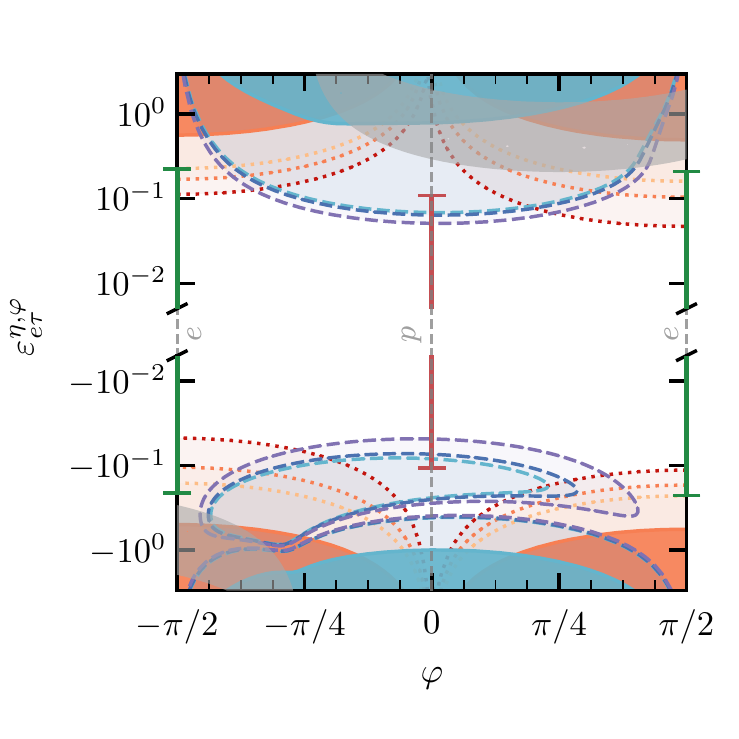}

    \vspace*{-6ex}
    \includegraphics[]{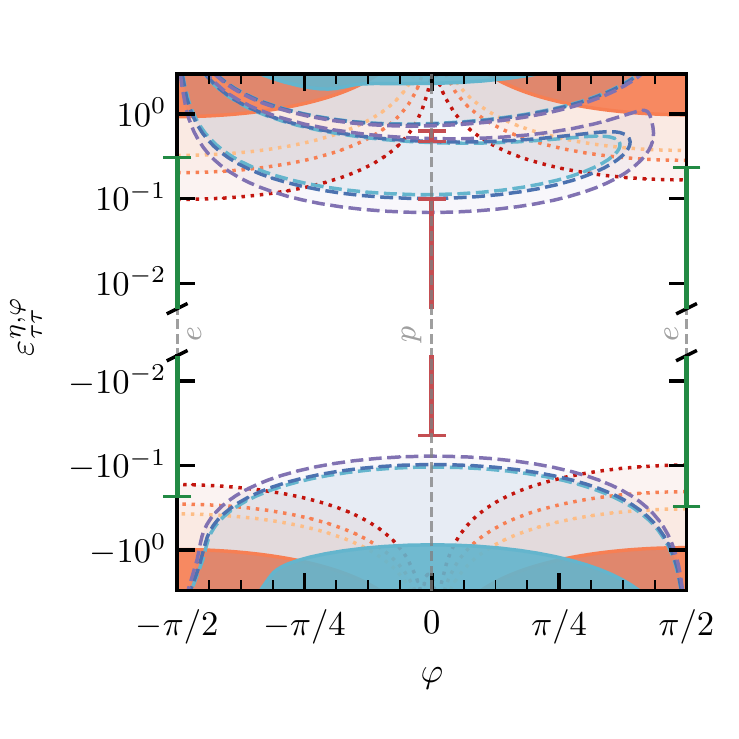}
    \includegraphics[]{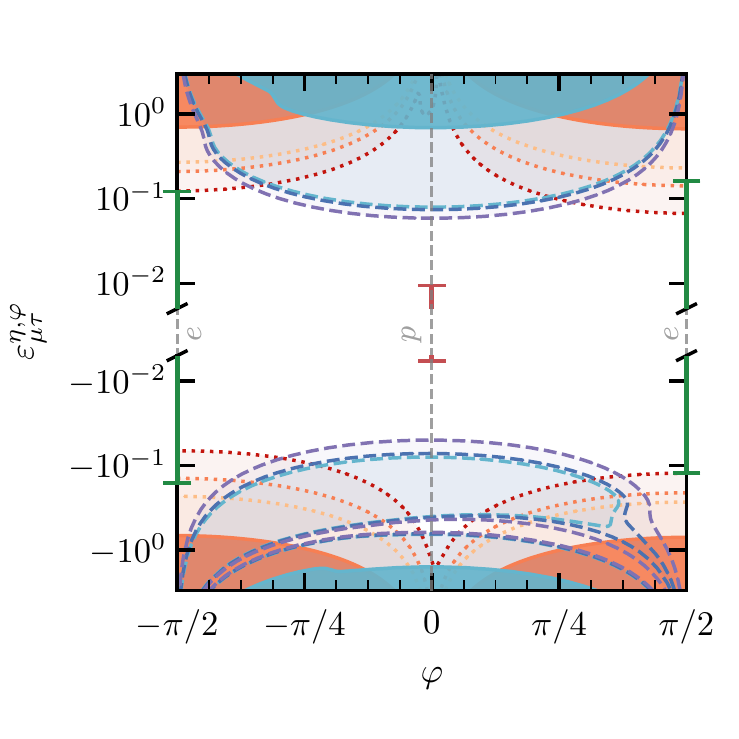}

    \vspace{-3.5ex}
    \caption{Same as \cref{fig:dd-nsi-cevns_NR} but now fixing the NSI to lie in the $(\varepsilon^{p}_{\alpha\beta} \varepsilon^{e}_{\alpha\beta})$-plane ($\eta=0$) and using both NRs and ERs. The colour scheme for the NR results is the same as in \cref{fig:dd-nsi-cevns_NR}. For the ER analyses, we show the limits derived from the first set of data from XENONnT \cite{XENONCollaboration:2022kmb} (dark orange), as well as projections for XENONnT (amber), LZ (light orange), and DARWIN (red). The bounds from the global analysis of Ref.~\cite{Coloma:2019mbs} (red bars) and the Borexino analysis of Ref.~\cite{Coloma:2022umy} (green bars) are shown for comparison. The grey regions show where the adiabatic limit breaks down ($\gamma <100$) for energies relevant to NRs (light grey) and ERs (dark grey).}
  \label{fig:dd-nsi-cevns_ER}
\end{figure}

While the \eves cross section will only be modified when $\nsi{\alpha\beta} \neq 0$ and $\varphi \neq 0$, even in the case of pure nuclear NSI couplings ($\varphi=0$), propagation effects within the solar medium can still alter the expected ER rate in DD experiments. Thus, since the charged plane contains proton NSI modifications, both ER and NR signals must be considered. There is only one direction in which a non-zero NSI will not affect the NR signal, and that is precisely the electron-only direction, $\varphi=\pm \pi/2$ and $\eta=0$. For this reason, in \cref{fig:dd-nsi-cevns_ER} we include both the NR and ER analysis to show the projected sensitivities of DD when NSI lie in the $(\varepsilon^{p}_{\alpha\beta} \varepsilon^{e}_{\alpha\beta})$-plane.

Currently, the world-leading DD constraint on ERs  comes from XENONnT~\cite{XENONCollaboration:2022kmb}, which has thus far reached an exposure of $\SI{1.16}{\ton\year}$. We have replicated this analysis by taking the efficiency function, expected backgrounds, and observed number of events from Ref.~\cite{XENONCollaboration:2022kmb}, where, for the background model $B_0$, we subtract their expected solar neutrino background (Fig.~4 of Ref.~\cite{XENONCollaboration:2022kmb}). In the signal region of $[0-140]\,\si{{\rm keV}}$, the SM counts predicted by \code is $274$, which is lower than the quoted $300\pm30$~\cite{XENONCollaboration:2022kmb}. This could be due to the fact that the neutrino signal in Ref.~\cite{XENONCollaboration:2022kmb} uses a simplified modelling of the neutrino spectrum. Specifically, \code uses the relativistic random phase approximation (RRPA) studied in Ref.~\cite{Chen:2016eab} along with a series of step-functions to model the effect of electron binding energies in xenon. This reduces the overall rate and introduces discontinuous jumps in the spectra when more electrons can be ionised above certain energies. One such discontinuity is around $\sim30\,{\rm keV}$, which does not appear to be present in Fig.~4 of Ref.~\cite{XENONCollaboration:2022kmb}. When we remove both RRPA and the step function approximation in \code to determine the expected number of solar neutrino events with the same setup as Ref.~\cite{XENONCollaboration:2022kmb}, we predict $299$ solar neutrino events.

Unlike in the NR case, the sizeable backgrounds for ERs mean that spectral information should be used to harness greater sensitivity. Following XENONnT, our analysis uses $2\,{\rm keV}$-width bins from $[0-30]\,{\rm keV}$. We have refrained from using the entire signal region because the backgrounds are at their lowest at low energies. Additionally, the backgrounds below $30\,{\rm keV}$ are dominated by one source, $^{214}$Pb, which has an associated uncertainty that we treat as a nuisance parameter. If we consider higher recoil energies, backgrounds such as $^{124}$Xe, $^{83{\rm m}}$Kr and $^{136}$Xe become important, all of which have different associated uncertainties. A dedicated experimental analysis would include all backgrounds and their uncertainties to perform a multivariate fit to the observed events. We believe such an in-depth study should be done in consort with the collaboration. The uncertainty we take for the $^{214}$Pb dominated background is $\sigma_{b}=12.5\%$~\cite{XENONCollaboration:2022kmb}. We consider this to be our dominant nuisance parameter and find that if we instead perform the fit assuming the $pp$ neutrino flux is the dominant nuisance parameter ($\sigma_{a}=1\%$~\cite{Coloma:2022umy}) our limits for XENONnT see a substantial improvement.

As mentioned above, the potential for future ER analyses relies primarily on the anticipated background reduction. For the full XENONnT run, we take the backgrounds from Ref.~\cite{XENON:2020kmp}, for LZ we use Ref.~\cite{LZ:2021xov}, and for DARWIN we use the predictions given in Ref.~\cite{Baudis:2013qla}. Unlike with the NR signal, the ER neutrino spectrum does not fall off sharply at $E_R \sim \rm{keV}$, so we do not extend the ER efficiency functions to lower energies as we did for our NR projections. We use the efficiency functions given in Ref.~\cite{LZ:2018qzl} for LZ, and in Ref.~\cite{XENON:2020rca} for XENONnT and DARWIN, where they reach $50\%$ at $1.46\,\rm{keV}_{ee}$ and $1.51\,\rm{keV}_{ee}$, respectively. For these projections, we also perform spectral analyses, binning with $2\,{\rm keV}$-width bins in the energy ranges $[0-60]\,{\rm keV}$ for XENONnT and DARWIN, but in the range $[0-30]\,{\rm keV}$ for LZ. LZ's maximum is limited by their reported efficiency function~\cite{LZ:2018qzl}. Additionally, we have assumed that these experiments will have a greater understanding of their backgrounds and therefore consider the dominant nuisance parameter to be the $pp$ neutrino flux, $\sigma_{a}=1\%$. We  believe this is an achievable goal for future DD experiments and see our projected sensitivities as an additional motivation for improved understanding and reduction of backgrounds. We also considered a far-future xenon detector with an exposure of $10^3~\si{\ton\year}$ as in Ref~\cite{Dutta:2020che} and found that the $pp$ flux uncertainty drives the projected sensitivity to the extent that even with five times the exposure of DARWIN and no backgrounds, only marginal improvements are made to the sensitivities.

\begin{table}[t]
\begin{center}
    \begin{tabular}{@{\extracolsep{\fill}}c  c  c c  c}
    \toprule
         & \textbf{L} & \textbf{R} & \textbf{V} & Ref.\\[1ex]
        \hline 
        \multirow{ 4}{*}{$\varepsilon^{e}_{ee}$} & $[-0.021, 0.052]$ & $[-0.18, 0.51]$ &-- &  SK \& KamLAND \cite{Bolanos:2008km}
        \\[0ex]
        & $[-0.046, 0.053]$ & $[-0.21, 0.16]$ &-- &  Borexino Phase-I \cite{Agarwalla:2012wf}
        \\[0ex]
        &-- &-- & $[-0.56, 0.24]$&  Borexino \& COHERENT \cite{Dutta:2020che} \\[0ex]
        & \begin{minipage}{3.0cm} $[-1.37, -1.29]\ \oplus$ \\ $[0.03, 0.06]$ \end{minipage}
        & $[-0.23, 0.07]$ & $[-0.09, 0.14]$&  \textcolor{borexino_fit}{Borexino Phase-II}  \cite{Coloma:2022umy} \\[0ex]
        & --
        & -- & $[-2.65, 0.78]$ &  \textcolor{xnt22}{XENONnT 2022 \textbf{(this work)}} \\[0ex]\hline
        \multirow{5}{*}{$\varepsilon^{e}_{\mu\mu}$} & -- & -- & $|\varepsilon^{e}_{\tau\tau}- \varepsilon^{e}_{\mu\mu}| < 0.097$ &  SK atm.~\cite{Gonzalez-Garcia:2011vlg}
        \\[0ex] 
        & $[-0.03, 0.03]$ & $[-0.03, 0.03]$ & -- & react. + acc. \cite{Davidson:2003ha,Barranco:2007ej}
        \\[0ex] 
        &-- &-- & $[-0.58, 0.72]$&  Borexino \& COHERENT \cite{Dutta:2020che}
        \\[0ex]
        & \begin{minipage}{3.0cm}$[-0.20, 0.13]\  \oplus$ \\ $[0.58, 0.81]$\end{minipage}
        & $[-0.36, 0.37]$ & $[-0.51, 0.35]$& \textcolor{borexino_fit}{Borexino Phase-II}   \cite{Coloma:2022umy} \\[0ex]
        & -- 
        & -- & $[-2.19, 2.34]$ &\textcolor{xnt22}{XENONnT 2022 \textbf{(this work)}} \\[0ex]\hline
        \multirow{6}{*}{$\varepsilon^{e}_{\tau\tau}$} & $[-0.12 ,0.060]$ & $[-0.99, 0.23]$ & --&  SK \& KamLAND \cite{Bolanos:2008km}
        \\[0ex] 
        & $[-0.23 ,0.87 ]$ & $[-0.98 ,0.73]$ &-- &  Borexino Phase-I \cite{Agarwalla:2012wf}
        \\[0ex]
        & -- &-- & $|\varepsilon^{e}_{\tau\tau}- \varepsilon^{e}_{\mu\mu}| < 0.097$ &  SK atm.~\cite{Gonzalez-Garcia:2011vlg}
        \\[0ex] 
        &-- &-- & $[-0.60, 0.72]$&  Borexino \& COHERENT \cite{Dutta:2020che}
        \\[0ex]
        & \begin{minipage}{3.0cm}$[-0.26, 0.26]\ \oplus$ \\  $[0.45, 0.86]$\end{minipage}
        & $[-0.58, 0.47]$ & $[-0.66, 0.52]$&  \textcolor{borexino_fit}{Borexino Phase-II}   \cite{Coloma:2022umy} \\[0ex]
        & -- 
        & -- & $[-2.09, 2.20]$ &\textcolor{xnt22}{XENONnT 2022 \textbf{(this work)}} \\[0ex]\hline
        \multirow{4}{*}{$\varepsilon^{e}_{e\mu}$} & $[-0.13, 0.13]$ &  $[-0.13, 0.13]$ &-- & react. + acc. \cite{Barranco:2007ej}
        \\[0ex]
        &-- &-- & $[-0.58, 0.60]$&  Borexino \& COHERENT \cite{Dutta:2020che}
        \\[0ex]
        & $[-0.17, 0.29]$ & $[-0.21, 0.41]$ & $[-0.34, 0.61]$&  \textcolor{borexino_fit}{Borexino Phase-II}   \cite{Coloma:2022umy} \\[0ex]
        & -- 
        & -- & $[-1.03, 1.41]$ &\textcolor{xnt22}{XENONnT 2022 \textbf{(this work)}} \\[0ex]\hline
        \multirow{5}{*}{$\varepsilon^{e}_{e\tau}$} & $[-0.33, 0.33]$ & 
        \begin{minipage}{3.0cm}$[ -0.28, -0.05]\ \oplus $ $[0.05, 0.28]$\end{minipage}
        & --& react. + acc. \cite{Barranco:2007ej}
        \\[0ex] 
        &-- & $[ -0.19, 0.19]$ &-- & TEXONO \cite{TEXONO:2010tnr}
        \\[0ex]
        & --& --& $[-0.60, 0.62]$&  Borexino \& COHERENT \cite{Dutta:2020che}
        \\[0ex]
        & $[-0.26, 0.23]$ & $[-0.35, 0.31]$ & $[-0.48, 0.47]$&  \textcolor{borexino_fit}{Borexino Phase-II}   \cite{Coloma:2022umy} \\[0ex]
        & -- 
        & -- & $[-1.26, 1.11]$ &\textcolor{xnt22}{XENONnT 2022 \textbf{(this work)}} \\[0ex]\hline
        \multirow{6}{*}{$\varepsilon^{e}_{\mu\tau}$} & -- &-- & $[- 0.035, 0.018]$ & SK atm.~\cite{Gonzalez-Garcia:2011vlg}
        \\[0ex] 
        & -- &-- & $[- 0.20,  0.07]$ & MINOS \cite{MINOS:2013hmj}
        \\[0ex] 
        & -- & --& $[- 0.018, 0.016]$ & IceCube \cite{Salvado:2016uqu}
        \\[0ex] 
        & -- & -- & $[-0.67, 0.62]$&  Borexino \& COHERENT \cite{Dutta:2020che}
        \\[0ex]
        & \begin{minipage}{3.0cm}$[-0.62, -0.52]\ \oplus $\\ $[-0.09, 0.14]$ \end{minipage}
        & $[-0.26, 0.23]$ & $[-0.25, 0.36]$&  \textcolor{borexino_fit}{Borexino Phase-II}   \cite{Coloma:2022umy} \\[0ex]
        & -- 
        & -- & $[-1.57, 1.50]$ &\textcolor{xnt22}{XENONnT 2022 \textbf{(this work)}} \\[0ex]\hline
    \end{tabular}
\end{center}

\vspace{-3ex}
\caption{
\label{table:nsi_el}  Limits on electron NSI, most extracted from Ref.~\cite{Farzan:2017xzy}. The limits derived from our analysis of the recent XENONnT ER results are highlighted in orange. The limits derived by Ref.~\cite{Coloma:2022umy} using the Borexino Phase-II data, which we compare to in \cref{fig:dd-nsi-cevns_ER}, are highlighted in green.
}
\end{table}

We show our results in \cref{fig:dd-nsi-cevns_ER}, where we have filled contours for ER (pink/red colours, dotted lines for projections) and NR (blue/purple colours, dashed lines for projections) analyses. In order to place these sensitivities in the wider experimental context, we take the recent results of Ref.~\cite{Coloma:2022umy}, which used the spectral data from Phase-II of the Borexino experiment \cite{Borexino:2017rsf} to constrain $\varepsilon^e_{\alpha \beta}$. As Ref.~\cite{Coloma:2022umy} does not mention the potential impact of either proton or neutron NSI on neutrino oscillations, we assume that they have only considered NSI with the electron, with no contribution from either the proton or the neutron. As a result, we set $\eta = 0$, and we place their bounds at $\varphi = \pm \pi/2$, corresponding to electron-only NSI.

We note that, while previous studies have also constrained electron NSI, most of them place individual bounds on the left- and right-handed components of the interaction \cite{Bolanos:2008km,Agarwalla:2012wf,Gonzalez-Garcia:2011vlg,Davidson:2003ha,TEXONO:2010tnr}. For comparison with the previous literature, we tabulate many of these results alongside the corresponding allowed intervals derived in this work from the XENONnT ER data in \cref{table:nsi_el}. In this context, it is worth noting that Ref.~\cite{Dutta:2020che} previously considered  the potential impact of including \eves data from DUNE and a theoretical high-exposure DD experiment on global fit results of electron NSI. We investigate this impact in more detail by computing the solar neutrino scattering rate via the coherent treatment of oscillation effects in the density matrix formalism in~\cref{eq:dr_gen}, considering both \cevns and \eves, and studying the non-trivial behaviour of DD sensitivities in the full plane of charged NSI $(\varepsilon^e_{\alpha\beta}\varepsilon^p_{\alpha\beta})$ by means of our parametrisation in~\cref{fig:dd-nsi-cevns_ER}. Finally, comparing our XENONnT limits and future projections to the limits derived using Borexino data in Refs.~\cite{Dutta:2020che} and~\cite{Coloma:2022umy}, from~\cref{table:nsi_el} we see that while current DD data sets are not able to yield competitive bounds,  next-generation and far-future DD experiments will be able to improve on current limits.

As in \cref{fig:nsi-borexino,fig:dd-nsi-cevns_NR}, we show the points of parameter space where the adiabatic limit may no longer be valid for neutrino propagation in the Sun. Since $\gamma$ is energy-dependent, this region is different for the values of $E_\nu$ probed by NR and ER analyses. In light grey, we show the regions relevant for NRs ($E_\nu=16\,{\rm MeV}$ as in \cref{fig:dd-nsi-cevns_NR}) and in dark grey we show the regions for ERs ($E_\nu = 1\,{\rm MeV}$), which roughly correspond to the highest energy of $^7$Be neutrinos. This is not the primary neutrino source for the \eves signal ($pp$), but it does contribute at higher energies. Taking this value  is a conservative choice since higher energies correspond to a greater violation of adiabaticity. This is reflected in the fact that the dark grey regions, if present, are contained within the light grey regions in \cref{fig:dd-nsi-cevns_ER}.

 \cref{fig:dd-nsi-cevns_ER} demonstrates that next-generation and far-future DD experiments will form powerful probes of electron NSI, with almost all of our projections cutting into portions of the bounds placed with the Borexino experiment. DARWIN can give us considerably more sensitivity to all NSI parameters, showcasing its excellent potential in searching for new physics in the neutrino sector. However, such potential exhorts substantial efforts in background modelling and reduction, something which is already well underway in the respective collaborations.

As in the \cevns case, the limits for the \nue case exhibit  blind spots where the predicted rate is indistinguishable from the SM expectation. Once again, this weakens  the limits at certain values of $\varphi$ and a series of bands where DD experiments appear to lose sensitivity. Here, the complementarity with the NR analysis can be seen explicitly, since at precisely $\varphi=0$, the NSI effect on the NR signal is maximal. However, there are two notable physical differences between the \nue case and the \cevns case, arising from both the different \cevns and \nue cross sections and the way in which non-standard matter effects enter.

Firstly, we have no strong cancellation in the ER limits. While one might expect a complete loss of sensitivity when $\varphi = 0$, where the \nue cross section is unchanged by the presence of NSI, neutrino oscillations are still impacted by  the NSI contribution to the matter Hamiltonian from the nucleons (in this case only the proton since $\eta = 0$). Thus, for high enough values of $\varepsilon_{\alpha\beta}^{\eta,\varphi}$, the effect of NSI on the neutrino flavour fractions is large enough to give us an observable deviation from the SM expectation. This is analogous to what we saw in our Borexino analysis of \cref{subsec:osc}. Consequently, while we do lose sensitivity in ERs as $\varphi$ approaches zero, our limits ultimately reach a finite value.\footnote{Note that this is only possible as the cross section for electron neutrinos contains the extra CC contribution, making it different from that of the muon and tau neutrinos. Changes in the electron neutrino fraction then lead to measurable changes in the total number of CC interactions in the detector; the NC interactions from all flavours, on the other hand, remain equal.}

Secondly, we have fewer bands of insensitivity in ER over $\varphi$ than we did for the NR case over $\eta$. The location of these bands can be calculated through identical arguments to the \cevns case, whereby those values of the NSI parameters where the NSI-augmented rate is equal to the expected SM rate, $\mathrm{d}R/\mathrm{d}E_R - \mathrm{d}R/\mathrm{d}E_R|_\mathrm{SM} =0$, are found. The derivation of these cancellation conditions again crucially hinges on the coherent treatment of the neutrino propagation via the density matrix formalism in~\cref{eq:dr_gen}.  Critically, the condition for cancellation in the off-diagonal NSI elements $\varepsilon_{\alpha\beta}$  is completely missed in the simplified treatment of the rate as the sum over the oscillation probabilities times scattering cross section, $\sum_\alpha P_{e\alpha}\, \mathrm{d}\sigma_{\nu_{\alpha T}}/\mathrm{d}E_R$.

In the case that only one diagonal NSI element, $\varepsilon_{\alpha\alpha}^{\eta,\varphi}$, is active, the cancellation equation for the differential rate for \nue reads,
\begin{multline}\label{eq:cancel_diag_er}
    \int_{E_{\nu}^{\mathrm{min}}}\frac{\mathrm{d}\phi_{\nu_{e}}}{\mathrm{d}E_{\nu}}\ \rho_{\alpha\alpha} \ \Bigg\{ \left(1 - \frac{E_{R}}{E_{\nu}}\left(1+\frac{m_e-E_R}{2E_\nu}\right)\right) \left[ 4 \,s_W^2 +\xi^{e}\,\varepsilon_{\alpha\alpha}^{\eta,\varphi}  \right] \,\xi^{e}\,\varepsilon_{\alpha\alpha}^{\eta,\varphi}  \\
    + \left(1-\frac{m_e\,E_R}{2\,E_\nu^2}\right) \left[4 \,s_W^2\,\frac{\rho_{ee}-\rho^\mathrm{SM}_{ee}}{\rho_{\alpha\alpha}} + \left(2\,\delta_{\alpha e} - 1 \right) \xi^{e}\,\varepsilon_{\alpha\alpha}^{\eta,\varphi}  \right] \Bigg\} \,\mathrm{d}E_{\nu}= 0\,.
\end{multline}
On the other hand, for off-diagonal NSI, $\varepsilon_{\alpha\beta}^{\eta,\varphi}$ with $\alpha\neq\beta$, the cancellation condition can be expressed as
\begin{multline}\label{eq:cancel_odiag_er}
    \int_{E_{\nu}^{\mathrm{min}}}\frac{\mathrm{d}\phi_{\nu_{e}}}{\mathrm{d}E_{\nu}}\ \Bigg\{ \left(1 - \frac{E_{R}}{E_{\nu}}\left(1+\frac{m_e-E_R}{2E_\nu}\right)\right)\left[ \left(\xi^{e}\,\varepsilon_{\alpha\beta}^{\eta,\varphi}\right)^2\,\left(\rho_{\alpha\alpha} + \rho_{\beta\beta}\right) + 8 \,s_W^2\,\xi^{e}\,\varepsilon_{\alpha\beta}^{\eta,\varphi}\,\rho_{\alpha\beta}\right] \\
    + \left(1-\frac{m_e\,E_R}{2\,E_\nu^2}\right)\,\left[4\, s_W^2\,\left(\rho_{ee} -\rho^\mathrm{SM}_{ee}\right) - \delta_{\alpha\mu} \delta_{\beta\tau}\ 2 \,\xi^{e}\,\varepsilon_{\alpha\beta}^{\eta,\varphi}\, \rho_{\alpha\beta}  \right] \Bigg\} \,\mathrm{d}E_{\nu}= 0  \,,
\end{multline}
where the last term in the second line is only present for $\alpha\beta=\mu\tau$. In the above expressions, $\rho^\mathrm{SM}$ refers to the density matrix obtained in the SM case (i.e.~$\varepsilon_{\alpha\beta}=0$) and $\rho$ to the one obtained with non-zero NSI elements. As can be seen in~\cref{fig:dd-nsi-cevns_ER}, for \eves we obtain insensitivity bands similar to those in the \cevns case for the projected limits; however we only see this for $\varepsilon_{ee}^{\eta,\varphi}$ and $\varepsilon_{e\mu}^{\eta,\varphi}$.  In most cases, DD sensitivities are not good enough to reach these cancellation regions where the SM scattering rate is recovered.

Finally, it is worth noting that, for very small but non-zero $\varphi$, the line of exact cancellation has a very sharp zero-transition from very large (positive) to very small (negative) values of $\varepsilon_{\alpha\beta}^{\eta,\varphi}$ (or vice versa) making it seem like there exists an asymptote at $\varphi=0$. This effect is similar to that exhibited by \cref{fig:nsi-borexino} in our analysis of Borexino data. In this region of parameter space, the NSI-only term is negligible since it is quadratic in $\xi^e$, and thus $\varphi$. The reason for this change in sign is a rapid flattening of the SM-NSI interference terms in~\cref{eq:cancel_diag_er,eq:cancel_odiag_er}, which are linearly proportional to $\xi^e$, and thus $\varphi$. This flattening leads to a rapid change in the value of $\varepsilon_{\alpha\beta}^{\eta,\varphi}$ where the interference term cancels off the residual SM-like term proportional to $\rho_{ee} -\rho^\mathrm{SM}_{ee}$ and hence restores the SM neutrino rate.

The bounds from the global analysis of Ref.~\cite{Coloma:2019mbs} are shown in \cref{fig:dd-nsi-cevns_ER} as they were in \cref{fig:dd-nsi-cevns_NR}, but now only the proton direction ($\eta=0, \varphi=0$) is visible. By plotting the NR analyses, we can see how the sensitivities of \cref{fig:dd-nsi-cevns_NR} extend into the $\varphi$ direction. We observe the same regions of nonzero $\varepsilon^{\eta,\,\varphi}_{\alpha\beta}$ where our xenon-based DD experiments lose sensitivity. For the diagonal elements, this region simply follows from \cref{eq:blind-band-on} and, for the off-diagonal elements, the more complicated behaviour is expressed in \cref{eq:blind-band-off}. Furthermore, the off-diagonal elements again exhibit some non-trivial behaviour in the form of `kinks' (see for example $\varepsilon^{\eta,\,\varphi}_{e\mu}\approx 1.0$ and $\varphi\approx -5/16\,\pi$). The appearance of these kinks is analogous to those observed in the NR sensitivities in the nucleon plane, as described at the end of~\cref{sec:nuc_plane}.

We re-iterate that the limits presented in \cref{fig:dd-nsi-cevns_NR,fig:dd-nsi-cevns_ER} have been calculated by switching on only one NSI parameter at a time. Due to potential interference effects between different NSI parameters, a global analysis that allows all NSI parameters to vary, before marginalising to compute the limits on any one parameter, would generally lead to weaker bounds \cite{Dutta:2020che,Coloma:2022avw}. However, the point of our study is to illustrate the potential of DD experiments in this direction. Our study makes a strong case for their inclusion in future global analyses.

\subsection{Final Remarks}
\label{subsec:discuss_results}

Having access to both the NR and ER signals in one experiment makes DD incredibly powerful as a probe for NSI. As far as we are aware, this is the only experimental technology that is able to perform such simultaneous analyses. For example, if a signal inconsistent with the SM is detected in the future, both channels would be pivotal for exploring the possible values of $\eta$ and $\varphi$, or, equivalently, the relative strength of NSI with electrons, protons and neutrons. This will come in tandem with other more traditional searches for new physics in the neutrino sector. However, given the number of parameters one is trying to constrain or fit, the addition of DD will provide important input complementary to that of oscillation and spallation source experiments.

Above, we have treated the NR and ER signals in DD experiments as separable. Indeed, in the name of background discrimination for DM searches, DD experiments are capable of this for large parts of the signal region. Taking into account experimental inputs, as described in \cref{eq:n_dd}, we are able to model NR and ER spectra accurately without resorting to a full Monte Carlo simulation of the detector responses in terms of S1 (scintillation) and S2 (ionisation) signals. Since detector responses from the point of interaction, be it NR or ER, have been well studied and calibrated within experimental collaborations, we are confident that introducing nonzero NSI will not alter the expectation that future experiments will be able to resolve S1 and S2 signals.

Interestingly, many DD collaborations also perform S2-only analyses, which has the benefit of lowering the experimental threshold $E_{\rm th}$, increasing the sensitivity to lighter DM values. However, this comes at the cost of losing NR/ER discrimination. In our NR projections for future experiments, we took the liberty of lowering $E_{\rm th}$ in a modest way, assuming that NR/ER discrimination was still possible, and indeed S2-only analyses boast much lower thresholds. As this choice implies, reducing $E_{\rm th}$ is beneficial for the NR signal, but not necessarily for the ER signal. This is because low-energy neutrinos are unable to impart sufficient energy to excite the bound electrons. It is likely then that an S2-only analysis will only improve the prospects of the NR signals, but one would then have to account for larger background rates.

Furthermore, in our analysis of DD experiments, we have not included argon-based liquid experiments. This direction is not without its potential, but we leave incorporation of such experiments for future work. As can be seen in Ref.~\cite{Amaral:2020tga}, the prospects for argon detectors are not as promising as those for xenon. However, Ref.~\cite{Amaral:2020tga} only considered the implications for specific BSM scenarios. The limiting factor for argon detectors seemed to be the experimental threshold and increased ER backgrounds, both of which tend to be much higher than their xenon counterparts. Recent progress from the DarkSide collaboration indicates that  argon detectors may be able to provide competitive bounds in the future~\cite{DarkSide:2018kuk,DarkSide:2022dhx,DarkSide-50:2022qzh}.

To our knowledge, this work is the first to derive dedicated limits on NSI from both \cevns and \eves in DD experiments from the recent first results XENONnT~\cite{XENONCollaboration:2022kmb} and LZ~\cite{LZ:2022ufs}. Our analysis of future multi-ton LXe detectors in this section has exposed the huge potential that DD experiments have in fully exploring the parameter space of NSI, especially due to their increased sensitivity to \eves. While there have been some initial studies considering the \eves signals for non-zero NSI~\cite{Dutta:2017nht,Dutta:2020che}, we take a comprehensive approach by considering  the solar \cevns and \eves signals, modelling the experimental setups as close to the experimental collaborations as possible, and treating solar neutrino propagation in the coherent density matrix approach. Combined with our convenient parametrisation of the NSI parameter space, this allows us to derive an accurate overview of DD sensitivities and blind spots in the entire NSI parameter space. Moreover, as we pointed out in the previous section, since the blind spots for \cevns and \eves do not coincide in the charged NSI plane, DD experiments using a combination of both signatures can effectively avoid these and remain sensitive for most of this region.

When presenting our extended parametrisation in \cref{sec:framework}, we introduced the axial-vector NSI coupling $\tilde{\varepsilon}_{\alpha\beta}^{f}$, only to set it to zero because it does not contribute to matter effects. Similarly, the effect of $\tilde{\varepsilon}_{\alpha\beta}^{f}$ on NRs will be minimal because of the coherent nucleon number enhancement that the vector current receives over the axial-vector. This enhancement is no longer present when one considers ERs, and would constitute an additional set of parameters that one could probe. The sensitivity of DD experiments to such axial-vector NSI is an interesting direction to be studied in future work.

Finally, we comment on the particle physics interpretation of the most promising projections we report in this work. Namely, the potential for a DARWIN-like experiment to probe NSI at the level of $\varepsilon^{\eta,\,\varphi}_{ee}\sim 10^{-2}$. Reinterpreting this value in a more canonical EFT approach implies $\Lambda_{\rm NP}/\sqrt{C}\sim 1.8\, {\rm TeV}$, where $C$ is the Wilson coefficient of the four-fermion operator. We see here that for $C>1$ these experiments have the capability to probe new physics above the TeV scale. In this context, it would presumably make the most sense to embed NSI analyses within the more general SMEFT framework. It would be interesting to study whether this can already be done in a consistent way. SMEFT observables are typically at collider scales, while NSI studies are much below, so the effective approach is appropriate for a greater range of $\Lambda_{\rm NP}$.

\section{Conclusions}
\label{sec:conclusions}

We have demonstrated that direct detection experiments will soon become powerful probes of neutrino non-standard interactions, testing the parameter space in a complementary way to spallation source and oscillation experiments. This owes to their simultaneous sensitivity to nuclear and electron recoils and their unique capability to test tau neutrinos from the solar neutrino flux.

To do so, we have developed an extension of an earlier NSI parametrisation, allowing for non-standard interactions with nucleons and electrons simultaneously. Our parametrisation captures the rich phenomenology that arises when one allows for NSI to impact both neutrino propagation and neutrino scattering. We have shown that previous NSI limits from spallation source experiments, such as CENNS-10 LAr, and oscillation experiments, such as Borexino, map non-trivially to this extended parameter space, demonstrating the importance of allowing for a variable NSI contribution from the proton and the electron.

We have derived current direct detection constraints and projected the sensitivities of future direct detection experiments on the NSI landscape using the expected solar neutrino rate. We have thoroughly studied the resulting bounds in the different NSI directions by taking into account both \cevns and \eves signals, accurately modelling the experimental setups and consistently treating the coherent neutrino propagation via the density matrix. Furthermore, we have identified the potential blind spots where sensitivity is lost due to cancellations in the expected rate.

While current leading constraints from LZ and XENONnT are not competitive in this landscape yet, we have shown that those from future experimental runs and the projected DARWIN detector will cut into new regions of the NSI parameter space. We believe that the conclusion is clear: upcoming multi-ton, LXe-based DD experiments are poised to make a considerable impact in the neutrino NSI landscape. We therefore recommend that they be included in future global NSI studies, incorporating a more complete treatment of the systematics.

\section*{Acknowledgements}

We want to thank Michele Maltoni for many insightful discussions on neutrino oscillations and non-standard interactions. We also thank Felix Kahlhoefer for the invaluable help with the LZ implementation, as well as Christropher Tunnell and Aar\'{o}n Higuera for valuable discussions regarding XENONnT.
Finally, we are also grateful to Pilar Coloma, Enrique Fern\'andez Mart\'\i nez, Danny Marfatia, Iv\'an Mart\'\i nez-Soler, Pablo Mart\'\i nez-Mirav\'e, Yuber F.~P\'erez-Gonz\'alez and Thomas Schwetz for helpful discussions during the preparation of this manuscript.

DA is supported by the National Science Foundation under award 2209444. 
DGC acknowledges support from the Spanish Ministerio de Universidades under grant SI2/PBG/2020-00005. 
AC is supported by the grant ``AstroCeNT: Particle Astrophysics Science and Technology Centre" carried out within the International Research Agendas programme of the Foundation for Polish Science financed by the European Union under the European Regional Development Fund.
PF would like to express special thanks to the Mainz Institute for Theoretical Physics (MITP) of the Cluster of Excellence PRISMA+ (Project ID 39083149), for its hospitality and support. The work of PF was partially supported by the UKRI Future Leaders Fellowship DARKMAP. 
This work is partially supported by the Spanish Agencia Estatal de Investigaci\'on through the grants PID2021-125331NB-I00 and CEX2020-001007-S, funded by MCIN/AEI/10.13039/501100011033.

\section*{APPENDICES}

\appendix

\section{Solar neutrino transition rate}
\label{sec:amplitude}

In writing~\cref{eq:dr_gen}, we automatically retain the full phase correlation of the different solar neutrino flavour states reaching the detector. The way to understand how this formula comes about is to consider the amplitude for the combined propagation $\nu_\alpha\to\nu_\gamma$ of solar neutrinos from the point of production to the detector, and the scattering $\nu_\gamma\, T\to \nu_f\, T$ of the propagated neutrino $\nu_\gamma$ with the target material $T$ into any final state neutrino $\nu_f$, which is then lost. 
In quantum field theory, asymptotic \textit{in}- and \textit{out}-states are definite momentum states of the free theory, and as such they are on mass-shell. Hence, they describe well-defined mass eigenstates in the far past and far future.
For our scattering of the neutrino $\nu_\gamma$ on  the target $T$, we thus have to sum over all possible mass eigenstates $\nu_i$ in the \textit{out}-state.\footnote{We would like to express our special thanks to Thomas Schwetz for clarifying discussions on this point.} Hence, the amplitude square of the process $\nu_\alpha \to \sum_i\nu_i$ can be written as
\begin{equation}  \left|\mathcal{A}_{\nu_\alpha\to\sum_i\nu_i}\right|^2 = \sum_i |\langle\nu_i|S|\nu_\alpha\rangle|^2 =  \sum_i \Bigg|\sum_\beta U^*_{\beta i}\,\langle\nu_\beta|S|\nu_\alpha\rangle\Bigg|^2\,,
\end{equation}
where we have factored out the nuclear part of the elastic scattering process. Here, $S$ is the $S$-matrix for the full process, and we have decomposed the neutrino mass eigenstate $\nu_i$ into its flavour components, $|\nu_i\rangle=\sum_\beta U_{\beta i} |\nu_\beta\rangle$, in terms of the PMNS matrix $U_{\beta i}$.
Then, the amplitude can be decomposed as
\begin{align}
    \left|\mathcal{A}_{\nu_\alpha\to\sum_i\nu_i}\right|^2 =&  \sum_i \Big|\sum_\beta U^*_{\beta i}\,  \langle  \nu_\beta | S_\mathrm{int} \left(\sum_\gamma| \nu_\gamma \rangle \langle \nu_\gamma| \right) S_\mathrm{prop} | \nu_\alpha \rangle \Big|^2 \\
    =& \sum_{\beta,\gamma,\delta,\lambda}\ \overbrace{\sum_i  U^*_{\beta i} U_{\lambda i}}^{\delta_{\beta\lambda}}\  \langle \nu_\beta | S_\mathrm{int} | \nu_\gamma \rangle \langle \nu_\gamma | S_\mathrm{prop} \left(\sum_\rho | \nu_\rho \rangle \langle \nu_\rho |\right) | \nu_\alpha \rangle \langle \nu_\alpha | \left(\sum_\sigma | \nu_\sigma \rangle \langle \nu_\sigma |\right) S_\mathrm{prop}^{\dagger} | \nu_\delta \rangle \notag\\
    & \qquad \quad \times\langle \nu_\delta | S_\mathrm{int}^{\dagger} | \nu_\lambda \rangle \\[8pt]
    =& \sum_{\gamma,\delta,\rho,\sigma} \underbrace{ (S_\mathrm{prop})_{\gamma\rho}\ \pi^{(\alpha)}_{\rho\sigma} \, (S_\mathrm{prop})^*_{\delta\sigma}}_{\equiv\rho^{(\alpha)}_{\gamma\delta}} \ \underbrace{\sum_{\beta} (S_\mathrm{int})^*_{\beta\delta} \, (S_\mathrm{int})_{\beta\gamma}  }_{\mathcal{M}^*(\nu_\delta\to f)\, \mathcal{M}(\nu_\gamma\to f)}\,,\label{eq:ampl_final}
\end{align}
where in the second to last line we have used the unitarity of the PMNS matrix, $\sum_i  U^*_{\beta i} U_{\lambda i}=\delta_{\beta\lambda}$.
Here, $\pi^{(\alpha)}$ is the projector onto the neutrino-flavour state $|\nu_\alpha\rangle$.
In the second line, we have separated the $S$-matrix into $S_\mathrm{prop}$,  describing the propagation of the initial neutrino $\nu_\alpha$ from the source to the detector, and $S_\mathrm{int}$,  describing  the interaction with the detector material.
Thus, decorating the expression~\cref{eq:ampl_final} with the relevant phase-space factors for the generalised cross section (cf.~\cref{eq:gen_xsec}), we finally find that
\begin{equation}
    \left|\mathcal{A}_{\nu_\alpha\to\sum_i\nu_i}\right|^2 \propto \mathrm{Tr}\left[\bb{\rho}^{(\alpha)}\, \diff{\bb{\zeta}}{ E_R}\right] \,.
\end{equation}

\newpage

\section{$\Delta \chi^2$ Plots for CENNS-10 LAr and Borexino}
\label{sec:app_delta}
\vspace*{-1cm}
\begin{figure}[h!]
    \includegraphics[]{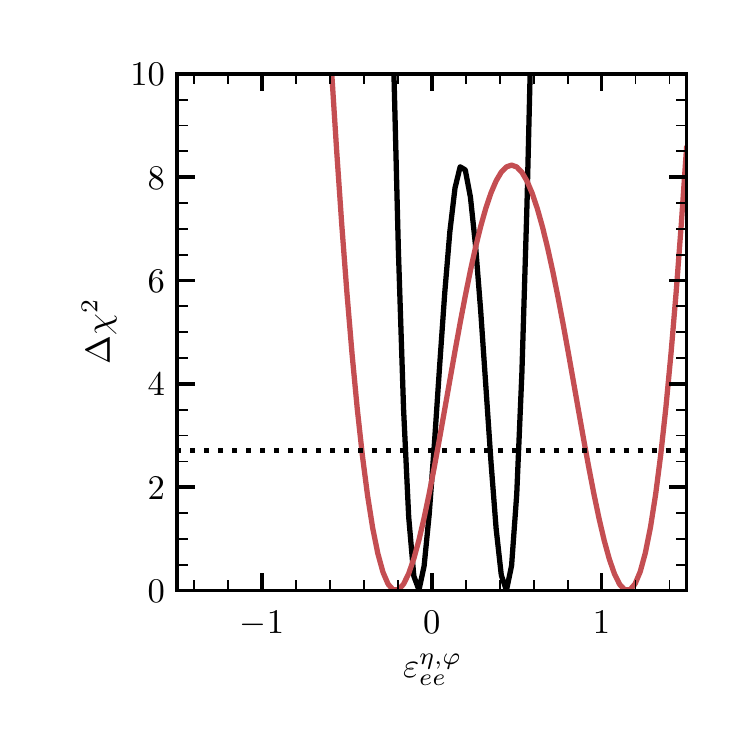}
    \includegraphics[]{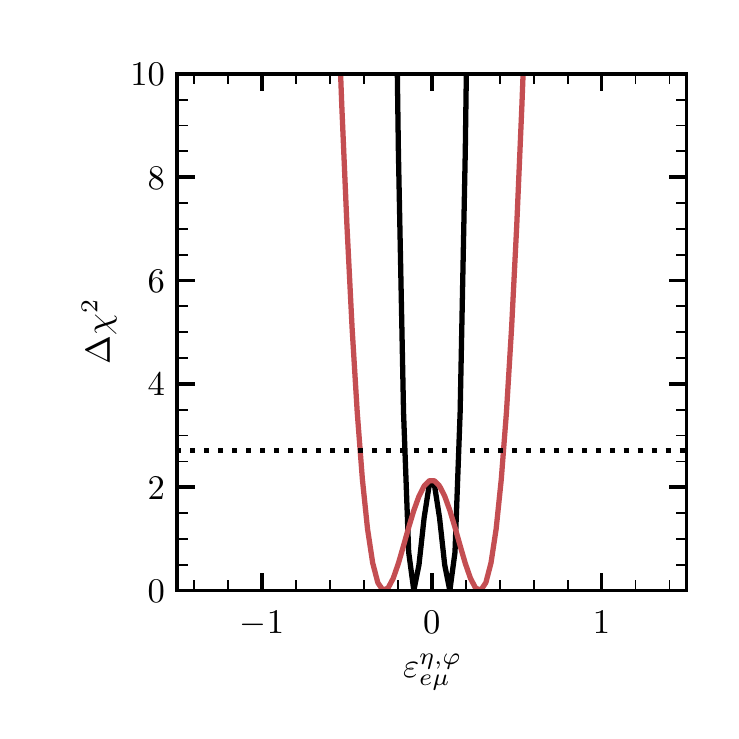}

    \vspace*{-7ex}
    \includegraphics[]{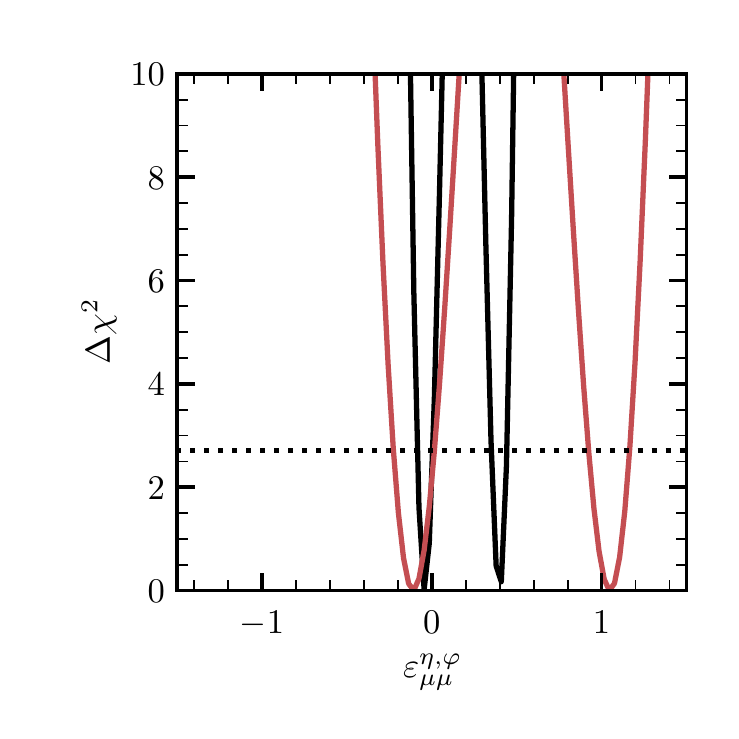}
    \includegraphics[]{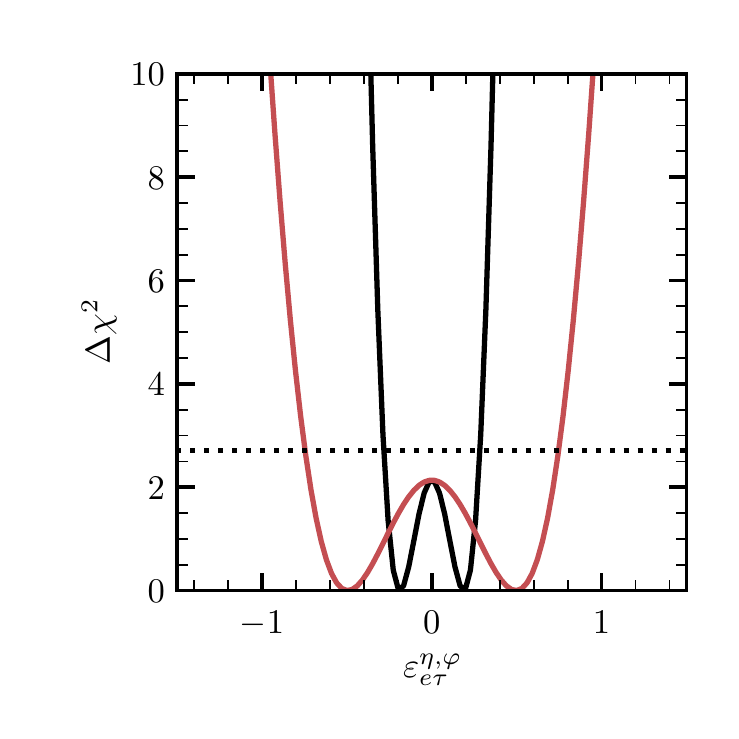}

    \vspace*{-7ex}
    \includegraphics[]{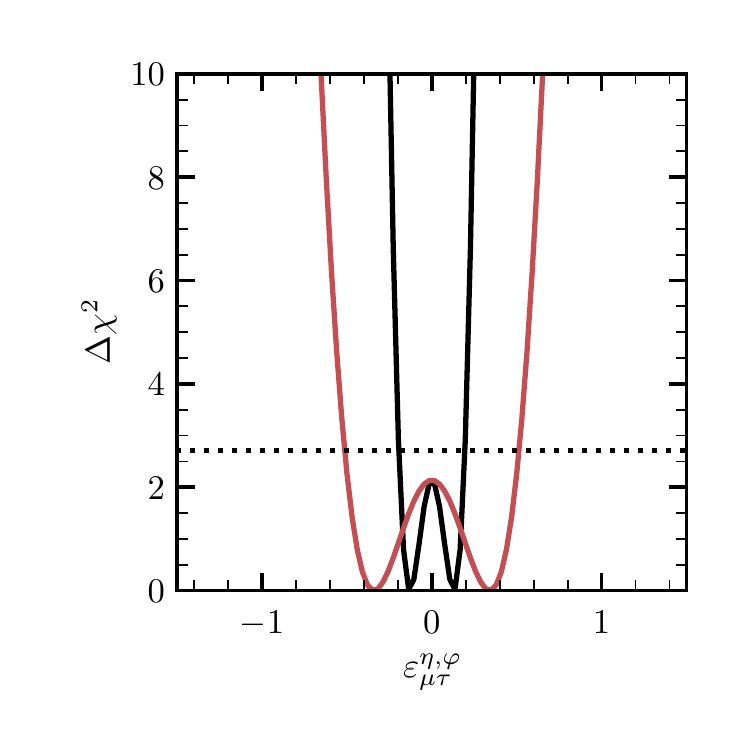}

    \vspace*{-30ex}
    \hspace*{70ex}
    \includegraphics[]{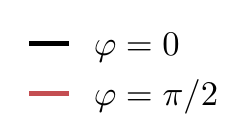}
    
    \vspace*{15ex}
    \caption{The variation in the $\Delta \chi^2$ statistic in our CENNS-10 LAr analysis under two assumptions for $\varphi$: $\varphi = 0$ (black) and  $\varphi = \pi/2$ (red). We have fixed $\eta = \tan^{-1}(1/2)$, corresponding to a pure up-quark NSI when $\varphi = 0$. The dashed line shows where $\Delta \chi^2 = 2.71$, where we draw our $90\%$ CL limit.}
    \label{fig:chi2_cenns}
\end{figure}

\begin{figure}[p!]
    \includegraphics[]{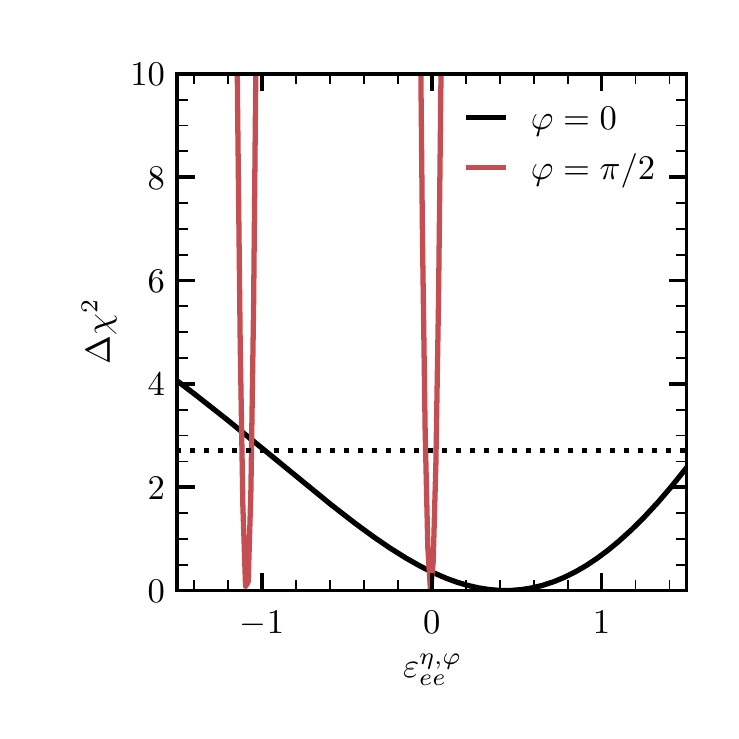}
    \includegraphics[]{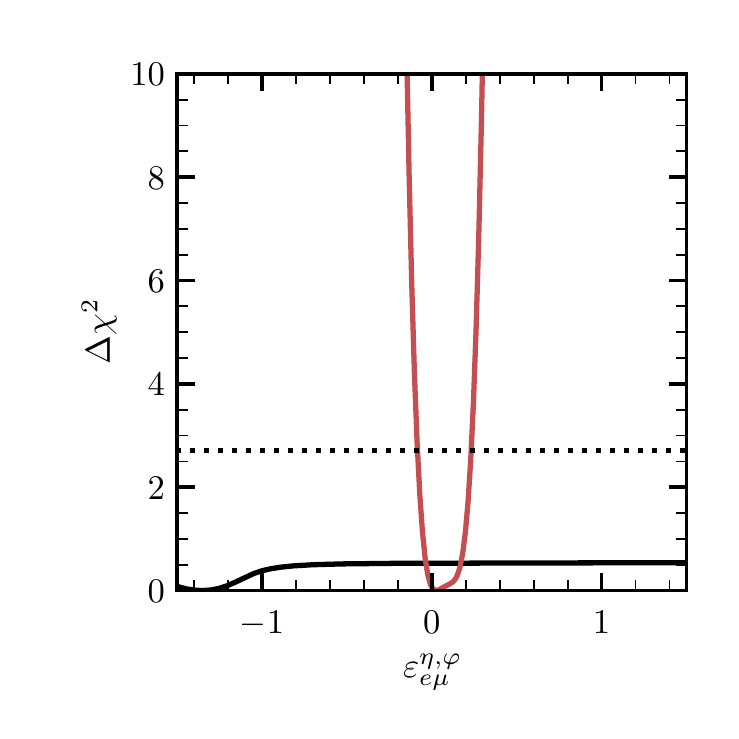}

    \vspace*{-7ex}
    \includegraphics[]{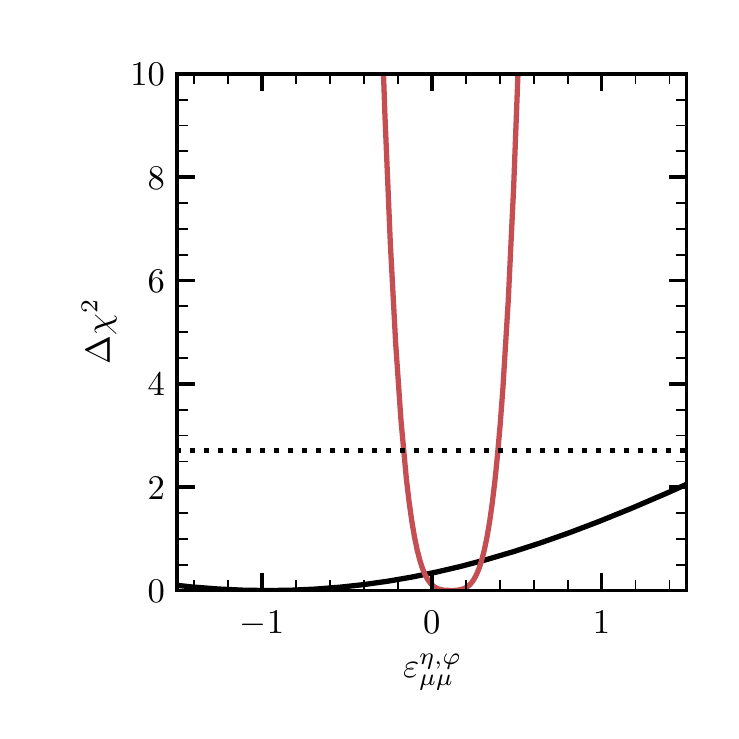}
    \includegraphics[]{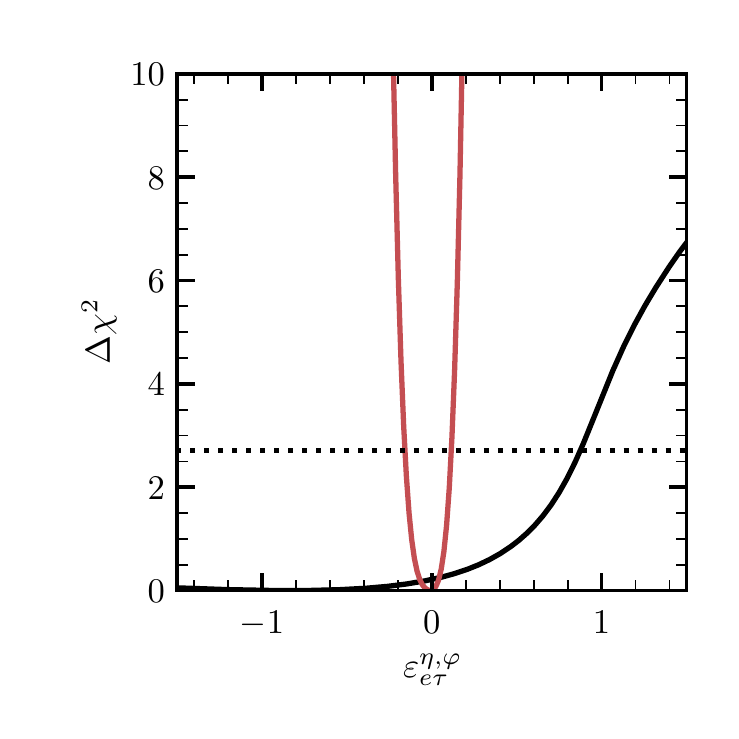}

    \vspace*{-7ex}
    \includegraphics[]{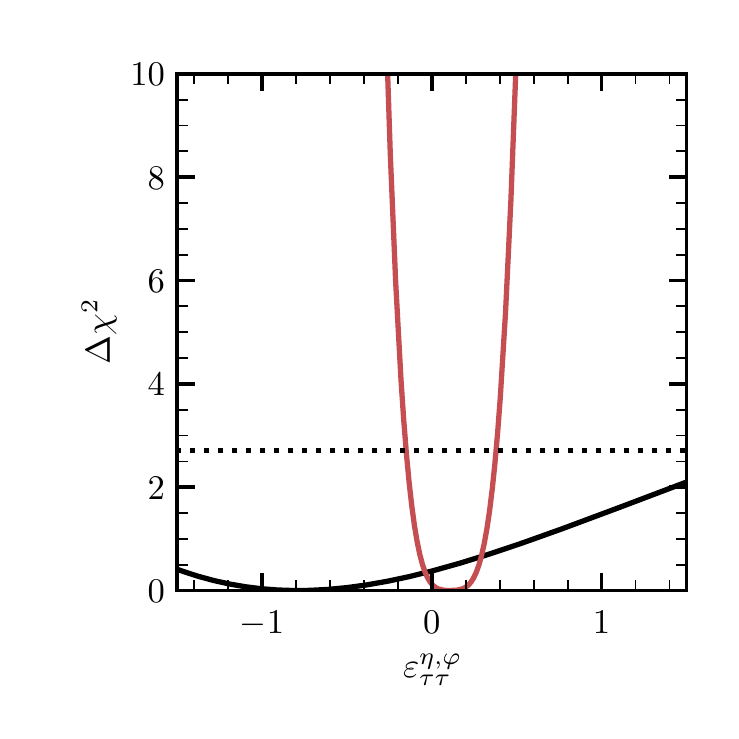}
    \includegraphics[]{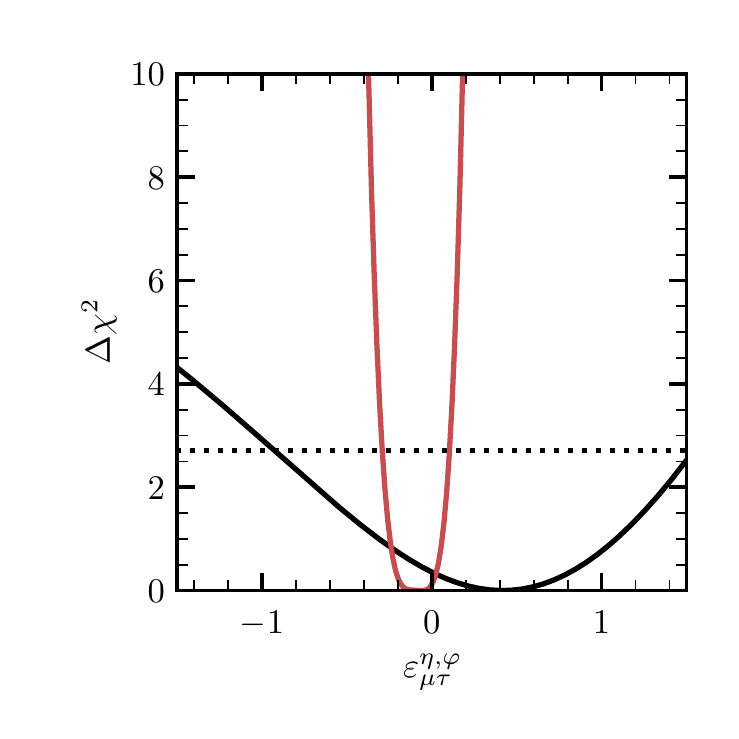}

    \caption{The variation in the $\Delta \chi^2$ statistic in our Borexino analysis under two assumptions for $\varphi$: $\varphi = 0$ (black) and $\varphi = \pi/2$ (red). We have fixed $\eta = 0$, corresponding to a pure proton NSI when $\varphi = 0$ and a pure electron NSI when $\varphi = \pi/2$. The dashed line shows where $\Delta \chi^2 = 2.71$, where we draw our $90\%$ CL limit.}
    \label{fig:chi2_borexino}
\end{figure}

\clearpage

\let\oldaddcontentsline\addcontentsline
\renewcommand{\addcontentsline}[3]{}
\bibliographystyle{JHEP}
\bibliography{literature}
\let\addcontentsline\oldaddcontentsline

\end{document}